\documentclass[11pt]{article} 
\usepackage{geometry}  
\usepackage{amsfonts}
\usepackage{xcolor}
\usepackage{graphicx}
\usepackage{amssymb}
\usepackage{amsmath}
\usepackage{txfonts}

\usepackage{enumerate}
\usepackage[round]{natbib}
\usepackage{hyperref}
\hypersetup{colorlinks=false}
\usepackage{url}
\usepackage{array}
\usepackage{multirow}
\usepackage{setspace}
\def\bSig\mathbf{\Sigma}

\newcommand{\mcl}[1]{\mathcal{#1}}

\def\log{\hbox{log}}

\def\pj2s{\phi^2_j(\Suu)}

\def\Suu{\boldSigma_{uu}}

\def\boxit#1{\vbox{\hrule\hbox{\vrule\kern6pt
          \vbox{\kern6pt#1\kern6pt}\kern6pt\vrule}\hrule}}

\def\bse{\begin{eqnarray*}}
\def\ese{\end{eqnarray*}}
\def\be{\begin{eqnarray}}
\def\ee{\end{eqnarray}}
\def\bq{\begin{equation}}
\def\eq{\end{equation}}
\def\bse{\begin{eqnarray*}}
\def\ese{\end{eqnarray*}}

\def\wh{\widehat}
\def\bbeta{\mbox{\boldmath $\beta$}}

\def\balpha{\mbox{\boldmath $\alpha$}}

\def\btheta{\mbox{\boldmath $\theta$}}
\def\bdelta{\mbox{\boldmath $\delta$}}

\def\bZ{\mbox{\boldmath $Z$}}

\def\bSigma{\mbox{\boldmath $\Sigma$}}

\def\bx{\mbox{\boldmath $x$}}
\def\ba{\mbox{\boldmath $a$}}

\def\bv{\mbox{\boldmath $v$}}

\def\bLambda{\mbox{\boldmath $\Lambda$}}

\def\bV{{\bf V}}
\def\bU{{\bf U}}
\def\bX{{\bf X}}

\def\boxit#1{\vbox{\hrule\hbox{\vrule\kern6pt
          \vbox{\kern6pt#1\kern6pt}\kern6pt\vrule}\hrule}}
 \def\bbE{\mathbb{E}}
\def\bbS{\mathbb{S}}
\def\bbP{\mathbb{P}}
 \newcommand{\qed}{\hfill \mbox{\raggedright \rule{.07in}{.1in}}}
\usepackage[figuresright]{rotating}

\begin{document}

\title{\Large \bf Control Function Assisted IPW  Estimation with a Secondary Outcome in Case-Control Studies 
}
\author
{Tamar Sofer$^{1}$,
Marilyn C. Cornelis$^2$, Peter Kraft$^{1,3}$, and 
 Eric J. Tchetgen Tchetgen$^{1}$ \\
$^{1}$Department of Biostatistics, Harvard School of Public Health, 
Boston, Massachusetts, U.S.A.\\
$^{2}$Department of Nutrition, Harvard School of Public Health,  Boston, Massachusetts, U.S.A.\\
$^{3}$Department of Epidemiology, Harvard School of Public Health,  Boston, Massachusetts, U.S.A.  }

\maketitle

\begin{abstract}
Case-control studies are designed towards studying associations between risk factors and a single, primary outcome. Information about additional, secondary outcomes is also collected,  but association studies targeting such secondary outcomes should account for the case-control sampling scheme, or otherwise results may be biased. Often, one uses inverse probability weighted (IPW) estimators to estimate population effects in such studies. However, these estimators are  inefficient relative to estimators that make additional assumptions about the data generating mechanism. 
We propose a class of estimators for the effect of risk factors on a secondary outcome in case-control studies,  when the mean is modeled using either the identity or the log link.  The proposed estimator combines IPW with a mean zero control function that depends explicitly on a model for the primary disease outcome. The efficient estimator in our class of estimators reduces to standard IPW when the model for the primary disease outcome is unrestricted, and is more efficient than standard IPW when the model is either parametric or semiparametric. 
\end{abstract}

%

\par\vfill\noindent


\newpage
\pagestyle{plain}
\setcounter{page}{1}
\doublespacing

\section{Introduction}
\label{sec:intro}

 Case-control studies are designed to study associations between exposures and a traditionally-rare, primary outcome. Recently, genome-wide association studies (GWAS) are routinely conducted using a  case-control study design, even when the primary disease outcome is relatively common, to increase power while maintaining relatively low cost. For instance, type 2 diabetes (T2D) is studied in a case-control GWAS study nested within the Nurses Health Study (NHS), and its prevalence  in the cohort is estimated to be 8.4\% \citep{Cornelis2012}. 
Such case-control studies typically collect information about additional, secondary outcomes, potentially associated with the primary disease. Specifically, Body Mass Index (BMI) measurements, which is well known to be associated with T2D, were collected in the T2D case-control study. We are interested in re-purposing the T2D GWAS data to study associations of Single Nucleotide Polymorphisms (SNPs) from the FTO gene, coding the Fat Mass and Obesity Protein, with BMI. 

As \citet{Nagelkerke1995} pointed out, and others later demonstrated \citep[for instance]{Jiang2006,Richardson2007,Wang2011}, applying standard regression methods to case-control data for  analysis of a secondary outcome can bias inference, and therefore analysts need to adapt analysis schemes. 

Several approaches have been proposed for the analysis of secondary outcomes from case-control studies. \citet{Nagelkerke1995}  suggested that using solely the control group will be valid if it is fairly representative of the general population. This happens when the disease is rare, but may not hold otherwise. 
\citet{Richardson2007} and \citet{Monsees2009} discussed using Inverse Probability Weighting (IPW), in which the contribution of each subject for the estimating equation is weighted by the inverse of its selection probability into the sample.  Using IPW  is  robust to the sampling bias, though it may be inefficient when, as typically the case in nested case-control studies, additional information can be obtained from the underlying cohort.  In such settings, Augmented inverse-probability (AIPW) can be used for efficiency gain \citep{Robins1994}. However, this is potentially true for other estimators as well, and we subsequently discuss ``stand-alone" case-control studies, i.e. assuming the underlying cohort from which cases and controls emanate, is not available to the analyst.
\citet{Lin2009} proposed to estimate model parameters by maximizing the retrospective likelihood,  taking into account case-control ascertainment. \citet{Li2012} generalized their approach and suggested an adaptively weighted estimate of the association between the exposure and a binary secondary outcome, via a weighted sum of two retrospective likelihood-based estimators that differ in their assumed disease model. \citet{Chen2013} proposed a bias correction formula for an estimated odds ratio parameter, so that one can fit a regression model for the marginal or conditional analysis of the secondary outcome, and correct the estimate using the result from regressing the primary outcome on the secondary outcome and the exposure. 
 Fewer methods are available for continuous secondary outcomes. \citet{Ghosh2013} also took a retrospective likelihood approach, extending the previous work mainly by incorporating auxiliary covariates. These likelihood based estimators rely heavily on distributional assumptions. \citet{Wei2013} modeled a continuous secondary outcome semiparametrically  and relaxed the distributional assumptions, but assumed that the primary disease is rare, which does not apply in many situations, including the T2D case-control study introduced earlier. 
\citet{TchetgenTchetgen2013} proposed a general model based on a nonparametric parameterization  for the secondary outcome conditional on disease status and covariates. One can use this framework to compute an estimator under parametric, semiparametric or nonparametric models. The estimator is semiparametric locally efficient, i.e. it achieves the semiparametric efficiency bound in the absence of model misspecification, and remains consistent and asymptotically linear even if the error distribution for the outcome is incorrectly specified, provided the specified mean structure is correct. Bias may result from an incorrect mean model.  The approach is developed for the identity, log and logit link functions.  An implication of the  parameterization proposed by \citet{TchetgenTchetgen2013} is that adding a disease  indicator to the regression design model (i.e., treating disease as a predictor or confounder) will bias effect estimates of the SNPs on the secondary outcome, unless either the conditional mean of the secondary outcome is almost identical in cases and in control (i.e. approximately no selection bias), or the primary disease is rare across all levels of the exposure, and the magnitude of the selection bias does not vary with covariates.


Current methodology  (1)  relies on distributional assumptions or (2) in the cases where fewer assumptions are made, proposed estimators are not necessarily efficient. Here, we use semiparametric theory to propose estimators for the population regression of the secondary outcome on covariates that are both robust and locally semiparametric  efficient.
We construct a control function in terms of a model for the primary disease risk conditional on covariates, and add it to the usual IPW estimating equation. We get a new estimating equation, which reduces to the usual IPW in the absence of any restriction on the model of disease risk given covariates. When this model is  (semi)parametric, our proposed estimator is more efficient than IPW. Interestingly, we show that the parameterization proposed by \citet{TchetgenTchetgen2013} is closely related to the new estimating equation. However, focusing on the identity and log links, our approach is more robust to certain forms of misspecification than the estimator of  \citet{TchetgenTchetgen2013}. We emphasize that the proposed approach is crucially different from AIPW. Specifically, in contrast with AIPW, here only data available in the case-control sample contribute information, so that our estimators retain an IPW form. However, we also note that in a nested case-control study, one could in principle augment the estimating equations developed in this paper for additional efficiency gains using AIPW theory.


This paper is organized as follows. In Section~\ref{sec:model} we describe the proposed class of estimators. In Section~\ref{sec:semi_theory} we introduce semiparametric theory that forms the basis for our suggested estimators, provide the semiparametric locally efficient estimator in the class of estimators, and asymptotic properties. Throughout,  we focus on  the identity link (continuous outcome) and the log link (count, or positive outcome) for modeling the outcome mean. In Section ~\ref{sec:simulations} we  present simulation results, empirically demonstrating the balance that our proposed estimators strike between robustness and efficiency, by comparing them to prevailing estimators in the literature.   We use our proposed estimator in Section~\ref{sec:data} in associating SNPs from the FTO gene with BMI, using the case-control, GWAS, T2D data set.  Finally, in Section~\ref{sec:discussion} we discuss our results.

\section{Model}
\label{sec:model}

Suppose the case-control study has $i=1, \ldots,n$ independent participants, with $D_i$ an indicator for the primary disease, so that $D_i=1$ if the $i$th participant is a case and $D_i=0$ otherwise. Let $Y_i$ denote the secondary outcome of interest, and $\bX_i$ the $q\times 1$ vector of covariates of subject $i$. Let $S_i$ be an indicator of inclusion in the case-control study. 

We assume that the probability of selection into the study depends solely on the disease status, $D_i$, and is denoted by $ p(S_i=1|D_i, Y_i,\bX_i) = \pi(D_i)$. Further, we assume that $\pi(D_i)$ is known by design.  Equivalently, we assume that $p(D_i=1)$ in the population is known. Denote by $p(\bX_i) = p(D_i=1|\bX_i)$ the conditional probability of  disease given covariates in the target population, and let 
\begin{equation}\label{eq:mean_link}
\mu(\bX_i;\bbeta) = g\big\{\bbE(Y_i|\bX_i)\big\}
\end{equation}
be the model for the mean after transformation using the link function $g(\cdot)$. In the case of a continuous outcome with the identity link, for instance, $  \mu(\bX_i; \bbeta) = \bbE(Y_i|\bX_i) $, and when the log link is used, $\exp\big\{ \mu(\bX_i;\bbeta)  \big\}  = \bbE(Y_i|\bX_i) $,  where expectations are taken over the entire population (rather than the case-control study population). Note that $\bbeta$ is the $q\times 1$ vector of population regression coefficients that we wish to estimate.  Let $\mathcal{M}$ denote the semiparametric model defined by the mean model specification  (\ref{eq:mean_link}) and the assumed model for $p(\bX)$.

 Hereafter, unless otherwise stated, all expectations are taken with respect to the case control study population.
Taking an estimating equations approach, parameter estimates are obtained by solving an equation of the form 
\begin{equation}\label{eq:basic}
\sum_{i=1}^n\bU_i (\bbeta)= 0
\end{equation}
for $\bbeta$, where $\bU_i(\bbeta)$ are $q\times 1$ functions, with $\bbE\{\bU_i(\bbeta)\} = 0$,  i.e. the estimating equation should be unbiased. A traditional approach for estimation in case-control studies, originating in the sample survey literature, is  Inverse Probability Weighting of each equation according to its probability of selection into the study.    IPW  to estimate the population mean model entails solving for $\bbeta$ equation (\ref{eq:basic}) with 
\begin{equation}\label{eq:ipw}
\bU_{ipw,i}(\bbeta) = \frac{S_i h(\bX_i)}{\pi(D_i)} \big[Y_i - g^{-1}\{\mu(\bX_i;\bbeta)\big\} \big],
\end{equation}
where $h(\bX_i)$ is a user specified $q\times 1$ function, such that $\bbE(\partial \bU_{ipw}/\partial\bbeta)$ is invertible. It is straightforward to see that this equation is unbiased, using the law of iterated expectations. 

 Suppose that the  probability of disease conditional on covariates $p(\bX)$ is  known. We can use such knowledge to extend IPW estimating equations. Consider adding a general {\it control function}  to the estimating equation to obtain:
\begin{equation}\label{eq:cont}
\bU_{cont} (\bbeta)= \sum_{i=1}^n\frac{S_i }{\pi(D_i)} \bigg(  h_1(\bX_i)\big[Y_i - g^{-1}\{\mu(\bX_i;\bbeta)\}\big]  -h_2(\bX_i,D_i)    \bigg)= 0,
\end{equation}
where  $h_2(\bX, D)$ is a $q\times 1$ vector control function that depends on the disease model and satisfies $\bbE\big\{Sh_2(\bX,D)/\pi(D) \big| \bX\big\} = 0$.  Control functions have  been used in econometrics as a mean to control for bias due to  specific forms of selection, see  \citet{Wooldridge2002,Petrin2010} for instance. Typically, a control function approach includes two stages of estimation, a first stage in which a subset of observed variables is employed to estimate the control function, which is subsequently used to augment a second stage regression model to identify the parameter of interest. In the present setting, we adopt a control function framework  for efficiency improvement.

Note that the  second term in (\ref{eq:cont}) is  inverse probability weighted, and has mean zero for all $h_2$, so that $\bU_{cont}(\bbeta)$ is unbiased.  The choice $h_2(\bX,D) = 0$ gives standard IPW. We aim to find $h_2(\bX,D)\ne 0$ such that the resulting estimator is asymptotically at least as efficient as IPW for a fixed $h_1(\bX)$. We subsequently will characterize the optimal choice of $h_1(\bX)$.


For any choice of $h_2(\bX,D)$, there exists a corresponding choice of $\tilde h_2(\bX)$ such that $h_2(\bX, D) = \tilde h_2(\bX)\{D - p(\bX)\}$, that is, the set of functions $h_2(\bX,D)$ satisfying the mean zero restriction is equivalent to the set of functions $\tilde h_2(\bX)\{D-p(\bX)\}$ (see  Appendix, Lemma 1). Using the second parameterization, it is clear that $\bU(\bbeta, \tilde h_2)$ is unbiased for all $\tilde h_2(\bX)$. In practice, $p(\bX)$ is unknown and must be estimated. Here, we use semiparametric theory to  study in a unified framework the semiparametric efficiency implications of positing a nonparametric, semiparametric or parametric  model for $p(\bX)$.

\section{Semiparametric theory}
\label{sec:semi_theory}

 In this section, we develop the semiparametric framework that serves as a basis for our methods. We first provide definitions of Regular and Asymptotically Linear (RAL) estimators and tangent spaces, and provide some examples. We then  characterize the RAL estimators  corresponding to a given disease model $p(\bX)$, and subsequently continue to discuss inference.  
\subsection{Asymptotically linear estimators } An estimator  $\wh\bbeta$ is said to be {\it asymptotically linear}  if one can write
\[
n^{1/2}(\wh \bbeta - \bbeta) = n^{-1/2}\sum_{i=1}^n\psi(Y_i, \bX_i,D_i;\bbeta) + o_p(1)
\]
where $\psi(Y_i, \bX_i,D_i;\bbeta)$ is a zero-mean function, called the $i$th {\it influence function} for $\bbeta$.  The asymptotic variance of the estimator $\wh \bbeta$ has a  simple form and is given by
\[
\mbox{Var}(\wh \bbeta) =  \bbE\{\psi(Y, \bX,D;\bbeta) \psi(Y, \bX,D;\bbeta)^T\}.
\] 
 The influence function is therefore very important, as the asymptotic distribution of an asymptotically linear estimator is completely identified from its influence function.  Influence functions were first introduced by Huber \citep{Huber1972} in the context of robust statistics. Later, they took on a different role in semiparametric theory, in the sense of \citet{Bickel1993}. In either context, they represent the influence of a single observation on the estimator.
 
An {\it  efficient} estimator is the one with associated influence function $\psi^{opt}$ satisfying 
\[
\ba^T \big[  \bbE\{\psi^{opt}(Y, \bX,D;\bbeta) \psi^{opt}(Y, \bX,D;\bbeta)^T\}  - \bbE\{\psi(Y, \bX,D;\bbeta) \psi(Y, \bX,D;\bbeta)^T\}    ]\ba \geq 0
\]
for all other influence functions $\psi$ and $q\times 1$ vectors $\ba$. In other words, the difference between the covariance matrices of the efficient estimator and another estimator is a positive semidefinite matrix.

In this paper we will restrict attention to the subset of asymptotically linear estimators that are also regular, i.e. that are locally uniformly consistent. Regularity is a desirable property for an estimator, since it ensures that its asymptotic behavior provides a good approximation for its finite sample behavior irrespective of the data generating mechanism within the model.
For completeness, local uniform consistency is formally defined in the  Appendix. A more technical definition of RAL estimators can be found in \citet{Bickel1993}. 

\subsection{Tangent spaces}
 Let $\mathcal{L}_2^0$ be the space of square integrable mean zero functions.  According to semiparametric theory, the  {\it tangent space} of a parametric model is the finite dimensional linear subspace of $\mathcal{L}_2^0$, spanned by its scores. The definition of a  tangent space of a parametric model generalizes to that of a semiparametric model, as  the  closed linear span  in $\mathcal{L}_2^0$, of scores of its parametric submodels. 
 The {\it nuisance tangent space} is the subspace of the tangent space containing all scores for {\it nuisance parameters}, i.e. any parameters indexing the observed data law, that are distinct from the parameter of scientific interest $\bbeta$. We now consider the tangent space $\Lambda_D$ of scores of $p(\bX)$, under parametric, semiparameteric, and nonparametric specifications.  Denote the probability of disease for individuals in the case-control study, conditional on covariates, by $p_{cc}(\bX) = p(D=1|\bX,S=1)$.  Let $\balpha$ be any set of parameters indexing the probability model $p(\bX)$. In the nonparametric model in which $p(\bX)$ is unrestricted, $\Lambda_D = \Lambda_{D,npar}$, where 
 \[
\Lambda_{D,npar} = \left\{ \frac{S}{\pi(D)}h(\bX)\{D- p(\bX; \balpha)\}, \mbox{ for any function } h(\bX) \right \}  \cap \mathcal{L}_2^0.
\] 
In the  parametric model for $p(\bX;\balpha) = \mbox{expit}(\balpha^T\bx)$ with an unknown parameter $\balpha$, 
\[
\Lambda_D = \Lambda_{D,\alpha} = \left \{ \frac{S}{\pi(D)}\frac{p_{cc}(\bX)}{p(\bX;\balpha)}C^T\bX\{D- p(\bX; \balpha)\}, \mbox{ for any conformable matrix } C \right \}  \cap \mathcal{L}_2^0.
\]
Consider now the semiparametric model 
$p(\bX;\balpha) = \mbox{expit}\{\alpha_1(\bx_1) +  \balpha_2^T\bx_2\}$, in which the function $\alpha_1(\bx_1)$ is unrestricted. Here
\begin{eqnarray*}
\Lambda_D = \Lambda_{D,\alpha_1,\alpha_2} = \bigg\{ \frac{S}{\pi(D)}\frac{p_{cc}(\bX)}{p(\bX;\balpha)}\big\{g(\bX_1) + C^T\bX_2\big\} \{D- p(\bX; \balpha)\}, \mbox{ for any conformable matrix } C \\
&& \hspace{-2.5in} \mbox{ and any function } g(\bX_1) \bigg\}  \cap \mathcal{L}_2^0.
\end{eqnarray*}
 We show in the  Appendix that the scaling factor $p_{cc}(\bX)/p(\bX)$ is required to appropriately account for retrospective sampling.
 In general, we will denote the tangent space of a parametric, semiparametric, or nonparametric submodel for $p(\bX)$ by $\bLambda_{D,sub}$.
\subsection{The RAL estimators for $\bbeta$ }

Let $\Pi(\bv |\Lambda)$ denote the orthogonal projection of the vector $\bv$ on the subspace $\Lambda$ of $\mathcal{L}_2^0$.

\noindent \textbf{Theorem 1.}\textit{
The set of influence function of $\bbeta$ is given by
\begin{eqnarray*}
\Gamma =\bigg\{ \frac{S}{\pi(D)}h_1(\bX)[Y - g^{-1}\{\mu(\bX,\bbeta)\}]  - \frac{Sh_2(\bX,D)}{\pi(D)} + \Pi\bigg( \frac{Sh_2(\bX,D)}{\pi(D)}  \bigg| \Lambda_{D,sub}   \bigg): \\
\bbE\big\{\frac{S}{\pi(D)}h_2(\bX, D)|\bX\big\} = 0  \bigg \} \cap \mathcal{L}_2^0
\end{eqnarray*}
up to a multiplicative constant.}

Theorem 1 characterizes all RAL estimators of $\bbeta$ in a semiparametric model $\mathcal{M}$ defined by $\mu(\bX,\bbeta)$ and a choice of model for $p(\bX)$.  The proof is in the  Appendix.  Interestingly, it states that if  
\[
\frac{Sh_2(\bX,D)}{\pi(D)} = \Pi\bigg( \frac{Sh_2(\bX,D)}{\pi(D)}  \bigg| \Lambda_{D,sub}\bigg),
\]
then all influence functions for $\bbeta$ are IPW influence functions. This equality holds, for instance, in the special case where the model $p(\bX)$ is saturated, or nonparametric.  In other words, even if one uses the estimator (\ref{eq:cont}), for any choice of $h_2(\bX,D)$ the asymptotic distribution of the estimator will mimic the IPW estimator and the estimator could not be made more efficient. The following Corollary 1 summarizes this observasion.

\noindent \textbf{Corollary 1.}\textit{
Consider  the model for $\mathcal{M}$ with $p(\bX)$ unrestricted. For a fixed choice of $h_1(\bX)$ in (\ref{eq:cont}), the optimal choice of function $h_2(\bX,D)$ is $h_2^{opt}(\bX,D)=0$, and the most efficient estimator for $\bbeta$ is the IPW estimator that solves the estimating equation
\begin{eqnarray*}
\bU_{ipw} (\bbeta)= \sum_{i=1}^n\frac{S_i }{\pi(D_i)} \bigg(  h_1(\bX_i)\big[Y_i - g^{-1}\{\mu(\bX_i;\bbeta)\}\big]   \bigg)= 0.
\end{eqnarray*}
}

 In the following sections we restrict $p(\bX)$ by posing modeling assumptions. We first focus on finding the  most efficient estimating equation for $\bbeta$ in $\Gamma$  for any fixed $h_1(\bX)$ in Section~\ref{sec:infer_h1_fix}, and then provide the optimal $h_1(\bX)$ and the  locally efficient estimator  in Section~\ref{sec:infer_h1_opt}.


\subsection{Inference for a restricted model $p(\bX)$ and a  fixed $h_1(\bX)$}\label{sec:infer_h1_fix}


 Suppose that the model for $p(\bX)$ is restricted. For a fixed $h_1(\bX)$, we wish to find the optimal $h_2(\bX,D)$, that minimizes the variance of the estimating equations in $\Gamma$ defined in Theorem 1. This function is given in the following Theorem 2 and the proof is in the  Appendix. 

\noindent \textbf{Theorem 2.}\textit{
Suppose that $h_1(\bX)$ is fixed. The function $h^{opt}_2(\bX,D)$ that minimizes the variance of $\wh \bbeta$ in model $\mathcal{M}$ is given by
\begin{eqnarray*}
h_2^{opt}(\bX,D) = h_1(\bX)\big[ \bbE(Y | \bX,D; \bbeta)  -g^{-1}\{\mu(\bX;\bbeta)\}\big].
\end{eqnarray*}
Denote $\tilde\mu(\bX, D;\bbeta) = g\big\{\bbE(   Y  | \bX, D; \bbeta)  \big\}$, which satisfies $\bbE\big\{ \bbE(   Y  | \bX, D; \bbeta  )\big| \bX\big\} = g^{-1}\{\mu(\bX;\bbeta)\}$. Then the $i$th  influence function corresponding to $h_2^{opt}(\bX,D)$, up to a multiplicative constant, is
\[
\frac{S_i h_1(\bX_i)}{\pi(D_i)} \left[Y_i -   g^{-1}\big\{\tilde\mu(\bX_i,D_i;\bbeta)  \big\} \right].
\]
}

\citet{TchetgenTchetgen2013} provided parameterizations of $\tilde\mu(\bX,D;\bbeta)$ in terms of $\mu(\bX;\bbeta)$ for the identity, log, and logit links. We use these parameterizations to construct feasible estimating equations $\bU_{ident}^{opt}$ and $\bU_{log}^{opt}$ based on Theorem 2.    Consider first the identity link function.  As was shown in \citet{TchetgenTchetgen2013}, $\bbE(Y|\bX,D;\bbeta)$ can be parameterized as 
$\bbE(Y|\bX,D;\bbeta) = \mu(\bX;\bbeta) +  \gamma(\bX)\{D - p(\bX)\}$, where $\gamma(\bX) = \bbE( Y|D=1,\bX) - \bbE(Y|D=0,\bX)$ is the ``selection bias function", resulting from sampling according to disease status. We have that
\begin{equation*}\label{eq:cont_identity_2}
\bU^{opt}_{ident}(\bbeta) = \sum_{i=1}^n\frac{S_i   h_1(\bX_i)}{\pi(D_i)} \bigg[Y_i - \mu(\bX_i;\bbeta)  -\gamma(\bX_i)\{D_i - p(\bX_i)\}    \bigg].
\end{equation*}
For the log link, it was shown in \citet{TchetgenTchetgen2013} that
\[
\tilde\mu(\bX,D;\bbeta) = \bbE(Y|\bX,D;\bbeta) = \exp\big(\mu(\bX;\bbeta) + \nu(\bX,D) - \log\ \bbE[\exp\{\nu(\bX,D)\}|\bX]     \big),
\]
where the selection bias function $\nu(\bX,D)$ is defined as
\[
\nu(\bX,D) = \log\bigg\{\frac{\bbE(Y|\bX,D)}{\bbE(Y|\bX,D=0)} \bigg\}
\]
and reflects the  log multiplicative association between $D$ and $Y$ given $\bX$, and note that the expectation in $\bbE[\exp\{\nu(\bX,D)\}|\bX] $ is taken over the population.   Therefore, we have that
\begin{equation*}
\bU^{opt}_{log}(\bbeta) = \sum_{i=1}^n\frac{S_i h_1(\bX_i)}{\pi(D_i)} \bigg\{Y_i -  \exp\big(\mu(\bX_i;\bbeta) + \nu(\bX_i,D_i) - \log\ \bbE[\exp\{\nu(\bX_i,D_i)\}|\bX_i]     \big)   \bigg\}. 
\end{equation*}
Note that these estimating equations are robust, in the sense   that even if the selection bias functions $\gamma$ and $\nu$ are misspecified, the estimating equations would remain unbiased as long as $\mu(\bX;\bbeta)$ and $p(\bX)$ are correctly modeled.

\subsection{The semiparametric locally efficient estimator}\label{sec:infer_h1_opt}

An estimator $\wh \bbeta$ of $\bbeta$ is called {\it locally efficient} at  a  submodel for   $f(Y,\bX,D,S)$ in a semiparametric model $\mathcal{M}$ if its asymptotic variance achieves the {\it semiparametric efficiency bound}  for $\mcl{M}$, and remains consistent and asymptotically normal (CAN) outside of the submodel \citep{Bickel1993}. For instance, in the identity link case, $\mcl{M}$ may be a model that specifies parametric models for $p(\bX;\balpha)$, $\mu(\bX;\bbeta)$, and its parameters could be estimated using an estimating equation $\bU_{ident}^{opt}(\bbeta)=0$.  Different choices of $h_1(\bX)$ will lead to estimators of $\wh \bbeta$, that are CAN but with different asymptotic variances.  The {\it semiparametric efficiency bound} is the smallest variance that can be obtained  by a regular estimator in $\mcl{M}$. This RAL estimator has the {\it efficient influence function}, which is the projection of any influence function of $\bbeta$ in $\mathcal{M}$ onto the tangent space for the model \citep{Bickel1993}.

Theorem 3 below characterizes the efficient influence function of $\bbeta$ in model $\mathcal{M}$  and the corresponding estimating equation, by giving the optimal  $h_1^{opt}(\bX)$.

\noindent \textbf{Theorem 3.}\textit{
The semiparametric efficient influence function for $\bbeta$ in model $\mcl{M}$  is given by
\[
\frac{S_i h_1^{opt}(\bX_i)}{\pi(D_i)} \bigg[Y_i - g^{-1}\big\{\tilde\mu(\bX_i,D_i;\bbeta)  \big\}   \bigg], 
\]
with 
\[
h_1^{opt}(\bX) = \bbE\bigg\{ \frac{1}{\pi(D)}\mbox{var}(Y|D,\bX) \bigg| \bX\bigg\}^{-1}\frac{\partial}{\partial \bbeta}\left[ g^{-1}\big\{\tilde\mu(\bX,D;\bbeta)  \big\}   \right].
\]
}
  The corresponding estimator $\wh \bbeta$ is locally efficient in the submodel of $\mathcal{M}$ in which $h_1(\bX)$ and $h_2(\bX,D)$ are correctly modeled. If these functions are misspecified, $\wh \bbeta$ will still be CAN, but less efficient.
Below we use the efficient influence function to define an estimating equation by substituting empirical estimates of all unknown nuisance parameters. The asymptotic distribution of the resulting estimator allowing for model misspecification is  given in Section~\ref{sec:asymptotics}.

%
\subsection{Asymptotic properties}\label{sec:asymptotics}

 We saw that $\wh \bbeta$ is a RAL estimator  in model $\mathcal{M}$ in which $p(\bX)$ is correctly specified. We  compute $\wh\bbeta$ by solving the estimating equation $\wh\bU_{cont}^{opt}(\bbeta)=0$, defined as $\bU^{opt}_{cont}(\bbeta)$ with $\wh h_1(\bX), \wh h_2(\bX,D)$, and $\wh p(\bX)$.  

 Let $\bdelta$ denote the parameters for the selection bias function, i.e. either $\nu(\bX,D; \bdelta)$ (log link) or $\gamma(\bX; \bdelta)$ (identity link). Let $\btheta = (\bbeta^T, \bdelta^T)^T$. It is convenient to estimate $\btheta$ jointly, by modifying the estimating equation $\bU^{opt}_{cont}(\bbeta)$ to define $\bU^{opt}_{cont}(\btheta)$ by taking
\[
h_1^{opt}(\bX) = \bbE\bigg\{ \frac{1}{\pi(D)}\mbox{var}(Y|D,\bX) \bigg| \bX\bigg\}^{-1}\frac{\partial}{\partial \btheta}\big[ g^{-1}\big\{\mu(\bX,D;\btheta)  \big\}   \big].
\]
In the  Appendix, we describe how to compute the estimator $\wh \btheta$. To find its asymptotic distribution (and calculate standard errors), we need to know its influence function. Its influence function  is found from the first order Taylor expansion of the estimating equation around the limiting value of $\wh\btheta = (\wh\bbeta^T, \wh\bdelta^T)^T$. We provide this derivation in  Appendix.   
Let $\bV(\balpha)$ be the estimating equation for $\balpha$. The influence function for $\btheta$ is given by

\begin{eqnarray*}
\psi(\btheta; \balpha)\hspace{-.1in}  &=& \hspace{-.1in}- \bigg[  \bbE\frac{\partial}{\partial \btheta}\bigg\{\bU_{cont}(\btheta; \balpha)    \bigg\} \bigg]^{-1}\times \\
&&\hspace{.8in} \left[  \bU_{cont}(\btheta; \balpha) - \bbE\left\{ \frac{\partial}{\partial \balpha}\bU_{cont}(\btheta; \balpha)   \right\} \bbE\left\{\frac{\partial}{\partial \balpha}\bV(\balpha)   \right\}^{-1}   \bV(\balpha)  \right]. \\
\end{eqnarray*}
A consistent estimator of the covariance matrix of the estimator $\wh\theta$ is given by 
\[
\wh\bSigma (\btheta) = \frac{1}{n}\sum_{i=1}^n \wh\psi_{i}\left(\wh\btheta; \wh\balpha\right)\wh\psi_{i}^T \left (\wh\btheta; \wh\balpha\right),
\]
where $\wh\psi_i$ is the influence function evaluated at the $i$th subject, with all  expectations in the expression $\psi(\btheta; \balpha)$ estimated by the corresponding sample means.

\noindent \textbf{Corollary 2.}\textit{
The estimator $\wh \btheta$ that solves $\bU_{cont}(\bbeta, \bdelta; \wh\balpha)$ under $\mathcal{M}$ is asymptotically normally distributed with asymptotic mean $\btheta$ and covariance
\[
\bSigma (\btheta)  = \bbE\left\{  \psi(\btheta; \balpha) \psi(\btheta; \balpha)^T   \right\}.
\]
Further more, in the submodel where $\wh h_1^{opt}(\bX) \to p\lim_{n\to \infty} h_1^{opt}(\bX)$, and \\ $ \wh h_2^{opt}(\bX,D) \to p\lim_{n\to \infty} h_2^{opt}(\bX,D)$, $\wh \bbeta$ is locally efficient. }

 Note that $\wh \btheta$ will be asymptotically normal with covariance matrix $ \bbE\left\{  \psi(\btheta^*; \wh\balpha) \psi(\btheta^*; \wh\balpha)^T   \right\}$, where $\btheta^*$ is $p\lim_{n\to \infty} \wh\btheta$, even if one of  $p(\bX),\mu(\bX;\bbeta)$, or both, are misspecified. In the case of misspecification, $\btheta^*$ is likely a biased estimate of the true $\btheta$.

\section{Simulations}\label{sec:simulations}

In this section, we demonstrate the robustness and efficiency of our proposed estimators compared to the prevailing estimators, when modeling the mean via the identity link.  We simulate case-control studies with continuous  secondary outcomes in two sets of simulations. The goal of the first set was to investigate the robustness and efficiency of the proposed control function estimator (`cont'), compared to multiple other prevailing estimators: the estimator that conditions on disease status, using disease indicator in the regression of the secondary outcome on covariates,   (denoted by Dind), the estimator that treats all observations equally, ignoring disease status (pooled), and the IPW estimator (IPW). The goal of the second set was to compare the performance of `cont' to the estimators proposed by \citet{Ghosh2013} and \citet{Lin2009}. In each  section below, we describe the simulations and provide results, where for cont,  we provide two sets of results: when the model for $\bbE\big(Y|\bX,D\big)$ is correctly specified, and when it is misspecified. For each scenario, we calculated the mean bias of the estimates $\frac{1}{n.sim}\sum_{k=1}^{n.sim}\wh\beta_k - \beta$, the mean squared error (MSE) $\frac{1}{n.sim}\sum_{k=1}^{n.sim}(\wh\beta_k - \beta)^2$, the sample standard deviation of the estimator $\{\frac{1}{n.sim}\sum_{k=1}^{n.sim}(\wh\beta_k - \bar{ \wh \beta})^2\}^{1/2}$, the mean of the estimated standard deviations in the simulations $\frac{1}{n.sim}\sum_{k=1}^{n.sim}\wh {\mbox{sd}(\beta_k)} $, and the Wald coverage probability.  Due to limited space, only some of the simulation results are presented in the main manuscript. Additional extensive simulation results are delegated to the  Appendix, including all summaries pertaining to the performance of the estimator ``Dind" and ``pooled".

The proposed cont estimators and the estimated standard deviations were calculated as described in the  Appendix. The IPW estimator and the estimated standard deviations were calculated using Newton-Raphson iterations of the estimating function $\bU_{ipw}$, with $h_1(\bX_i) = \bX_i$, with the robust (sandwich) covariance matrix. The na\"ive estimators Dind and pooled were calculated from linear regression. 

All simulation scenarios included 500 cases and 500 controls, and were run 1000 times. The prevalence of the disease $D$ in the population (the primary case-control outcome) was fixed at $0.12$, i.e. the disease is relatively common. 

 We conducted other simulation studies, under a variety of plausible  scenarios. First, we performed a simulation study for the identity link with a single exposure variable, in which we also considered the estimator proposed by \citet{TchetgenTchetgen2013}. Second, we performed
simulations for the log link, and lastly, we carried out another identity link simulation study, closely mimicking the observed data distribution in the T2D sample. Results for these additional scenarios are provided in the  Appendix. In general, they support the conclusions of the simulations presented here.

\subsection{Simulation set 1 - studying robustness and efficiency}\label{sec:sims_cont_1}
To design the simulations, we first note that we need to sample data from the distribution $f(Y,D|\bX)$, in such a way that the parameter of interest $\bbE[Y|\bX]$ is defined.  We consider the decomposition $f(Y,D|\bX) = f(Y|D,\bX)p(D|\bX)$, and generate the data according to the two parts of the likelihood, $p(D|\bX)$, and $f(Y|D,\bX)$. Note that this decomposition always holds, and makes no assumption on the underlying model. We use the nonparametric decomposition of $\bbE[Y|\bX,D]$ proposed by \citet{TchetgenTchetgen2013}, that allows specifying the two parts of the likelihood. First, exposures/covariates variables $\bX$ were sampled. Then, disease probabilities were calculated for each subject, based on exposure values. The intercept for the disease model $p(\bX)$ was set so that disease prevalence was 0.12. Disease statuses were obtained from disease probabilities, and the secondary outcomes $Y$ were generated based on exposure values and disease status. 

In more details, we simulated two covariates, $X_1$ and $X_2$ where  $X_1\sim \mathcal{N}(2,4)$,  and $X_2\sim \text{Binary}(0.1)$. The primary disease probability was calculated by 
\[
\text{logit}\left\{p(D=1|\bX)\right\} = -3.2+ 0.3X_1 + X_2,
\] and disease status was sampled. The conditional mean of the secondary outcome was:
\[
\bbE(Y|\bX,D) = 50+ 4X_1 + 3X_2 + 3X_1X_2 + \{D-p(\bX)\}(3 + 2X_1 + 2X_2 + 2X_1X_2),
\]
so that $\mu(\bX,\bbeta) = \bX^T\bbeta$ with $\bX = (1, X_1,X_2, X_1X_2)^T$ and $\bbeta = (50,4,3,3)^T$, and $\gamma(\bX) = \bX^T\balpha$ with $\balpha = (3,2,2,2)^T$.
The residuals were sampled by $\epsilon \sim \mathcal{N}(0,4)$. The design matrix for $\gamma(\bX)$ was $\bX = (1, X_1,X_2, X_1X_2)^T$ when the model was correctly specified. We studied the following forms of misspecification of the design matrix of $\gamma(\bX)$. The estimator `cont-mis1' had the design matrix $\bX = (1, X_1,X_2)^T$ (no interaction term), `cont-mis2' had $\bX = (1, X_1)^T$, `cont-mis3' had $\bX = (1, X_2)^T$, and `cont-mis4' accounted only for an intercept, i.e. design matrix $\bX = 1$.

A table providing comprehensive simulation results for this simulation is provided in the  Appendix. Figure~\ref{fig:identity2} compares between the estimated bias, MSE, and coverage probabilities of the cont estimators (correctly specified and misspecified), and the usual IPW. We do not provide graphic results for the estimators that where heavily biased: Dind, and pooled.

\begin{figure}
 \centerline{\includegraphics[scale = 0.7]{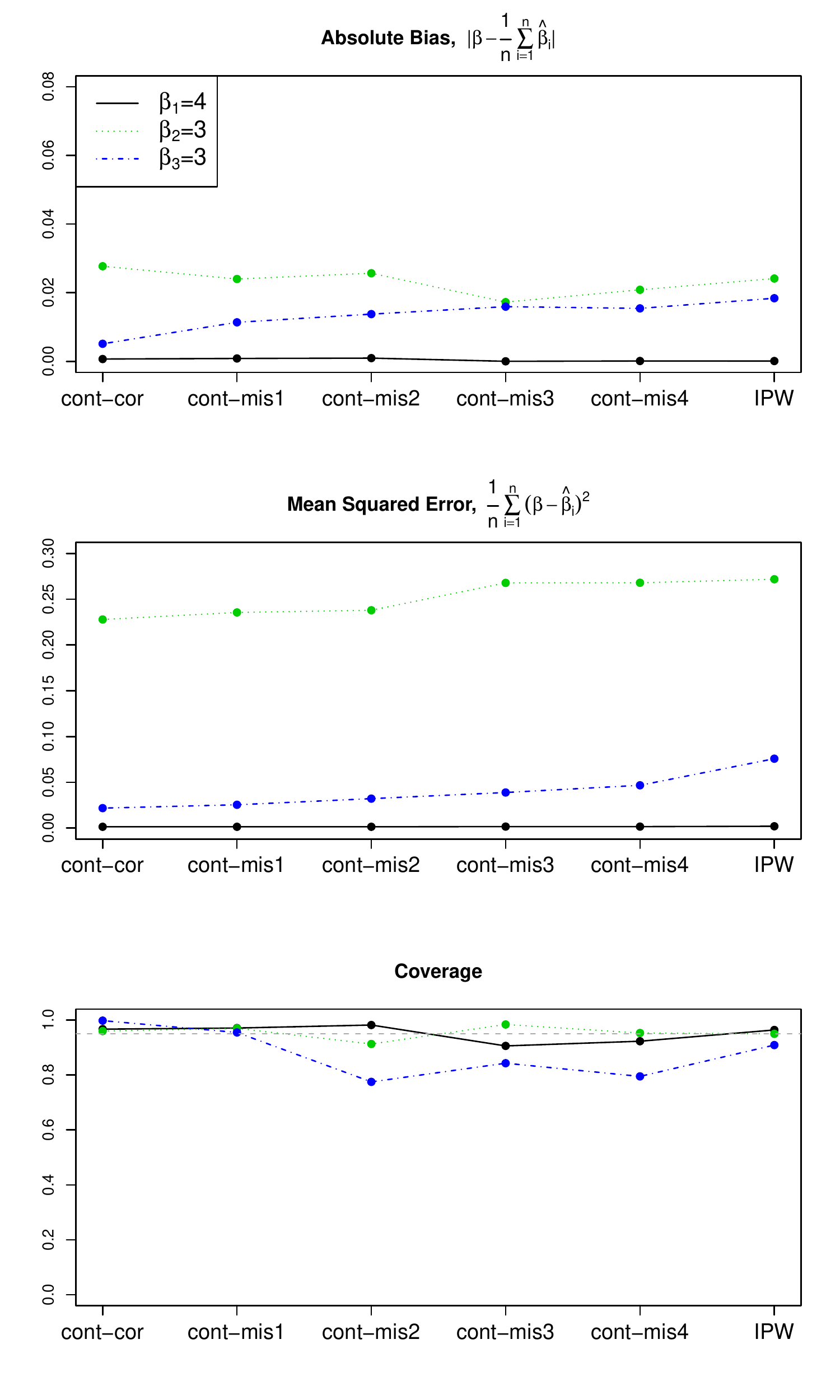}} 
\caption{Results from Identity link simulations set 2, second settings, two covariates. Estimated bias, MSE, and coverage probability of the control function under correct and misspecification of the selection bias function (cont-cor, cont-mis1, $\ldots$, cont-mis4), and IPW, in estimating population effects of $X_1,X_2$ and their interaction.}
\label{fig:identity2}
\end{figure}

As shown in Figure~\ref{fig:identity2} and in the Appendix, 
the estimated bias of the cont estimator is usually smaller than that of the IPW estimator and is very small when the model for $\gamma(\bX)$ is correctly specified, and slightly larger when the model for $\gamma(\bX)$ is misspecified.

Both the MSE  and the empirical standard deviation of the cont estimator were higher when the model for $\gamma(\bX)$ was misspecified, yet interestingly, the MSE of  cont was always smaller than that of the IPW.  In fact, the relative efficiency of cont-cor  was from 17\% ($\beta_2$) to 71\% ($\beta_3$) lower than that of the IPW. Even cont-mis4, the estimator with only intercept used in the selection bias function, had relative efficiency from 1.5\% to 38\% lower than that of the IPW. Coverage probabilities were always correct for the coefficients of $X_1$ and $X_2$,  but sometimes too small for the coefficient of the interaction term $X_1X_2$ when $\gamma(\bX)$ was misspecified (more specifically, when $X_1$ was not in the design matrix of $\gamma(\bX)$). In comparison, the coverage probability of the IPW estimator was always very close to the desired 95\%. Dind and the pooled estimator again yielded biased estimates with, usually, very low coverage probability.  Interestingly, the bias of pooled was usually lower than the bias of Dind.

\subsection{Simulation set 2 - comparison to another recently proposed method}
Here we compare our estimator `cont', and the IPW, to the pseudo-likelihood estimator proposed by \citet{Ghosh2013}, and the retrospective likelihood estimator proposed by \citet{Lin2009}.   We followed the simulation scenario performed in \citet{Ghosh2013}, using code shared by the authors. We also adapted our simulations from Section~\ref{sec:sims_cont_1} to their assumed data structure. 

First, we ran 1000 simulations in \citet{Ghosh2013} simulation settings and compared the estimators. In their simulations, they focused on a single coefficient, namely the effect of a single nucleotide polymorphism (SNP)  $G$ on the outcome $Y$. $G$ had a minor allele frequency (MAF) 0.25. There were two covariates $Z$, one continuous and one binary, with probability 0.45. The disease and the secondary outcome were modeled by a bivariate normal distribution and thresholding, so that the disease model is  dependent on $G$ and $\bZ$ via a logistic model. However, it is unclear how to specify correctly $\gamma(\bX)$. We use a linear model of the form $\gamma(\bX) = \bX\bdelta$, though this is likely incorrect. The outcome $Y$ had variance $1$, and disease prevalence was 0.05. The effect of interest was 0.1. We used 500 cases and 500 controls. More details can be found in \citet{Ghosh2013}.
The results of these simulations are presented at the top part of Table~\ref{tab:identity_3}. 

Then, we ran 1000 simulations in settings adapted from our simulations from Section~\ref{sec:sims_cont_1}. Here, we had the same $G,\bZ$ variables, with $Z_1$ continuous and $Z_2$ binary. $Z_1 \sim \mathcal{N}(0,4)$, and $Z_1 \sim \text{Binary}(0.2)$. The primary disease probability was calculated by 
\[
\text{logit}\left\{p(D=1|\bX)\right\} = -3.8+ 0.3X_1 + X_2,
\] and disease status was sampled. Note that the intercept value was selected to that 
disease prevalence was roughly 0.05, as in \citet{Ghosh2013}.  The SNP $G$ has minor allele frequency 0.3. The  conditional mean model was:
\[
\bbE(Y |\bX,D) = 3 + 0.7Z_1 + 0.5Z_2 + 0.3G + \{D -p(\bX)\}(1 + 0.5Z_1 + 0.3Z_2).
\]
500 cases and 500 controls were sampled from the simulated population. We compared the estimation of the effect of $G$ on $Y$. The results of these simulations are presented at the bottom part of Table~\ref{tab:identity_3}. 

In the first simulation set, the estimators Ghosh2013, IPW and cont were unbiased, and achieved the nominal coverage level, while Lin2009 was heavily biased.  Note that cont likely  misspecified the model $\gamma(\bX)$. The estimator of \citet{Ghosh2013} had slightly lower MSE than the IPW and control function estimators, as expected, since this estimator is based on the same model used to produce the simulated data. In the second set of simulations, in which the data were sampled by specifying models for $p(\bX), \gamma(\bX)$, and $\mu(\bX;\bbeta)$, both estimators Ghosh2013 and Lin2009 were biased (both biases about -0.4) and had low coverage of 0.67\% (Ghosh2013) and 50\% (Lin2009). In contrast, the IPW and cont estimators performed similarly, with low bias and correct coverage probability. These results demonstrate the robustness of IPW and cont, which use fewer modeling assumptions.  

The estimator Lin2009 was biased under both simulation scenarios. As other likelihood-based estimators, it assumes a certain probability model, and it is biased when this model is fails to hold. The model proposed by \citet{Ghosh2013} is different than the one proposed by \citet{Lin2009}. 

\begin{table}[htbp] 
\caption{Simulation results for estimating the effect of a SNP on a normally distributed secondary outcome. We compare results for the usual IPW estimator, the proposed  control function estimator (`cont'), the pseudo-likelihood estimator of \citet{Ghosh2013} (`Ghosh 2013), and the retrospective likelihood estimator of \citet{Lin2009} (`Lin2009'). The top part of the table provides results of simulations in the settings in \citet{Ghosh2013}, and the bottom part is of simulations designed according to the conditional mean model $\bbE(Y|\bX,D)$. For each estimator and each estimated parameter the table reports the estimator's mean bias, MSE, empirical standard deviation over all simulations, mean estimated standard deviation using the appropriate formula, and coverage probability.  }
\label{tab:identity_3}
\begin{center}
\begin{tabular}{lrrrrr}
\hline\hline
\multicolumn{1}{l}{estimator/value}&\multicolumn{1}{c}{bias}&\multicolumn{1}{c}{MSE}&\multicolumn{1}{c}{emp sd}&\multicolumn{1}{c}{est sd}&\multicolumn{1}{c}{coverage}\tabularnewline
\hline
\multicolumn{6}{c}{Settings 1 (Ghosh, 2013). $\beta = 0.1$} \\ \hline
Ghosh2013&$ 0.000$&$0.003$&$0.056$&$0.057$&$0.961$\tabularnewline
cont&$ 0.000$&$0.004$&$0.067$&$0.067$&$0.952$\tabularnewline
IPW&$ 0.000$&$0.004$&$0.067$&$0.067$&$0.953$\tabularnewline 
Lin2009& -0.765 & 0.588 & 0.049 & 0.055& 0.000 \\ \hline
\multicolumn{6}{c}{Settings 2. $\beta = 0.7$}\\ 
\hline
Ghosh2013&$-0.402$&$0.228$&$0.258$&$0.261$&$0.666$\tabularnewline
cont&$ 0.002$&$0.002$&$0.043$&$0.042$&$0.946$\tabularnewline
IPW&$ 0.002$&$0.002$&$0.043$&$0.042$&$0.948$\tabularnewline
Lin2009& -0.394& 0.194 & 0.197 & 0.200 & 0.505\\
\hline
\end{tabular} 
\end{center}
\end{table}

\section{Analysis of Type 2 diabetes GWAS}\label{sec:data}
 We analyzed the case-control GWAS study of T2D, with the goal of identifying SNPs in the FTO gene region,  associated with BMI. There were 3080 female participants in this data, genotyped on the affymetrix 6.0 array, with 1326 cases and 1754 controls \citep{Cornelis2012}. There were 152 genotyped SNPs from the region on chromosome 16 spanning the FTO variants. There are a few SNPs from the FTO gene associated with BMI \citep{Speliotes2010}, and validated on large cohorts. In particular, the SNP rs1558902 has the strongest association with log-BMI. This SNP is not  in the data, but  other SNPs in high Linkage Disequilabrium (LD) with it are.   The population prevalence of T2D  was 8.4\% \citep{Cornelis2012}. We compared the usual IPW, the control function estimator `cont', the pooled estimator ignoring disease status, the estimator Dind with disease indicator in the design matrix, and the estimator of \citet{Lin2009} dubbed Lin2009. 
  
 All analyses were adjusted to age, binary smoking status (current versus past or never), binary alcohol intake measure according to less or more than 10 grams a day, physical activity (above or under the median) and to the first four principal components of the genetic data. The outcome, BMI, was log transformed, as is usually done with BMI. For the analysis using the estimator cont, all models, i.e. the mean model of BMI, the model for disease probability $p(D=1|\bX)$, and the selection bias model $\gamma(\bX)$ used the same covariates. All SNPs were analyzed in the additive mode of inheritance. 

\begin{figure}
 \centerline{\includegraphics[angle=270,scale = 0.6]{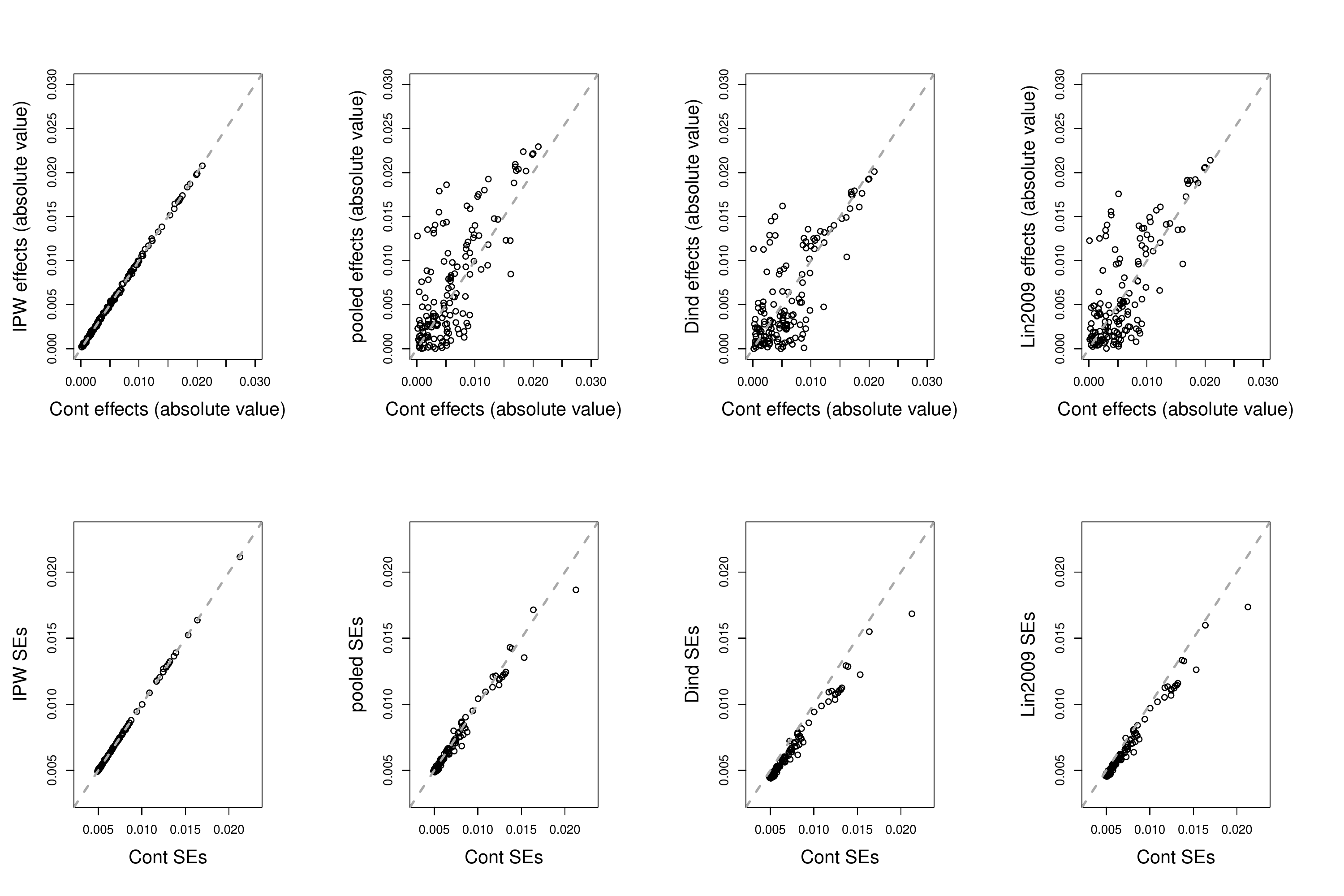}} 
\caption{Comparison of effect estimates for the SNPs in the FTO gene on log-BMI, and their standard errors. Estimates of the control function estimator (`cont') and their SE were compared to the usual IPW, the estimator ignoring disease status (pooled), the estimator using disease indicator in its design matrix (`Dind') and the estimator of \citet{Lin2009} (`Lin2009'). Every point in the plot represent a SNP. If a point falls on the diagonal - its associated effect (SE) estimate is equal in cont and the compared estimator. If it falls below the diagonal, its estimated effect (SE) is smaller in cont compared to the other estimator.  }
\label{fig:compare_fto}
\end{figure}

Figure~\ref{fig:compare_fto} compares between the estimated effect sizes and their respective standard errors (SEs), of all 152 SNPs in the FTO gene, between cont, and the other estimators under consideration. The cont estimator yielded roughly identical results to that of the IPW. This is in agreement with the simulation study imitating the effect sizes in the T2D data set (see  Appendix), and is expected since both T2D and BMI are complex traits, and no single SNP highly affects them. Thus, incorporating the disease and selection bias models in the estimation cannot improve it much. 
 Although the standard errors appear to be ``the same" when looking at the plot, in fact that are small differences, such that the p-values and adjusted p-values of the cont estimates are smaller than those of the IPW, as is seen in Table~\ref{tab:compare_fto}. Effect estimates of other estimators are quite different than those of cont, while their SEs are usually smaller. That is since these estimators make more assumptions on the data distribution, resulting in lower SEs. 

\begin{figure}
 \centerline{\includegraphics[scale = 0.6]{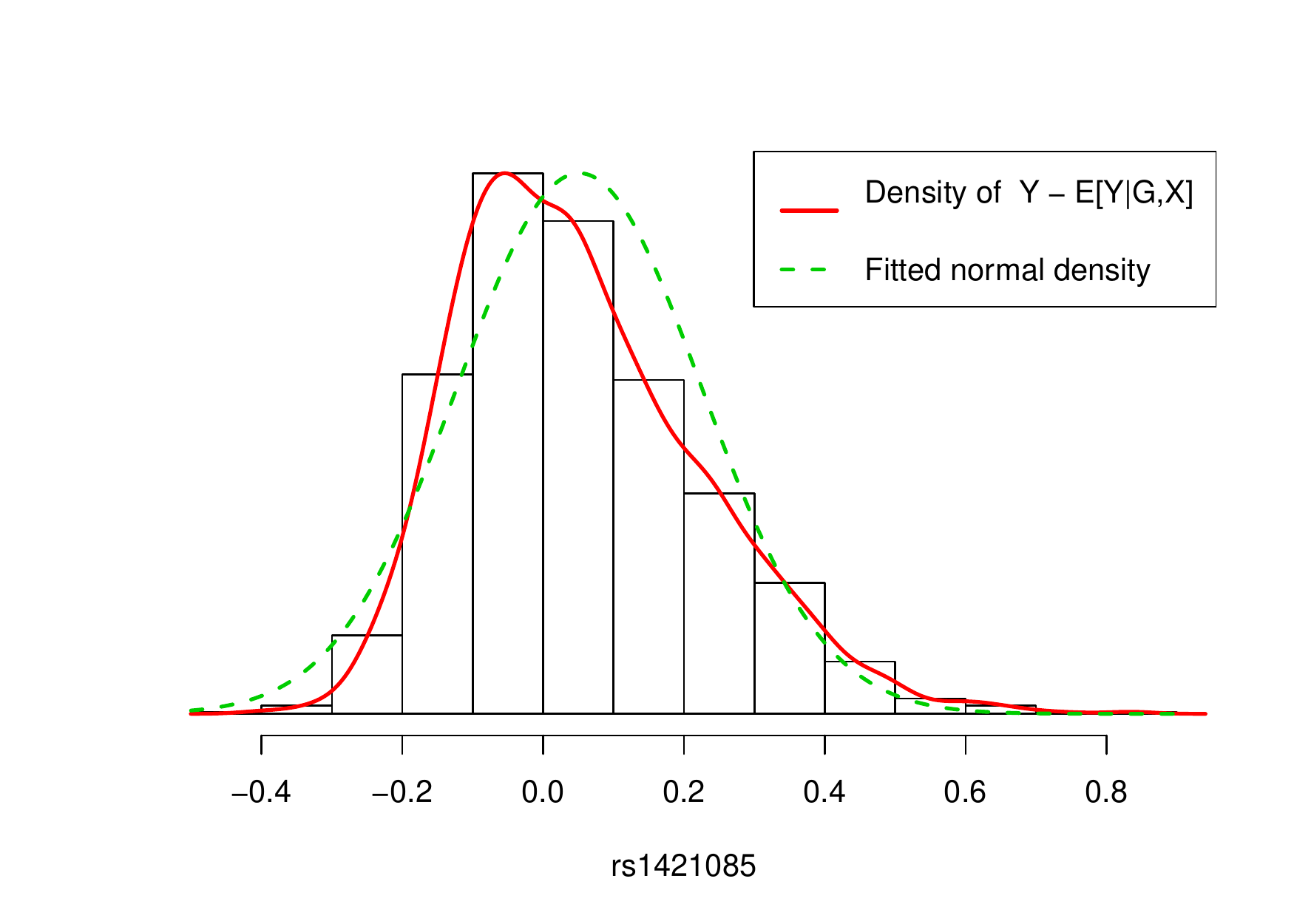}} 
\caption{ Histogram, and overlaid empirical and fitted normal densities to the residuals of log-BMI after removing estimated population mean.  }
\label{fig:compare_normal}
\end{figure}

There were ten SNPs with Holm's adjusted p-value $\leq$0.05 by the pooled estimator, which yielded the lowest p-values. As they were all in high LD, we selected the SNP that is in highest LD with rs1558902, namely, rs1421085 \citep{Johnson2008}.
Table~\ref{tab:compare_fto} compares between the various analyses results on this SNP. As the effects are relatively low ($\sim -0.02$), all estimates are within a rang of $0.04$ of each other. Consistent with the plot, cont and IPW gave identical effect estimates (after rounding) while other estimates were usually different. The effect estimate is largest (in absolute value) in the pooled estimator. Since pooled and Dind are likely biased estimators (as supported by the simulations mimicking the T2D diabetes, reported in the  Appendix), we now consider Lin2009. This estimator properly accounts for case-control sampling, but assumes that the outcome is normally distributed around the population mean. To study the appropriateness of this assumption, we compared the density of the residuals of log-BMI  after removing the population mean estimated by IPW. Figure~\ref{fig:compare_normal} provides this comparison, suggesting that the normality assumption does not hold and that the estimator is potentially biased. 


\begin{table}[htbp] 
\caption{Effect estimates, and their respective SEs and p-values for the SNP rs1421085 from the FTO gene.  The values were obtained by the control function estimator (`cont'), the usual IPW, the pooled estimator ignoring disease status, and the disease with disease indicator in the design matrix (`Dind').   }
\label{tab:compare_fto}
\begin{center}
\begin{tabular}{lrrrr}
\hline\hline
\multicolumn{1}{l}{Estimator}&\multicolumn{1}{c}{effect}&\multicolumn{1}{c}{SE}&\multicolumn{1}{c}{p-value (raw)}&\multicolumn{1}{c}{p-value (adj)}\tabularnewline
\hline
\multicolumn{5}{c}{rs1421085} \\ \hline
cont&$-0.017$&$0.0054$&1.7e-$3$&$0.247$\tabularnewline
IPW&$-0.017$&$0.0054$&1.9e-$3$&$0.273$\tabularnewline
pooled&$-0.021$&$0.0050$&4.2e-$5$&$0.006$\tabularnewline
Dind&$-0.018$&$0.0046$&9.3e-$5$&$0.014$\tabularnewline
Lin2009&$-0.019$&$0.0047$&4.7e-5&$0.007$\tabularnewline

\hline
\end{tabular} 
\end{center}
\end{table}

\section{Discussion}
\label{sec:discussion}

In this work we provide semiparametric, efficient and robust estimators for the population mean effects of covariates on secondary outcomes in case-control studies. The main idea behind the proposed estimators is the addition of an inverse probability weighted, control function that preserves unbiasedness of the estimating equation when a model for disease probability given covariates is correctly specified. Additional required assumptions are correct specification of the population mean model and known sampling fractions for the case-control study. No other distributional assumptions are made about the outcome.  We propose estimators that may be used with identity and log links. This approach could potentially extend to the logit link, which presents a challenge for future research. 

 The control function estimator is unbiased under correct specification of the disease model given covariates, even if the model for the selection bias function is misspecified. We recommend evaluating the disease model fit with respect to the model predictions (estimated disease probabilities). One can use Area Under the operating Curve (AUC) and cross validation as measures that give indications of fit due to good or poor prediction. For a comprehensive review of such methods see \citet{Harrell1996}. It is also useful to compare the control function effect estimate to the IPW, as the IPW is robust to misspecification of the disease model. Under correct specification of the disease model we expect to see similar effect estimates for both IPW and control function estimators, with smaller standard errors for the later. 

  Robustness, as is here attributed to the IPW and control function estimator, is somewhat different than the robustness property in  robust statistics. Operating in the semiparametric theory framework, we carefully define the modeling assumptions required to produce unbiased estimators. For instance, correct specification of the population mean, and of the disease model. Robustness here refers to the fact that our estimator will be consistent for all data generating mechanisms in which these assumption hold (e.g. normal errors, but also t-distributed errors, or even non-symmetric errors). In comparison, in robust statistics  the focus is on estimators that are protected against outliers, which is (conceptually) more oriented towards smaller sample sizes.

 In  recent work, especially that relying on the retrospective likelihood \citep{Lin2009,Li2012,Chen2013,Ghosh2013}, the primary disease probability is modeled in a logistic regression, with both the exposure and the secondary outcome, and sometimes their interaction, as predictors. Our formulation does not explicitly use the secondary outcome in the disease model. However, efficient control function estimator incorporates a selection bias function, which encodes the association between the secondary outcome and the case control status conditional on covariates. Hence, as in any likelihood based approach, this association is accounted for, while more general specifications of this association are readily applied. Although the control function estimator  is far more general and relies on fewer assumptions, it is guaranteed to be most efficient if all models are correctly specified.  We also note that in many settings, the secondary outcome may occur on the causal pathway between the exposure and the primary outcome (e.g. mammographic density and breast cancer, or smoking and lung cancer) in which case  the model for the D adjusting for X and Y is difficult to interpret.

\section*{Acknowledgements}

This research was supported by NIH grants R21ES019712 and R01ES020337.  \vspace*{-8pt}



\bibliographystyle{ims}

\section*{Appendix}

\section{What are regular estimators?}
According to the definition in \citet{Tsiatis2007}, it is assumed that there is a true distribution generating the data, indexed by a parameter $\theta$. In practice, a sampled data set is distributed according to  $\theta_n$ where $\theta_n$  is $\sqrt{n}$-consistent for the true $\theta$. This process by which data are sampled from a $\sqrt{n}$ perturbation of the truth is called a Òlocal data generating processÓ. Regularity, or ``local uniform consistency", means that the estimator for $\bbeta$ (some parameter of the distribution indexed by $\theta$) does not depend on the local data generating process.

\section{Mathematical derivations.}

\renewcommand{\theequation}{B. \arabic{equation}} 
 
\setcounter{equation}{0} 

\subsection{The tangent space of a model for $p(\bX)$ in a case-control study}
We here show that the tangent space, or the collection of scores, for $p(\bX)$ (disease probability in the general population) in a case-control study is related to the tangent space for $p_{cc}(\bX)$ (disease probability in the case-control study population) via the ``scaling factor" $p_{cc}(\bX)/p(\bX)$. Or in other words, a score for $p_{cc}(\bX)$ evaluated in a case-control study is multiplied by this scaling factor to obtain a score for $p(\bX)$. A general score for the disease probability $p_{cc}(\bX)$ in the case control study is given by:
\[
Sh(\bX)\big\{D - p_{cc}(\bX)\big\}.
\]
Recall the identity 
\begin{eqnarray*}
\mbox{logit}p(\bX) &= &\mbox{logit} p_{cc}(\bX) + \log\bigg[\frac{p(D=1)\big\{1- p(D=1|S=1)\big\}}{p(D=1|S=1)\big\{1-p(D=1)\big\}}  \bigg] \\
\Downarrow\hspace{0.8in}&&\\
\frac{p(\bX)}{1-p(\bX)} &=& \frac{p_{cc}(\bX)}{1-p_{cc}(\bX)}\bigg[\frac{p(D=1)\big\{1- p(D=1|S=1)\big\}}{p(D=1|S=1)\big\{1-p(D=1)\big\}}  \bigg].
\end{eqnarray*}
We make a few transformations in order to write this score in terms of $\big\{D - p(\bX)\big\}$:
\begin{eqnarray*}
Sh(\bX)\big\{D - p_{cc}(\bX)\big\} &=& Sh(\bX)\frac{\big\{D - p_{cc}(\bX)\big\}}{p_{cc}(\bX)\big\{1-p_{cc}(\bX)  \big\}}p_{cc}(\bX)\big\{1-p_{cc}(\bX)  \big\} \\
&=& Sh(\bX)\frac{(-1)^{1-D}}{p_{cc}^D(\bX)\big\{1-p_{cc}(\bX)  \big\}^{1-D}}p_{cc}(\bX)\big\{1-p_{cc}(\bX)  \big\} \\
&=& Sh(\bX)\frac{(-1)^{1-D}p_{cc}(\bX)}{\left\{ \frac{p_{cc}(\bX)} {1-p_{cc}(\bX) } \right\}^D} \\
&=& Sh(\bX)\frac{(-1)^{1-D}p_{cc}(\bX)}{\left( \bigg[\frac{p(\bX)} {1-p(\bX) } \bigg] \bigg[\frac{p(D=1|S=1)\left\{1-p(D=1)\right\}}{p(D=1)\left\{1- p(D=1|S=1)\right\}}  \bigg] \right)^D} 
\end{eqnarray*}
Since:
\[
\frac{(-1)^{(1-D)}}{p(\bX)^{D-1}\{1-p(\bX)\}^{-D}} = D- p(\bX),
\]
we get:
\begin{eqnarray*}
Sh(\bX)\big\{D - p_{cc}(\bX)\big\} 
&=& Sh(\bX)\frac{p_{cc}(\bX)}{p(\bX)}\big\{D -p(\bX)\big\}  \bigg\{\frac{p(D=1)p(D=0|S=1)}{p(D=1|S=1)p(D=0)}  \bigg\}^D\\
&&\hspace{-1.5 in}\propto\ \ Sh(\bX)\frac{p_{cc}(\bX)}{p(\bX)}\big\{D -p(\bX)\big\}  \bigg\{\frac{p(D=1)p(D=0|S=1)}{p(D=1|S=1)p(D=0)}  \bigg\}^D\cdot\frac{p(D=0)}{p(D=0|S=1)p(S=1)}\\
&=&\frac{S}{p(S=1|D)}\frac{p_{cc}(\bX)}{p(\bX)}h(\bX)\big\{D -p(\bX)\big\}.
\end{eqnarray*}
As required.

In the main manuscript, we showed that scores of the tangent space in the nonparametric model are 
\[
\frac{S}{p(S=1|D)}h(\bX)\big\{D -p(\bX)\big\},
\]
and this holds since the function $h(\bX)$ can be written as $\tilde h(\bX)p_{cc}(\bX)/p(\bX)$ for some $\tilde h(\bX) = h(\bX)p(\bX)/p_{cc}(\bX)$. However, in the parametric and nonparametric cases, $C^T\bX \ne p_{cc}(\bX)/p(\bX) \tilde C^T\bX$, since $p_{cc}(\bX)/p(\bX)$ is not fixed. 


\subsection{Proof of Theorem 1}\label{Appendix:thm1}
Before approaching this proof, Lemma 1 provides  the form of $h_2(\bX,D)$ in each of the link functions under consideration.

\noindent \textbf{Lemma 1}\textit{
\begin{enumerate}[(a)] 
\item Any function $h_2(\bX,D)$ such that $\bbE\{h(\bX,D)|\bX\} = 0$, where the expectation is taken in the general population, can be written as 
\[
h_2(\bX,D) = \gamma(\bX)\{D- p(\bX)\},
\]
for any function $\gamma(\bX)$. This parametrization will be used in the linear link ca se. 
\item $h_2(\bX,D)$ can equivalently be written in the form 
\[
h_2(\bX,D) = h(\bX)\big[ 1- \exp\left ( \nu(\bX,D) - \log \bbE\left[\exp \big\{\nu(\bX,D)\big\} |\bX\right] \right) \big ],
\]
where $h(\bX)$ is any function of $\bX$, and $\nu(\bX,D)$ is such that $\nu(\bX,0 ) = 0$. We will use this parametrization in the log link case. 
\end{enumerate}
}

\noindent \textbf{Proof of Lemma 1}
\begin{enumerate}
\item Define the two sets $\mathcal{A}_1 = \{h_2(\bX,D): \bbE\{h(\bX,D)|\bX\} = 0   \}$ (where the expectation is taken in the general population) and $\mathcal{A}_2 = \{ \gamma(\bX)\{D-p(\bX)\}:\gamma(\bX) \mbox{ any function of } $\bX$ \}$. We show that the two sets are equal. The first direction, $\mathcal{A}_2 \subseteq \mathcal{A}_1$ is trivial, by noting that $\bbE(D|\bX) = p(\bX)$. To show that $\mathcal{A}_1\subseteq\mathcal{A}_2$, let $h_2(\bX,D)$ be an element of $\mathcal{A}_1$. We show that it is also an element of $\mathcal{A}_2$.  Choose $\gamma(\bX) = h_2(\bX,1) - h_2(\bX,0)$. Then we can verify that for this choice of $\gamma(\bX)$, indeed $h_2(\bX,D) = \gamma(\bX)\{D- p(\bX)\} = \{h_2(\bX,1) - h_2(\bX,0)\}\{D-p(\bX)\}$.

For $D=1$, we have that $h_1(\bX,1) = \gamma(\bX)\{1-p(\bX)\}$ yields $h_2(\bX,0) = -\{h_2(\bX,1) -h_2(\bX,0)\}p(\bX)$, and for $D=0$, $h_1(\bX,0) = \gamma(\bX)\{0-p(\bX)\}$ also gives $h_2(\bX,0) = -\{h_2(\bX,1) -h_2(\bX,0)\}p(\bX)$. This equality is true: $\bbE\{h_2(\bX,D)|\bX\} = 0 = h_2(\bX,0)\{1 - p(\bX)\} + h_2(\bX,1)p(\bX)$.

\item First, rewrite 
\begin{eqnarray*}  h(\bX)\left\{ 1- \exp\left( \nu(\bX,D) - \log \bbE\left[\exp \big\{\nu(\bX,D)\big\} \big|\bX\right] \right)\right \}  & =& \\
&& \hspace{-2in} \frac{h(\bX)}{\bbE[\exp \{\nu(\bX,D)\} |\bX]}\left( \bbE[\exp \{\nu(\bX,D)\} |\bX]- \exp\big \{ \nu(\bX,D)  \big\}\right)  .
 \end{eqnarray*}
 We show that 
 \[  \bbE[\exp \{\nu(\bX,D)\} |\bX]- \exp\big \{ \nu(\bX,D)  \big\}   = \{p(\bX) - D\}[\exp\{\nu(\bX,1)\} - \exp\{\nu(\bX,0)\}],\]
 and therefore $\gamma(\bX) = h(\bX)/\big(  \bbE[\exp \{\nu(\bX,D)\} |\bX] [\exp\{\nu(\bX,1)\} - \exp\{\nu(\bX,0)\}]  \big)$. To show the required equality, notice that since $D$ is binary: 
 \[\exp\{\nu(\bX,D)\} = D\big[\exp\{\nu(\bX,1)\} - \exp\{\nu(\bX,0)\}\big] + \exp\{\nu(\bX,0)\}.\]
 Writing $\bbE[\exp \{\nu(\bX,D)\} |\bX]$ using simple algebra, the results follows.  $\qed$
\end{enumerate}

\noindent \textbf{Proof of the theorem.}
Recall 
\[
\bU_{cont}(\bbeta) =\sum_{i=1}^n\frac{S_i}{\pi(D_i)}\bigg( h_1(\bX_i)\big[Y_i - g^{-1}\{\mu(\bX_i;\bbeta) \}   \big] - h_2(\bX_i,D_i)    \bigg).
\]
Consider the parametric submodel $f_t(O) = f_t(Y|D,S=1,\bX)f(S=1|D)f_t(D|\bX)f_t(\bX)$, where $f_{t=0}(O) = f(O)$ is the true law. Denote by $\bbS^{sub}(O) = \bbS^{sub}(Y|D,S=1,\bX) + \bbS^{sub}(D|\bX) + \bbS^{sub}(\bX)$ the scores in the submodel  (e.g. $\bbS^{sub}(O) = \partial/\partial t\ \log\{ f_t(O)\}$, etc.) 

Let:
\[
\Psi_t(\bbeta, h_1, h_2) =\bbE_t\bigg\{  \frac{S}{\pi(D)}\bigg( h_1(\bX)\big[Y - g^{-1}\{\mu(\bX;\bbeta) \}   \big] - h_2(\bX,D)    \bigg)\bigg\}
\]
under the submodel, where $\bbeta$ may not be the true value $\bbeta_0$. Then, (assuming that integration and differentiation are exchangeable),

\begin{eqnarray*}
\frac{\partial \Psi_t(\bbeta, h_1, h_2)}{\partial t}\bigg|_{t = 0} &=& \frac{\partial}{\partial t}  \bbE_t\bigg\{  \frac{S}{\pi(D)}\bigg( h_1(\bX)\big[Y - g^{-1}\{\mu(\bX;\bbeta) \}   \big] - h_{2,t}(\bX,D)    \bigg)\bigg\} \\
&=&  \bbE\bigg\{ \bbS(O)  \frac{S}{\pi(D)}\bigg( h_1(\bX)\big[Y - g^{-1}\{\mu(\bX;\bbeta) \}   \big] - h_{2}(\bX,D)    \bigg)   \bigg\} \\
&& +  \frac{\partial}{\partial t}  \bbE\bigg\{  \frac{S}{\pi(D)}\bigg( h_1(\bX)\big[Y -g^{-1}\{\mu(\bX;\bbeta) \}   \big] - h_{2,t}(\bX,D)    \bigg) \bigg\}.
\end{eqnarray*}
Consider the second argument. 
\begin{eqnarray*}
 \frac{\partial}{\partial t}  \bbE\bigg\{  \frac{S}{\pi(D)}\bigg( h_1(\bX)\big[Y - g^{-1}\{\mu(\bX;\bbeta) \}   \big] - h_{2,t}(\bX,D)    \bigg)  \bigg\} &=& 
- \frac{\partial}{\partial t}  \bbE\bigg\{  \frac{S}{\pi(D)} h_{2,t}(\bX,D)    \bigg\}.
 \end{eqnarray*}

From Lemma 1 (a), with the log link function we have:

\begin{eqnarray*}
 \frac{\partial}{\partial t}  \bbE\bigg\{  \frac{S}{\pi(D)}\bigg( h_1(\bX)\big[Y - g^{-1}\{\mu(\bX;\bbeta) \}   \big] - h_{2,t}(\bX,D)    \bigg)  \bigg\} &=& \\
&& \hspace{-4in}  -\frac{\partial}{\partial t}  \bbE\bigg\{ \frac{S}{\pi(D)}  h_{2,t}(\bX,D)    \bigg\} =  -\frac{\partial}{\partial t}  \bbE\bigg\{ h(\bX) \exp\big ( \nu(\bX,D) - \log \bbE_t[\exp \{\nu(\bX,D)\} |\bX] \big)  \bigg\}\\
&& \hspace{-3in}=   -\bbE\bigg[  h(\bX) \exp \{\nu(\bX,D)\}   \frac{\partial}{\partial t}   \bbE_t[\exp \{\nu(\bX,D)\} |\bX]^{-1}   \bigg]\\
&& \hspace{-3in}=  \bbE\bigg(  h(\bX) \bbE\big[ \exp \{\nu(\bX,D)\} \big| \bX\big]    \bbE[\exp \{\nu(\bX,D)\} |\bX]^{-2}\frac{\partial}{\partial t}   \bbE_t[\exp \{\nu(\bX,D)\} |\bX]  \bigg)\\
&& \hspace{-3in}=  \bbE\bigg(  h(\bX)   \bbE\left [\exp \{\nu(\bX,D)\} \big|\bX\right]^{-1} \bbE\left [\exp \{\nu(\bX,D)\}\bbS^{sub}(D|\bX) \big |\bX\right ]   \bigg)\\
&& \hspace{-3in}=  \bbE\bigg[  h(\bX)   \bbE\left \{\exp\big( \nu(\bX,D)-\log  \bbE\left [\exp \{\nu(\bX,D)\} |\bX\right]\big)\bbS^{sub}(D|\bX) \big |\bX\right \}  \bigg]\\
&& \hspace{-3in}=  \bbE\bigg\{  h(\bX) \exp\big( \nu(\bX,D)-\log  \bbE[\exp \{\nu(\bX,D)\} |\bX]\big)\bbS^{sub}(D|\bX)    \bigg\}\\
%
%
%
&& \hspace{-3in}=  \bbE\bigg\{  h_2(\bX,D)  \bbS^{sub}(D|\bX)  \bigg\} =  \bbE\bigg\{ \frac{S}{\pi(D)} h_2(\bX,D) \bbS^{sub}(D|\bX)   \bigg\}.
\end{eqnarray*}


We now show the same result for the identity link. We use Lemma 1 (b) and get: 
\begin{eqnarray*}
 \frac{\partial}{\partial t}  \bbE\bigg\{  \frac{S}{\pi(D)}\bigg( h_1(\bX)\big[Y - g^{-1}\{\mu(\bX;\bbeta) \}   \big] - h_{2,t}(\bX,D)   \bigg) \bigg\}  &=& 
- \frac{\partial}{\partial t}  \bbE\bigg\}  \frac{S}{\pi(D)}  h_{2,t}(\bX,D)    \bigg\} \\
&&\hspace{-3.5in}=  -\frac{\partial}{\partial t}  \bbE\bigg[  \frac{S}{\pi(D)}\gamma(\bX) \{ D-p_t(\bX)  \} \bigg]
 =  \frac{\partial}{\partial t}\bbE \bigg\{ \frac{S}{\pi(D)} \gamma(\bX)p_t(\bX)  \bigg\} \\
&&\hspace{-3.5in}=  \bbE \bigg[ \frac{S}{\pi(D)} \gamma(\bX)\bbE\big\{ D\bbS^{sub}(D|\bX)   \big|\bX\big\}  \bigg] 
=   \bbE \bigg( \frac{S}{\pi(D)} \gamma(\bX)\bbE\big[ \{D - p(\bX)\}\bbS^{sub}(D|\bX)   \big|\bX\big]  \bigg) \\
&& \hspace{-3.5in}    =  \bbE \bigg[ \frac{S}{\pi(D)} \gamma(\bX) \{D - p(\bX)\}\bbS^{sub}(D|\bX)     \bigg]
 =  \bbE\bigg\{ \frac{S}{\pi(D)} h_2(\bX,D) \bbS^{sub}(D|\bX)   \bigg\}.
\end{eqnarray*}
Recall that $p(\bX)$ is restricted via some nonparametric, semiparameteric or parameteric model, and denote its  tangent space  by 
$\Lambda_{D,sub} \subseteq \Lambda_{D,npar}  $ where $\Lambda_{D,npar} $ is the tangent space in the unrestricted model for $p(\bX)$. 
The score $\bbS^{sub}(D|\bX)$ satisfies $\bbS^{sub}(D|\bX) \in \Lambda_{D,sub}$, since this tangent space is spanned by all scores of in the submodel for $p(D|\bX)$. Therefore, $\bbS^{sub}(D|\bX)$ is orthogonal to the orthocomplement of the submodel tangent space $ \Lambda_{D,sub}^{\perp}$. Denote the projection of a vector $v$ on a space $\bU$ by $\Pi\big( v\big|\bU \big)$. We can decompose
\begin{eqnarray*}
\frac{S}{\pi(D)} h_2(\bX,D) &=&  \Pi\bigg(  \frac{S}{\pi(D)} h_2(\bX,D) \bigg| \Lambda_{D,sub}  \bigg)  + \Pi\bigg(  \frac{S}{\pi(D)} h_2(\bX,D) \bigg| \Lambda_{D,sub}^{\perp}  \bigg)
\end{eqnarray*}
and the latter term is orthogonal to $\bbS^{sub}(D|\bX)$. Thus,
\[
\bbE \bigg\{ \frac{S}{\pi(D)} h_2(\bX,D)\bbS^{sub}(D|\bX)     \bigg\} = \bbE \bigg\{ \Pi \bigg( \frac{S}{\pi(D)} h_2(\bX,D)\bigg| \Lambda_{D,sub} \bigg)\bbS^{sub}(D|\bX)     \bigg\}.
\]
It follows that
\begin{eqnarray}\nonumber
\hspace{-1in}\frac{\partial}{\partial t}\Psi(\bbeta, h_1, h_2)\bigg|_{t=0}
&=&   \bbE\bigg\{ \bbS^{sub}(O) \frac{S}{\pi(D)}\bigg( h_1(\bX)\big[Y- g^{-1}\{\mu(\bX;\bbeta)\}\big]- h_2(\bX,D)  \bigg)   \bigg\}\\ \label{eq:appA_1}
&& -\bbE\bigg\{ \Pi\bigg(  \frac{S}{\pi(D)} h_2(\bX,D) \bigg| \Lambda_{D,sub}  \bigg)     \bbS^{sub}(D|\bX)      \bigg\}  .
\end{eqnarray}
To complete, it suffices to note that 
\begin{equation}\label{eq:appA_2}
 \bbE \bigg[\Pi\bigg( \frac{S}{\pi(D)} h_2(\bX,D) \bigg| \Lambda_{D,sub} \bigg) \bigg\{ \bbS^{sub}(\bX) + \bbS^{sub}(Y|D,\bX)\bigg\}     \bigg] = 0.
\end{equation}
Combining identities (\ref{eq:appA_1}) and (\ref{eq:appA_2}), and since every influence functions $\psi$ in the restricted model satisfies the following equation:
\[
\frac{\partial \Psi_t(\bbeta, h_1, h_2)}{\partial t}\bigg|_{t = 0}  = \bbE\{  \bbS^{sub}(O)  \psi^T \} ,\]
it follows that every influence function in the restricted model is of the form
\[
\frac{Sh_1(\bX)}{\pi(D)}\bigg[ Y- g^{-1}\{\mu(\bX; \bbeta)\}\bigg]  - \frac{S}{\pi(D)} h_2(\bX,D)  +  \Pi\bigg(  \frac{S}{\pi(D)} h_2(\bX,D) \bigg| \Lambda_{D,sub}  \bigg). \hspace{1.1in} \qed
\]

\subsection{Proof of Corollary 1}
Corollary 1 follows since if $p(\bX)$ is unrestricted, then so is the tangent space unrestricted $ \Lambda_{d,sub}  = \Lambda_{D,npar}$, and the projection of a vector on the submodel tangent space does not change the vector, i.e. 
\[
 \Pi\bigg(  \frac{S}{\pi(D)} h_2(\bX,D) \bigg| \Lambda_{D,sub}  \bigg) = \frac{S}{\pi(D)} h_2(\bX,D),
\]
so that the influence function has to be
\[
\frac{Sh_1(\bX)}{\pi(D)}\bigg[Y- g^{-1}\{\mu(\bX; \bbeta)\}\bigg],
\]
the IPW influence function. $\qed$

\subsection{Proof of Theorem 2}

Suppose that $h_1(\bX)$ is fixed, and $p(\bX)$ is known. We here find the function $h_2(\bX,D)$ that minimizes the variance over all functions in the submodel tangent space. First, note that we can write the influence functions for $\bbeta$ in the form:
\begin{eqnarray*}
\psi(\bbeta) &=& \frac{Sh_1(\bX)}{\pi(D)}\bigg[ Y- g^{-1}\{\mu(\bX; \bbeta)\}\bigg]  -  \Pi\bigg(  \frac{S}{\pi(D)} h_2(\bX,D) \bigg| \Lambda^{\perp}_{D,sub}\cap \Lambda_{D, npar}  \bigg)\\
&=& \Pi\bigg(  \frac{Sh_1(\bX)}{\pi(D)}\bigg[ Y- g^{-1}\{\mu(\bX; \bbeta)\}\bigg]  \bigg| \Lambda^{\perp}_{D,sub}\cap \Lambda_{D, npar}  \bigg)   -   \Pi\bigg(  \frac{S}{\pi(D)} h_2(\bX,D) \bigg| \Lambda^{\perp}_{D,sub}\cap \Lambda_{D, npar}  \bigg)\\
&& +  \Pi\bigg(  \frac{Sh_1(\bX)}{\pi(D)}\bigg[ Y- g^{-1}\{\mu(\bX; \bbeta)\}\bigg]  \bigg| \Lambda_{D,sub}  \bigg) ,
\end{eqnarray*}
so that minimizing the variance of $\psi(\bbeta)$ over functions $h_2(\bX,D)$ is  equivalent to  minimizing the variance of the first two terms (since the third term is orthogonal to the term involving $h_2(\bX,D)$). Consider finding 
$h_2^{opt}(\bX,D)$ that satisfies the normal equations:
\begin{eqnarray*}
0&=& \bbE\bigg\{  \frac{S}{\pi(D)} h_2(\bX,D)    \bigg(  \frac{Sh_1(\bX)}{\pi(D)}[Y-  g^{-1}\{\mu(\bX; \bbeta)\}] -   \frac{S}{\pi(D)} h_2^{opt}(\bX,D) \bigg)  \bigg\}\\
&=& \bbE\bigg\{  \frac{S}{\pi(D)} h_2(\bX,D)    \bigg( \frac{Sh_1(\bX)}{\pi(D)}[\bbE(Y|\bX,D)-  g^{-1}\{\mu(\bX; \bbeta)\}] -   \frac{S}{\pi(D)} h_2^{opt}(\bX,D) \bigg)  \bigg\}.
\end{eqnarray*}
This equality is satisfied by
\[
h_2^{opt}(\bX,D) = h_1(\bX)\big [\bbE(Y|\bX,D)- g^{-1}\{\mu(\bX;\bbeta) \}\big], 
\]
as required. $\qed$

\subsection{Proof of Theorem 3}
We here find the function $h_1^{opt}(\bX)$ that using it in the estimating equation $\bU_{cont}(\bbeta)$ yields the most efficient (with minimal variance) estimator of $\wh \bbeta$.
 According to the generalized information equality \citep{Newey1994}, for every function $h_1(\bX)$:
\[
- \bbE\bigg[ \frac{\partial \bU^{opt}_{cont}\big\{ \bbeta; h_1(\bX)\big\}}{\partial \bbeta}\bigg|_{\bbeta = \bbeta_0}   \bigg] = 
\bbE\bigg[ \bU^{opt}_{cont}\big\{ \bbeta; h_1(\bX)\big\}\bU_{cont}\big\{ \bbeta; h_1^{opt}(\bX)\big\}^T \bigg|_{\bbeta = \bbeta_0} \bigg].
\]
Then:
\begin{eqnarray*}
 \bbE\bigg[ \frac{S}{\pi(D)}h_1(\bX)\frac{\partial g^{-1}\{\tilde \mu(\bX,D; \bbeta)\}}{\partial \bbeta}\bigg|_{\bbeta = \bbeta_0}\bigg] &=&\bbE\bigg[ h_1(\bX)\frac{\partial g^{-1}\{\tilde\mu(\bX,D; \bbeta_0)\}}{\partial \bbeta}\bigg] \\
 &&\hspace{-2in}= 
\bbE\bigg\{ h_1(\bX)h^{opt}_1(\bX) \bbE\left(\frac{1}{\pi(D)} \big[ Y - g^{-1}\{\tilde\mu(\bX,D; \bbeta_0) \}\big]^2    \bigg| \bX\right) \bigg\}.
\end{eqnarray*}
This equation is satisfied by:
\begin{eqnarray*}
\frac{\partial g^{-1}\{\tilde\mu(\bX,D; \bbeta_0)\}}{\partial \bbeta} =
 h^{opt}_1(\bX) \bbE\bigg(\frac{1}{\pi(D)} \big[ Y - g^{-1}\{\tilde\mu(\bX,D; \bbeta_0) \}\big]^2  \bigg| \bX\bigg) .\\
\end{eqnarray*}
Recall that $\tilde\mu(\bX,D; \bbeta) = g\{\bbE(Y|\bX,D)\}$. We can then write:
\begin{eqnarray*}
h^{opt}_1(\bX) & =&
\bbE\bigg[\frac{1}{\pi(D)} \big\{ Y - \bbE(Y|\bX,D)\big\}^2  \bigg| \bX\bigg]^{-1} \frac{\partial g^{-1}\{\tilde\mu(\bX,D; \bbeta_0)\}}{\partial \bbeta}  \\
&=& \bbE\bigg\{\frac{1}{\pi(D)} \mbox{Var}\big( Y\big |\bX,D\big)  \bigg| \bX\bigg\}^{-1} \frac{\partial g^{-1}\{\tilde\mu(\bX,D; \bbeta_0)\}}{\partial \bbeta},
\end{eqnarray*}
as required. $\qed$

\subsection{Deriving the locally semiparametric efficient influence function}\label{Appendix:influence_function}
We first derive the estimating equation for $\btheta$ accounting for the estimation of $\balpha$, and then provide the corresponding influence function for $\btheta$. Denote the true value of $\balpha$ by $\balpha_0$, and recall that $\bV(\balpha)$ is the estimating equation for $\balpha$, and denote for simplicity $\bU(\btheta;\balpha) = \bU_{cont}^{opt}(\btheta)$. (In fact, the following derivation holds for any estimating equation $\bU(\btheta; \balpha)$, in particular to any functions $h_1(\bX), h_2(\bX,D)$, not just the optimal ones).  Let $\bV_i(\balpha), \bU_i(\btheta; \balpha)$ be the contributions of the $i$th subject to the estimating equations.
To estimate $\balpha, \btheta$, one solves $\bbP_n\bV_i(\balpha) = 0, \ \bbP_n\bU_i(\btheta; \balpha) = 0$, where $\bbP_n(x_i) = 1/n\sum_{i=1}^nx_i$.

 Consider the following expansions of the estimating equations around $\balpha_0$:
\begin{eqnarray*}
\sqrt{n}\bbP_n\bU_i(\btheta; \wh\balpha) &=& \sqrt{n}\bbP_n\bU_i(\btheta; \balpha)\bigg|_{\balpha  = \balpha_0} + 
 \sqrt{n}\bbP_n\frac{\partial}{\partial \balpha}\bU_i(\btheta; \wh\balpha)\bigg|_{\balpha = \balpha_0}(\wh\balpha - \balpha_0) + o_p(1) \\
 &=&  \sqrt{n}\bbP_n\bU_i(\btheta; \balpha_0)  + 
E\left\{ \frac{\partial}{\partial \balpha}\bU(\btheta; \balpha_0) \right\}\sqrt{n}(\wh\balpha - \balpha_0) + o_p(1) .
\end{eqnarray*}
Similarly,
\begin{eqnarray*}
\sqrt{n}\bbP_n \bV_i(\wh\balpha) &=&  \sqrt{n}\bbP_n \bV_i(\balpha_0) +
E \left\{\frac{\partial}{\partial \balpha} \bV(\balpha_0)\right\} \sqrt{n}(\wh\balpha - \balpha_0) + o_p(1) .\\
\end{eqnarray*}
From the estimation procedure, we have that by definition $\sqrt{n}\bbP_n \bV_i(\wh\balpha) =0$. Therefore, combining these two equations we get:
\begin{eqnarray*}
\sqrt{n}\bbP_n\bU_i(\btheta; \wh\balpha) &=&
 \sqrt{n}\bbP_n\bigg[      
\bU_i(\btheta; \balpha_0) 
  - 
E\left\{ \frac{\partial}{\partial \balpha}\bU(\btheta; \balpha_0) \right\}E \left\{\frac{\partial}{\partial \balpha} \bV(\balpha_0)\right\}^{-1}\bV_i(\balpha_0) 
  \bigg] + o_p(1).
\end{eqnarray*}
So that there is an additional term, namely  $ \sqrt{n}\bbP_nE\big\{ \frac{\partial}{\partial \balpha}\bU(\btheta; \balpha_0) \big\}E \big\{\frac{\partial}{\partial \balpha} \bV(\balpha_0)\big\}^{-1}\bV_i(\balpha_0) $, that accounts for the estimation of $\balpha$.    Notice that in order to estimate $\btheta$ we do not in fact need to use this estimating equation, since 
$ \sqrt{n}\bbP_n\bV_i(\balpha_0) $ is estimated by $ \sqrt{n}\bbP_n\bV_i(\wh \balpha)  = 0$. However, for the purpose of variance estimation, it is important to use this estimating equation and account for the estimation of $\balpha$. 

Using the same technic, we obtain
\begin{eqnarray*}
\sqrt{n}\bbP_n \bU_i(\wh\btheta;\wh \balpha) &=&  \sqrt{n}\bbP_n \bU_i(\btheta_0;\wh\balpha) +
E \left\{\frac{\partial}{\partial \btheta} \bU(\btheta_0; \wh\balpha)\right\} \sqrt{n}(\wh\btheta - \btheta_0) + o_p(1), \\
\end{eqnarray*}
and since $\sqrt{n}\bbP_n \bU_i(\wh\btheta;\wh \balpha) = 0$, we get:
\[
\sqrt{n}(\wh\btheta - \btheta_0) = \sqrt{n}\bbP_n\bigg[ E \bigg\{\frac{\partial}{\partial \btheta} \bU(\btheta_0; \wh\balpha)\bigg\}^{-1}  \bU_i(\btheta_0;\wh\balpha)   \bigg] 
+ o_p(1),
\]
and we see that $\wh\btheta$ is an asymptotically linear estimator with the $i$th influence function given by
\[
\psi_i(\btheta;\balpha) = E \bigg\{\frac{\partial}{\partial \btheta} \bU_i(\btheta_0; \wh\balpha)\bigg\}^{-1}  \bU_i(\btheta_0;\wh\balpha).
\]
Notice that 
\begin{eqnarray*}
\frac{\partial}{\partial \btheta} \bU_i(\btheta_0; \wh\balpha) &=& \frac{\partial}{\partial \btheta} \bigg[  \bU_i(\btheta; \balpha_0) 
  - 
E\bigg\{ \frac{\partial}{\partial \balpha}\bU(\btheta; \balpha_0) \bigg\}E \bigg\{\frac{\partial}{\partial \balpha} \bV(\balpha_0)\bigg\}^{-1}\bV_i(\balpha_0) 
   \bigg] \\ 
&=&  \frac{\partial}{\partial \btheta}   \bU_i(\btheta; \balpha_0) 
  - 
E\bigg\{ \frac{\partial}{\partial \balpha}\bU(\btheta; \balpha_0) \bigg\}E \bigg\{\frac{\partial}{\partial \balpha}  \bV(\balpha_0)\bigg\}^{-1}\frac{\partial}{\partial \btheta} \bV_i(\balpha_0) 
 \\
&=&  \frac{\partial}{\partial \btheta}   \bU_i(\btheta; \balpha_0) ,
\end{eqnarray*}
since $\bV_i(\balpha_0) $ does not depend on $\btheta$.

\subsection{Proof of Corollary 2}
Under the standard regularity conditions found in \citet{van2000},  the asymptotic normality of $\wh\btheta$ follows  from the central limit theorem, and its mean and covariance are as indicated since we assume that the models for $\wh \btheta = (\bbeta, \bdelta)$ are correctly specified, so that $\psi(\btheta;\balpha)$ has mean zero. Local efficiency follows from Theorem 3, in which we provide the efficient influence function, and from the definition of local efficiency \citep{Newey1990}. $\qed$

\section{Computation of the control function estimator}
\renewcommand{\theequation}{C. \arabic{equation}} 
 
\setcounter{equation}{0} 

Here we describe how to compute  estimators of $\bbeta$ for  the identity and log links,  when $p(\bX)$ is modeled parametrically with $p(\bX; \balpha)$. In general, to find the estimator $\wh\bbeta$ we need to solve the estimating equation $\wh\bU_{cont}^{opt}(\bbeta)=0$, defined as $\bU^{opt}_{cont}(\bbeta)$ with $\wh h_1(\bX), \wh h_2(\bX,D)$, and $\wh p(\bX)$. This can be performed using the Newton-Raphson (NR) algorithm. 

 Let $\bdelta$ denote the parameters for either $\nu(\bX,D; \bdelta)$ (log link) or $\gamma(\bX; \bdelta)$ (identity link). Let $\btheta = (\bbeta^T, \bdelta^T)^T$. It is convenient to estimate $\btheta$ jointly, by modifying the estimating equation $\bU^{opt}_{cont}(\bbeta)$ to define $\bU^{opt}_{cont}(\btheta)$ by taking
\[
h_1^{opt}(\bX) = \bbE\bigg\{ \frac{1}{\pi(D)}\mbox{var}(Y|D,\bX) \bigg| \bX\bigg\}^{-1}\frac{\partial}{\partial \btheta}\big[ g^{-1}\big\{\mu(\bX,D;\btheta)  \big\}   \big].
\]

The estimation procedure takes the  following  steps:
\begin{enumerate}
\item Estimate the parameters of $p(D=1| \bX, S = 1)$, the probability of disease conditional on covariates in the case-control study population, using logistic regression with an offset, by exploiting the known relationship between disease probability in the population to disease probability in the case-control sample:
\begin{equation}\label{eq:compute_1}
\mbox{logit}p(\bX) = \mbox{logit} p(D=1|\bX, S=1) + \log\bigg[\frac{p(D=1)\big\{1- p(D=1|S=1)\big\}}{p(D=1|S=1)\big\{1-p(D=1)\big\}}  \bigg]
\end{equation}
where $p(D=1)$ is the disease prevalence in the general population, and $p(D=1|S=1)$ is the fraction of cases in the case-control sample.

\item Obtain {\it starting values} for $\btheta$ according to the specifics given below. 
\item Plug $\wh\balpha$ into $\bU_{cont}^{opt}(\btheta)$ and solve $\wh \bU^{opt}_{cont}( \wh\btheta)=0$ using NR.
\end{enumerate}
This procedure is implemented in the R package RECSO \citep{Recso}. 
Note that the estimating equation $\bU^{opt}_{cont}(\btheta)$ is geared towards increasing the efficiency of  the estimator for $\bbeta$, so $\wh \bdelta$ may not be an efficient estimator. 

Next we detail the estimation procedure for $p(D=1|\bX,S=1)$, and $\btheta$ for each choice of link function.  

\subsection{Computation of  $\wh p(D=1|\bX,S=1)$ using a parametric model}

Let $\bV(\balpha)$ be the estimating equation for parameters $\balpha$ of $p(D=1|\bX,S =1; \balpha)$. In the simple logistic model, it is given by:
\[
\bV(\balpha) = \sum_{i = 1}^nS_i\bx_i\{D_i -p(D_i=1|\bX_i, S_i=1)\}
\]
where $p(D_i=1|\bx_i, S_i=1)$ is modeled through the inverse of the logit transformation, i.e. $p(D_i=1|\bx_i, S_i=1) = \exp(\balpha^T\bx_i)/\{1 + \exp(\balpha^T\bx_i)\}$. Here, we do not correct for the sampling bias resulting from the case-control ascertainment (e.g. we do not use IPW), but to later obtain estimates of the disease probability $\wh p(\bX)$ we use the correction (\ref{eq:compute_1}).

\subsection{Identity link}
To solve the estimating equation $\wh\bU_{ident}(\bbeta) = 0$ for $\bbeta$,  we follow the steps described above. First, we estimate $\wh\balpha$. Second, we calculate staring values for $\btheta$. 
 For staring values of $\bdelta$, we can estimate $\wh\gamma(\bX)$ by regressing  $Y$ on the covariates $\bX$ in the cases and control groups separately, calculating the predicted means for each subject under the two models, and taking the difference.  Initial estimators of $\bdelta$ can then be obtained by regressing the calculated differences on a given design matrix, say, if a linear model is assumed. Starting value for $\bbeta$ could be obtained as the IPW estimator. At the third step we solve
\[
0 = \sum_{i=1}^n \frac{h_1^{opt}(\bX_i)S_i}{\pi(D_i)}\bigg[ \big\{Y_i - \mu(\bX_i;\bbeta)   \big\}  -  \big\{D_i-\wh p(\bX_i) \big\} \gamma(\bX_i; \bdelta) \bigg]
\]
using NR iterations, by which we update the estimated $\wh\btheta$ until convergence.

 Note that for $h_1^{opt}(\bX)$ we need to estimate $\mbox{Var}\big(Y\big|D,\bX\big)$. When the outcome is continuous, it is convenient to assume homoscedasticity , in which case $h_1^{opt}(\bX_i)$ can be chosen 
 \[ \sum_{d\in\{0,1\}}\bigg[ \frac{1}{\pi(d)}\frac{1}{n}\sum_{j=1}^n\big(y_i - \mu(\bX_i;\wh\bbeta) -\wh\gamma(\bX_i)\{d - \wh p(\bX_i)\big\}^2  \bigg] p(D_i = d) \]
 The estimate  $\wh\bbeta$ will remain consistent even if the homoscedasticity assumption does not hold.

\subsection{Log link}

As in the identity link case, we start by estimating $\wh\balpha$, as described earlier. At the second step, we calculate starting values for $\btheta$. $\wh\nu(\bX,D)$ could be estimated, for instance, by estimating the parameters of a generalized linear model with the log link function, of $Y$ on the covariates $\bX$ in the cases and controls separately,  calculating the predicted means $\bbE(Y_i|\bX_j, D_i=0) $ and $\bbE(Y_i|\bX_j, D_i=1)$ for every subject $i$, and plugging-in to the equation for $\nu(\bX,D)$ for each subject. We can then estimate an initial $\wh\bdelta$ based on a model.  A starting value for $\bbeta$ could be the IPW estimator. We can  proceed to the third step and solve
\[
0 = \sum_{i = 1}^n\frac{h_1^{opt}(\bX_i)S_i}{\pi(D_i)}\bigg( Y_i - \exp\big[\mu(\bX_i;\bbeta)  + \nu^{opt}(\bX_i,D_i; \bdelta)- {\bar{\nu}}\{\bX_i; \bdelta, \wh{p}(\bX_i)\}  \big]  \bigg)
\]
for $\btheta$ using NR iterations.

 Note that at the $k$th iteration, we also need to estimate $h_1^{opt}(\bX)$. We can either update the estimate of $h_1^{opt}(\bX)$ at the $k$th iteration, using   the estimated $\wh\btheta$ from the $(k-1)$th iteration, or we can use a plug-in estimator based on the initial estimator  of $\btheta$. Usually the latter option is more stable (updating $h_1^{opt}(\bX)$ may lead to convergence problems). Note that for $h_1^{opt}(\bX)$ one needs an estimate of 
\[
\bbE\bigg\{ \frac{1}{\pi(D_i)}\mbox{Var}(Y_i|\bX_i,D_i)   \bigg| \bX_i\bigg\}= \sum_{d\in\{0,1\}}\bigg\{ \frac{1}{\pi(D_i)}\mbox{Var}(Y_i|\bX_i,D_i)p(D_i=d|\bX_i)    \bigg\}
\]
for each subject $i, i = 1,\ \ldots, n$. In the case of a Poisson model, we can simply use the predicted means, as $\wh{\mbox{Var}}(Y|\bX,D) = \wh{\bbE}(Y|\bX,D_i) =  \exp\big\{\mu(\bX;\wh\bbeta)  + \nu^{opt}(\bX,D;\wh \bdelta)- {\bar{\nu}}(\bX; \wh\bdelta)  \big\}$. As before, these predicted means could be updated at each iteration or be based on the initial estimators (the more stable option).

\section{Identity link simulations - additional information}
\subsection{Simulation study with a single exposure variable}

The simulation study described here, is similar to the identity link simulation study presented in the manuscript (Section 4.1), but simpler, so that only a single exposure variable is used. In this simulation we implemented and compared the estimator TT of \citet{TchetgenTchetgen2013}, which this estimator is not presented in the more complex simulation studies in the manuscript, as it then suffered from convergence problems. The TT estimator was calculated using maximum likelihood, and the robust standard error estimators. Results are provided under correct specification of the selection bias function $\gamma(\bX)$ (TT-cor) and under misspecification (TT-mis). 

The simulation was generated as follows. As in the simulation study presented in the main manuscript, first, an exposures variables $X_1$ was sampled with distribution  $X_1\sim \mathcal{N}(2,4)$.
 Then, disease probabilities were calculated for each subject, from the model 
\[
\text{logit}\left\{p(D=1|\bX)\right\} = -3.2+ 0.3X_1,
\]
and disease status was sampled. Then, the conditional mean of the secondary outcome was set to 
\[
\bbE(Y|\bX,D) = 50+ 4X_1+ \{D-p(\bX)\}(3 + 2X_1),
\]
so that the population mean is $\mu(\bX,\bbeta) = \bX^T\bbeta$ with $\bX = (1, X_1)^T$ and $\bbeta = (50,4)^T$, and $\gamma(\bX) = \bX^T\balpha$ with $\balpha = (3,2)^T$.
Finally, the residuals were normally distributed, so that $Y_i$ was sampled from:
\[
Y_i =  \bbE(Y|\bX_i,D_i)  + \epsilon_i, \mbox{ with } \epsilon_i\sim \mathcal{N}(0,4).
\]

All estimators  estimated the sample mean based on the full design matrix, i.e. with $\bX = (1, X_1)^T$. TT and the  control function estimator estimated $\gamma(\bX)$. When the model was correctly specified, the design matrix in the model for $\gamma(\bX)$ was taken to include all the terms $\bX = (1, X_1)^T$, but when the model was incorrectly specified, it only had the intercept, i.e.  $\bX = 1$.

\begin{figure}
 \centerline{\includegraphics[width=7in]{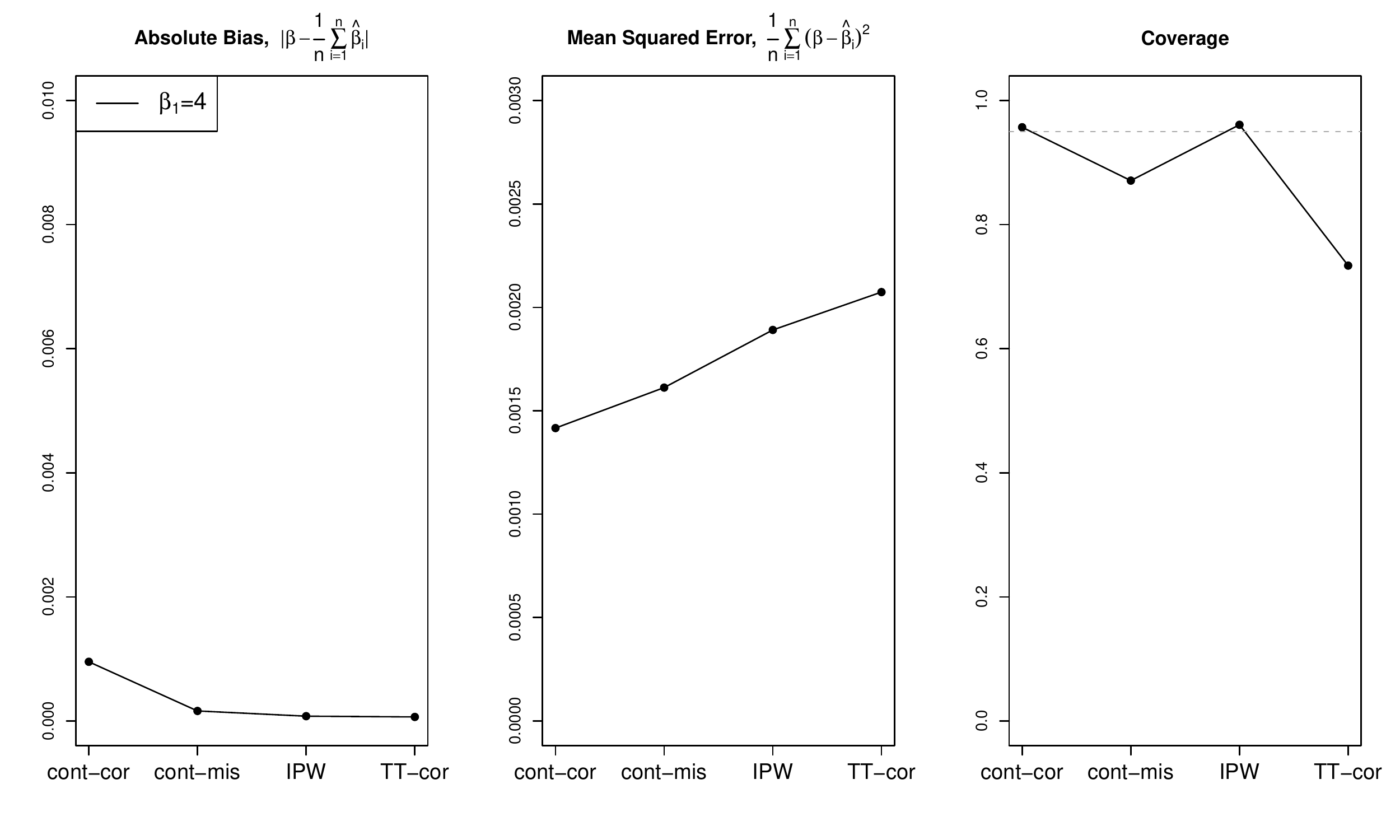}} 
\caption{Results from Identity link simulations  in the simple settings with a single covariate. Estimated bias, MSE, and coverage probability of the control function under correct and misspecification of the selection bias function (cont-cor, cont-mis, respectively), IPW and TT (correctly specified) estimators, in estimating the population effect of $X_1$.}
\label{fig:identity1}
\end{figure}

Table~\ref{tab:identity1} provides comprehensive simulation results (i.e. all summary statistics for all estimators under investigations), while Figure~\ref{fig:identity1} provides graphical results, comparing the bias, MSE and coverage of the unbiased estimators cont-mis, cont-cor, IPW and TT-cor. 

\begin{table}[htbp]
\begin{center}
\caption{Simulation results for estimating the effect of covariates on a normally distributed secondary outcome using the identity link function, in the first, simple settings (a single covariate). We report results for the usual IPW estimator, the proposed estimator with the control function, when the model for $\nu(\bX, D)$ is correctly specific (`cont-cor') and when the model is misspecified (`cont-mis1'), the na\"ive conditional and pooled estimators (Dind and pooled) with and without disease status in the regression model, respectively, and the estimator proposed by \citet{TchetgenTchetgen2013} (TT).   }
\begin{tabular}{lrrrrr}
\hline\hline
\multicolumn{1}{l}{Estimator}&\multicolumn{1}{c}{bias}&\multicolumn{1}{c}{MSE}&\multicolumn{1}{c}{emp sd}&\multicolumn{1}{c}{est sd}&\multicolumn{1}{c}{coverage}\tabularnewline \hline 
\multicolumn{6}{c}{Intercept, $\beta_0 = 50$}\\ \hline
cont-cor&$ 0.000$&$0.018$&$0.136$&$0.133$&$0.942$\tabularnewline
cont-mis&$-0.001$&$0.018$&$0.136$&$0.142$&$0.957$\tabularnewline
IPW&$-0.002$&$0.019$&$0.137$&$0.134$&$0.939$\tabularnewline
pooled&$ 1.640$&$2.711$&$0.145$&$0.200$&$0.000$\tabularnewline
Dind&$-1.450$&$2.126$&$0.151$&$0.165$&$0.000$\tabularnewline
TT-cor&$ 0.000$&$0.018$&$0.135$&$0.130$&$0.942$\tabularnewline
TT-mis&$-0.930$&$0.889$&$0.157$&$0.158$&$0.000$\tabularnewline
\hline
\multicolumn{6}{c}{$X_1$, $\beta_1 = 4$}\\ \hline
cont-cor&$-0.001$&$0.001$&$0.038$&$0.039$&$0.957$\tabularnewline
cont-mis&$ 0.000$&$0.002$&$0.040$&$0.032$&$0.871$\tabularnewline
IPW&$ 0.000$&$0.002$&$0.044$&$0.045$&$0.961$\tabularnewline
pooled&$ 0.753$&$0.568$&$0.036$&$0.036$&$0.000$\tabularnewline
Dind&$ 0.149$&$0.024$&$0.041$&$0.030$&$0.009$\tabularnewline
TT-cor&$ 0.000$&$0.002$&$0.046$&$0.026$&$0.734$\tabularnewline
TT-mis&$ 0.497$&$0.249$&$0.039$&$0.035$&$0.000$\tabularnewline
\hline\end{tabular}\label{tab:identity1}
\end{center}
\end{table}

\subsection{Table summarizing the identity link simulations provided in Section 4.1 in the manuscript}

The following Table~\ref{tab:identity2} provide comprehensive simulation results  for the simulation study described in Section 4.1 in the paper.

\clearpage
 \thispagestyle{empty}
\begin{table}[htbp]
\begin{center}
\caption{Simulation results for estimating the effect of covariates on a normally distributed secondary outcome using the identity link function, in the second settings (two covariates, interaction term in the population regression and selection bias models). We report results for the usual IPW estimator, the proposed estimator with the control function, when the model for $\nu(\bX, D)$ is correctly specific (`cont-cor') and when the model is misspecified (`cont-mis1') and the na\"ive conditional and pooled estimators (Dind and pooled) with and without disease status in the regression model, respectively.  }
\begin{tabular}{lrrrrr}
\hline\hline
\multicolumn{1}{l}{Estimator}&\multicolumn{1}{c}{bias}&\multicolumn{1}{c}{MSE}&\multicolumn{1}{c}{emp sd}&\multicolumn{1}{c}{est sd}&\multicolumn{1}{c}{coverage}\tabularnewline \hline 
\multicolumn{6}{c}{Intercept, $\beta_0 = 50$}\\ \hline
cont-cor&$ 0.007$&$0.019$&$0.138$&$0.139$&$0.958$\tabularnewline
cont-mis1&$ 0.007$&$0.019$&$0.138$&$0.139$&$0.959$\tabularnewline
cont-mis2&$ 0.007$&$0.019$&$0.138$&$0.141$&$0.961$\tabularnewline
cont-mis3&$ 0.006$&$0.019$&$0.139$&$0.150$&$0.971$\tabularnewline
cont-mis4&$ 0.006$&$0.019$&$0.139$&$0.153$&$0.973$\tabularnewline
IPW&$ 0.006$&$0.019$&$0.139$&$0.141$&$0.964$\tabularnewline
pooled&$ 1.520$&$2.332$&$0.150$&$0.227$&$0.000$\tabularnewline
Dind&$-1.577$&$2.515$&$0.165$&$0.183$&$0.000$\tabularnewline
\hline
\multicolumn{6}{c}{$X_1$, $\beta_1 = 4$}\\ \hline
cont-cor&$-0.001$&$0.001$&$0.038$&$0.042$&$0.967$\tabularnewline
cont-mis1&$-0.001$&$0.001$&$0.038$&$0.044$&$0.971$\tabularnewline
cont-mis2&$-0.001$&$0.001$&$0.038$&$0.047$&$0.982$\tabularnewline
cont-mis3&$ 0.000$&$0.002$&$0.040$&$0.034$&$0.906$\tabularnewline
cont-mis4&$ 0.000$&$0.002$&$0.040$&$0.035$&$0.923$\tabularnewline
IPW&$ 0.000$&$0.002$&$0.045$&$0.047$&$0.964$\tabularnewline
pooled&$ 0.724$&$0.526$&$0.038$&$0.041$&$0.000$\tabularnewline
Dind&$ 0.077$&$0.008$&$0.042$&$0.034$&$0.398$\tabularnewline
\hline
\multicolumn{6}{c}{$X_2$, $\beta_2 = 3$}\\ \hline
cont-cor&$ 0.028$&$0.228$&$0.477$&$0.491$&$0.960$\tabularnewline
cont-mis1&$ 0.024$&$0.236$&$0.485$&$0.526$&$0.970$\tabularnewline
cont-mis2&$ 0.026$&$0.238$&$0.487$&$0.431$&$0.913$\tabularnewline
cont-mis3&$ 0.017$&$0.268$&$0.517$&$0.656$&$0.984$\tabularnewline
cont-mis4&$ 0.021$&$0.268$&$0.517$&$0.536$&$0.953$\tabularnewline
IPW&$ 0.024$&$0.272$&$0.521$&$0.521$&$0.950$\tabularnewline
pooled&$ 2.256$&$5.461$&$0.608$&$0.648$&$0.051$\tabularnewline
Dind&$-0.061$&$0.419$&$0.645$&$0.479$&$0.852$\tabularnewline
\hline
\multicolumn{6}{c}{$X_1X_2$, $\beta_3 = 3$}\\ \hline
cont-cor&$ 0.005$&$0.022$&$0.148$&$0.207$&$0.998$\tabularnewline
cont-mis1&$ 0.011$&$0.025$&$0.159$&$0.164$&$0.955$\tabularnewline
cont-mis2&$ 0.014$&$0.032$&$0.179$&$0.116$&$0.775$\tabularnewline
cont-mis3&$ 0.016$&$0.039$&$0.197$&$0.146$&$0.843$\tabularnewline
cont-mis4&$ 0.015$&$0.047$&$0.216$&$0.146$&$0.795$\tabularnewline
IPW&$ 0.018$&$0.076$&$0.275$&$0.247$&$0.909$\tabularnewline
pooled&$ 0.317$&$0.126$&$0.160$&$0.117$&$0.297$\tabularnewline
Dind&$ 0.366$&$0.156$&$0.148$&$0.086$&$0.085$\tabularnewline
\hline\end{tabular}\label{tab:identity2}
\end{center}
\end{table}

\clearpage

\section{Simulation study: log link}
We compared the control function estimator to pooled and Dind, that were calculated using generalized linear models in standard software.
We simulated two covariates, $X_1$ and $X_2$, with  $X_1\sim \mathcal{N}(1,0.2)$ and $X_2\sim \mathcal{N}(1.5,0.2)$. Primary disease probability was calculated by 
\[
\text{logit}\left\{p(D=1|\bX)\right\} = -2.12+ 0.3X_1 + X_2,
\] so that disease prevalence is 0.12. Disease statuses were sampled from the calculated probabilities. 
The secondary outcome mean was calculated by:
\begin{eqnarray*}
\bbE(Y|\bX,D)& =& \exp\big\{ 3+ 0.7X_1 + (0.3 + 0.5X_1 + 0.5X_1X_2)D \big \}\\
&& \times \exp\big[ - \log\{\exp(0.5 + 0.3X_1 + 0.3X_2 + 0.3X_1X_2)p(D=1|\bX)  + p(D=0|\bX)\}\big], 
\end{eqnarray*}
so that the population mean is $\exp\big\{\mu(\bX,\bbeta)\big\} = \exp(\bX^T\bbeta)$ with $\bX = (1, X_1,X_2, X_1X_2)^T$ and $\bbeta = (3,0.7,0.5,0.5)^T$, and $\nu(\bX,D) = D\bX^T\balpha$ with $\balpha = (0.5,0.3,0.3,0.3)^T$.
Then  $Y$ was sampled from Poisson distributed, i.e.  $ Y \sim \text{Poisson}\{\bbE(Y|\bX,D)\}$. 1000 cases and controls were sampled from the generated population. 

All estimators  estimated the sample mean based on the full design matrix, i.e. with $\bX = (1, X_1,X_2, X_1X_2)^T$. The  control function estimator estimated $\nu(\bX,D)$. When the model was correctly specified, the design matrix was taken to include all of $\bX$. To study the effect of misspecification, we implemented the control function estimator with the following misspecifications of the selection bias function $\nu(\bX,D)$: cont-mis1 had design matrix $\bX = (1,X_1,X_2)$. cont-mis2 had design matrix $\bX = (1,X_1)^T$, cont-mis3 had $\bX = (1,X_2)^T$, and cont-mis4 had only intercept.

\begin{figure}
 \centerline{\includegraphics[scale = 0.7]{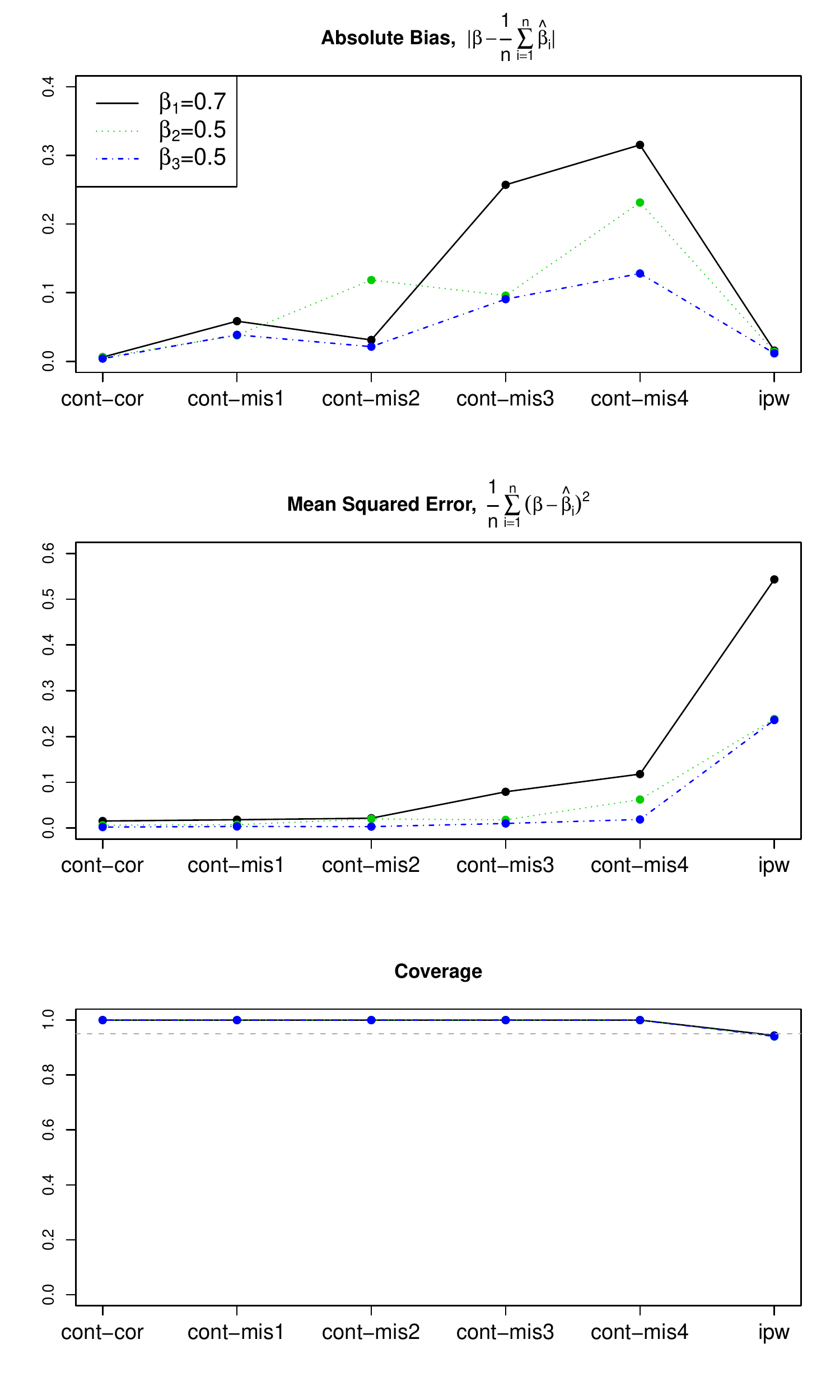}} 
\caption{Results from log link simulations. Estimated bias, MSE, and coverage probability of the control function under correct and misspecification of the selection bias function (cont-cor, cont-mis1, $\ldots$, cont-mis4), and IPW, in estimating population means.}
\label{fig:log}
\end{figure}

Figure~\ref{fig:log}, provides the bias, MSE and coverage probabilities of the IPW and the control function estimators, calculated over the 1000 simulations.  Table~\ref{tab:log} reports, for each estimator and each estimated parameter, the estimator's mean bias, MSE, empirical standard deviation over all simulations, mean estimated standard deviation, and coverage probability.
The bias of the control function estimator is small under correct specification of the selection bias function, but increases as more information is lost in various forms of misspecification. For instance, consider the estimator $\beta_1$, the coefficient of $X_1$. When the interaction term, or both interaction term and $X_2$, are not included in the design matrix for $\nu(\bX,D)$ (cont-mis1, cont-mis2), it becomes slightly biased. When both interaction term and $X_1$, or all covariates, are not included in the design matrix (cont-mis3, cont-mis4), its bias more than doubles. 
However, surprisingly, the MSE of the control function estimators is superior to the IPW, and performs well even when the model for $\nu(\bX,D)$ is misspecified.   When the model for $\nu(\bX,D)$ is misspecified, the bias, the MSE and the empirical standard deviation of the control function estimator  were higher than under correct specification. 
Coverage probability was inflated and very close to 1, both when the model for $\nu(\bX,D)$ was correctly specified and when it was misspecified. In comparison, the coverage probability of the IPW estimator was  accurate. 
Finally, as in the identity link simulations,  the na\"ive estimators Dind and pooled yielded biased  estimators with, substantially lower than nominal, coverage probability.

 \clearpage
 \thispagestyle{empty}

\begin{table}[htbp]
\caption{Simulation results for estimating the effect of covariates on a Poisson distributed secondary outcome using the log link function. We report results for the usual IPW estimator, the proposed estimator with the control function, when the model for $\nu(\bX, D)$ is correctly specific (`cont-cor') and when the model is misspecified, under four forms of misspecification  (`cont-mis1', $\ldots$, `cont-mis4'), and the na\"ive conditional and pooled estimators (Dind and pooled) with and without disease status in the regression model, respectively.   }
\begin{center}
\begin{tabular}{lrrrrr}
\hline\hline
\multicolumn{1}{l}{Estimator}&\multicolumn{1}{c}{bias}&\multicolumn{1}{c}{MSE}&\multicolumn{1}{c}{emp sd}&\multicolumn{1}{c}{est sd}&\multicolumn{1}{c}{coverage}\tabularnewline \hline
\multicolumn{6}{c}{Intercept, $\beta_0 = 3$}\\ \hline
ours-cor&$-0.009$&$0.023$&$0.151$&$0.645$&$1.000$\tabularnewline
ours-mis1&$ 0.057$&$0.025$&$0.148$&$0.642$&$1.000$\tabularnewline
ours-mis2&$-0.181$&$0.059$&$0.163$&$0.637$&$1.000$\tabularnewline
ours-mis3&$-0.272$&$0.101$&$0.165$&$0.638$&$1.000$\tabularnewline
ours-mis4&$-0.482$&$0.264$&$0.178$&$0.642$&$1.000$\tabularnewline
IPW&$-0.020$&$0.546$&$0.739$&$0.730$&$0.944$\tabularnewline
pooled&$ 0.025$&$0.444$&$0.666$&$0.064$&$0.135$\tabularnewline
Dind&$-0.484$&$0.245$&$0.104$&$0.064$&$0.002$\tabularnewline
\hline
\multicolumn{6}{c}{$X_1$, $\beta_1 = 0.7$}\\ \hline
ours-cor&$ 0.006$&$0.015$&$0.124$&$0.641$&$1.000$\tabularnewline
ours-mis1&$-0.059$&$0.018$&$0.122$&$0.639$&$1.000$\tabularnewline
ours-mis2&$ 0.031$&$0.022$&$0.144$&$0.635$&$1.000$\tabularnewline
ours-mis3&$ 0.257$&$0.079$&$0.115$&$0.627$&$1.000$\tabularnewline
ours-mis4&$ 0.315$&$0.118$&$0.136$&$0.628$&$1.000$\tabularnewline
IPW&$ 0.016$&$0.543$&$0.737$&$0.725$&$0.944$\tabularnewline
pooled&$ 0.095$&$0.447$&$0.662$&$0.059$&$0.150$\tabularnewline
Dind&$-0.078$&$0.016$&$0.097$&$0.060$&$0.623$\tabularnewline
\hline
\multicolumn{6}{c}{$X_2$, $\beta_2 = 0.5$}\\ \hline
ours-cor&$ 0.006$&$0.006$&$0.079$&$0.424$&$1.000$\tabularnewline
ours-mis1&$-0.038$&$0.007$&$0.077$&$0.423$&$1.000$\tabularnewline
ours-mis2&$ 0.119$&$0.020$&$0.078$&$0.416$&$1.000$\tabularnewline
ours-mis3&$ 0.096$&$0.018$&$0.094$&$0.421$&$1.000$\tabularnewline
ours-mis4&$ 0.231$&$0.062$&$0.094$&$0.419$&$1.000$\tabularnewline
IPW&$ 0.014$&$0.238$&$0.488$&$0.481$&$0.940$\tabularnewline
pooled&$ 0.044$&$0.193$&$0.437$&$0.041$&$0.148$\tabularnewline
Dind&$-0.348$&$0.126$&$0.067$&$0.041$&$0.002$\tabularnewline

\hline
\multicolumn{6}{c}{$X_1X_2$, $\beta_3 = 0.5$}\\ \hline
ours-cor&$-0.004$&$0.002$&$0.047$&$0.422$&$1.000$\tabularnewline
ours-mis1&$ 0.039$&$0.004$&$0.046$&$0.420$&$1.000$\tabularnewline
ours-mis2&$-0.021$&$0.003$&$0.052$&$0.415$&$1.000$\tabularnewline
ours-mis3&$-0.091$&$0.010$&$0.044$&$0.413$&$1.000$\tabularnewline
ours-mis4&$-0.128$&$0.019$&$0.049$&$0.409$&$1.000$\tabularnewline
IPW&$-0.012$&$0.236$&$0.486$&$0.477$&$0.941$\tabularnewline
pooled&$-0.049$&$0.190$&$0.433$&$0.038$&$0.141$\tabularnewline
Dind&$-0.002$&$0.004$&$0.063$&$0.038$&$0.758$\tabularnewline
\hline\end{tabular}\label{tab:log}
\end{center}
\end{table}

 \clearpage

\section{Simulation study mimicking the T2D case-control study data set}
\renewcommand{\theequation}{E. \arabic{equation}} 
\setcounter{equation}{0}

The goal of these simulations was to study the performance of the control function estimator in simulations mimicking the T2D data set, by using the same 
variable types, as well as effect sizes, as seen in the data. We considered a few forms of misspecification of the selection bias function, to glean into the plausible effects of misspecification on estimation. 

First, we took two SNPs that were found to be significantly associated with log-BMI an entire GWAS data analysis. These SNPs, dubbed SNP1 and SNP2, had very low Minor Allele Frequency (MAF), about 3\%. We estimated the logistic disease model with the predictors: smoking status, alcohol measure, physically active status, and SNP1 and SNP2.  We also estimated the regression model $\bbE[Y|\bX]$ of log-BMI with age, smoking status, physically active status, SNP1,  SNP2, and the interaction between SNP1 and physical activity status as predictors. In addition, we estimated a regression model for the selection bias function with smoking status and SNP1 as predictors. Note that for simplicity, we did not adjust for the principal components of the genetic data in these analysis. We used the estimated effects, rounded to the third digit, as effect values in the simulations. We then employed a few variations. We now describe the sampling and generation of the simulated data, and then the different variations of the simulation study.

\subsection{Data sampling and generation:}
We simulated a super population of 15,000 individuals. Then sampled cases and controls from this population, based only on disease status. For each of 1000 simulations, the super population was simulated as follows:
\begin{itemize}
\item SNP1 and SNP2 were sampled with replacement from the true SNP data. 
\item Binary smoking status as well as physically active status were sampled from a binary distribution, with parameter $p$ estimated from the diabetes data set (for simplicity, ignoring case-control sampling).
\item Alcohol measures and age were sampled form the case-control study data, with replacement. 
\item Disease probability was calculated by the inverse of the logistic model with parameters as estimated from the data, with adaptation of the intercept to have disease prevalence of about $8.4\%$, and possible variation as described later. 
\item Log-BMI values were simulated from a normal distribution, using the mean and variance parameters estimated from the diabetes data set, with possible variations as described later. 
\end{itemize}
We sampled 500 cases and 500 controls from the super population.

\subsection{Variations of the simulation}
To study the effect of some properties of the data on the estimators, we applied the following variations, so that the simulations were ran with all combinations of the following options:
\begin{enumerate}
\item SNP1 and SNP2 where either the SNPs with very low MAF used to estimate the model parameters, or other two SNPs with high MAF (closer to 50\%). 
\item The effect of SNP1 on disease was set to a `high' effect of 1.3 (instead of -0.04). 
\item The effect of SNP1 on the selection bias function was set to a `high' effect of  -1 (instead of -0.053). 
\end{enumerate}

\subsection{Misspecification of the selection bias function}
We studied the control function estimator when the selection bias function is correctly specified, and also when it is misspecified, in the following ways. Recall that a correct specification refers
to a linear model with an intercept, SNP1, and smoking status. The effect sizes were:\\
$\alpha_{intercept} = -0.158$\\
$\alpha_{smoke} = 0.022$\\
$\alpha_{snp1} = -0.053$ or (if set to `high') $\alpha_{snp1} = -0.2$. \\
We allowed for the following misspecifications of the selection bias function:
\begin{enumerate}
\item cont-mis1: no SNP1 effect (just intercept and smoking status).
\item  cont-mis2: no smoking status effect (just intercept and SNP1). 
\item cont-mis3: neither SNP1 nor smoking status (just intercept). 
\end{enumerate}

\subsection{Conclusions}
In the following, figures and tables provide the simulations results. The figures focus on the various control-function estimates, and IPW (which can also be thought of as type of control-function estimator with the selection bias misspecified and equal to zero), and compare between the bias and MSE of the SNP effects. The tables provide comprehensive simulation results for all measures and estimators used. 
\begin{enumerate}
\item The control function estimator improves over IPW when the effect of SNP1 (or more generally, covariates or exposures) on either the disease model or the selection bias model is high, and it is in fact included in the disease/selection bias model. In other words, cont-mis2 performs better than cont-mis1 and cont-mis3, that do not include effects of SNP1. Also, its performance is almost identical to the cont-cor and better than the usual IPW. 
\item The improvement seen in the control function estimator was in the effect (bias or MSE) estimate of SNP1 and the interaction SNP1 and being physically active. The various control function estimators (i.e. under the different forms of misspecification) had similar behavior with respect to the estimation of SNP2 effect. 
\item The control function estimators were never worse than IPW in terms of MSE. 
\item When the MAF of the SNPs was low (rare SNP), coverage probabilities of all estimators were reduced, compared to when the MAF was relatively high (common SNP). 
\end{enumerate}

\subsection{Figures and tables summarizing the results}

\begin{figure}\caption{Comparison between the estimated bias of SNP1 effect, over 1000 simulations, of the control-function estimator under various forms of mispecification (mis1, mis2, mis3) and under correct specification (cor) of the selection bias function, and of the IPW.  We compare between all combinations in which SNP1 and SNP2 have either low or high MAF, the effect of SNP1 on the disease model is either low or high, and the effect of SNP1 on the selection bias model ($\gamma(\bX)$) is either low or high. }
\includegraphics[scale = 0.6]{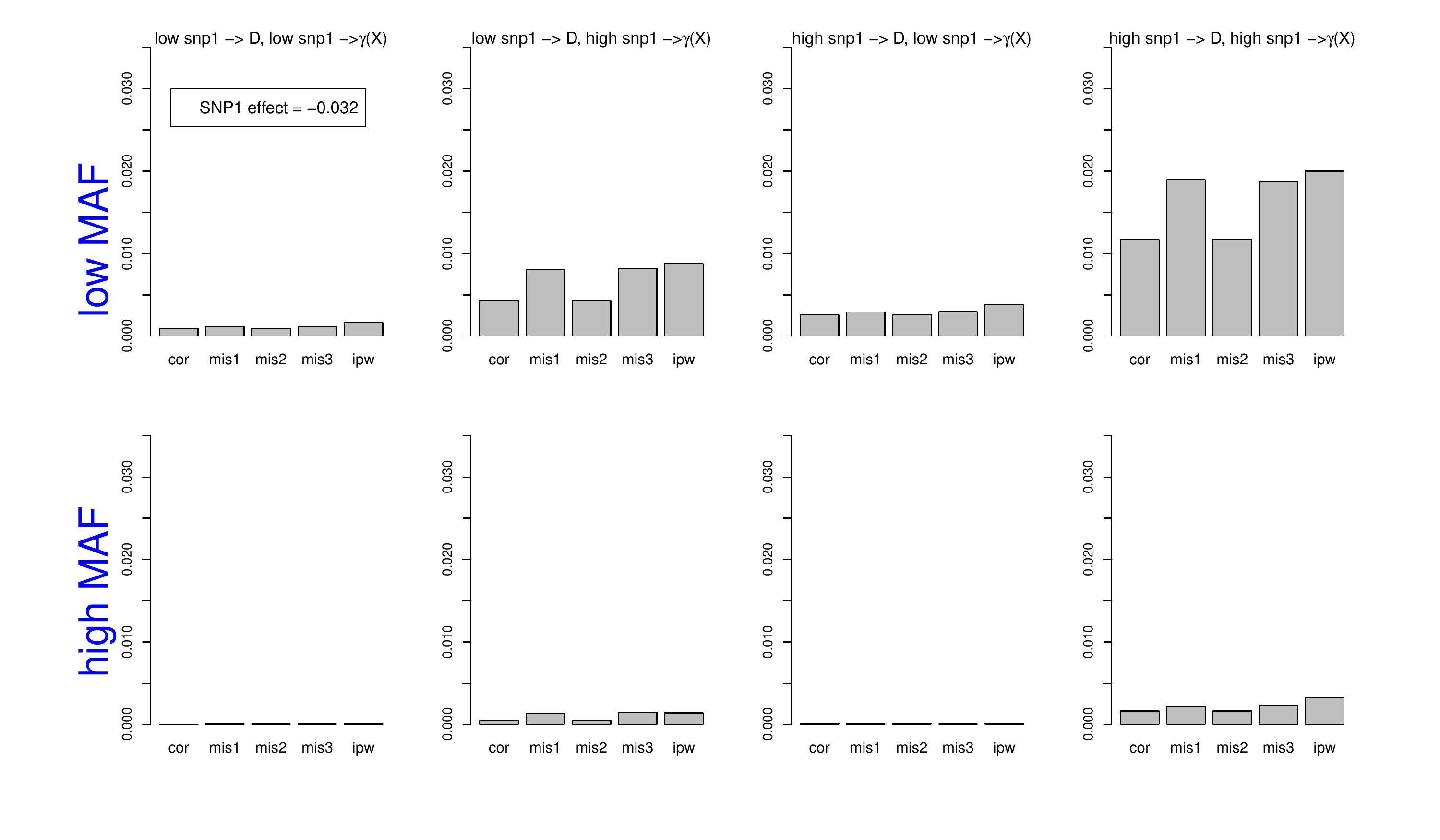}
\end{figure}

\begin{figure}\caption{Comparison between the Mean Square Error (MSE) of SNP1 effect, over 1000 simulations, of the control-function estimator under various forms of mispecification (mis1, mis2, mis3) and under correct specification (cor) of the selection bias function, and of the IPW.  We compare between all combinations in which SNP1 and SNP2 have either low or high MAF, the effect of SNP1 on the disease model is either low or high, and the effect of SNP1 on the selection bias model ($\gamma(\bX)$) is either low or high. }
\includegraphics[scale = 0.6]{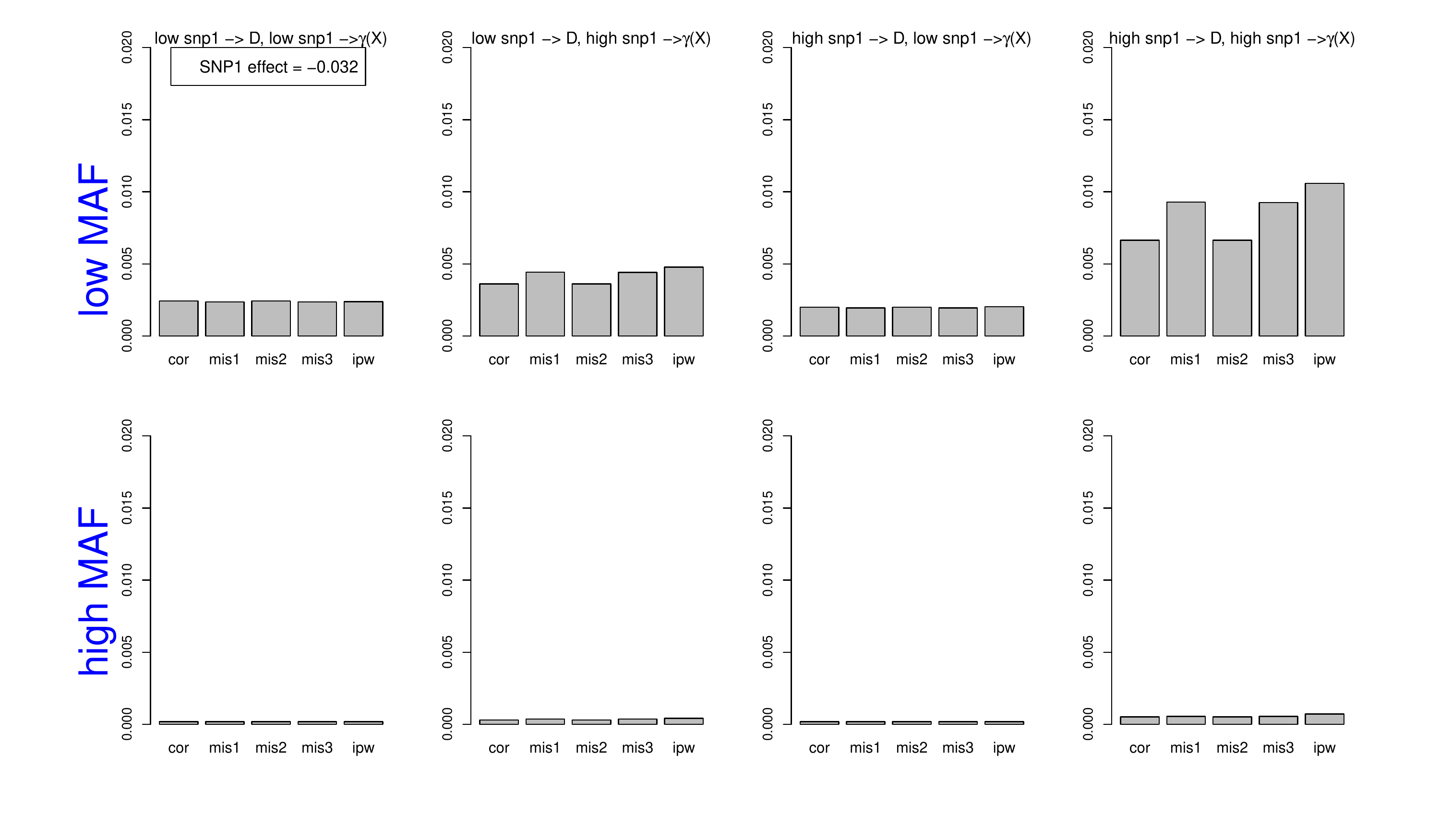}
\end{figure}

\begin{figure}\caption{Comparison between the estimated bias of the effect of the interaction Active$\times$SNP1 effect, over 1000 simulations, of the control-function estimator under various forms of mispecification (mis1, mis2, mis3) and under correct specification (cor) of the selection bias function, and of the IPW.  We compare between all combinations in which SNP1 and SNP2 have either low or high MAF, the effect of SNP1 on the disease model is either low or high, and the effect of SNP1 on the selection bias model ($\gamma(\bX)$) is either low or high. }
\includegraphics[scale = 0.6]{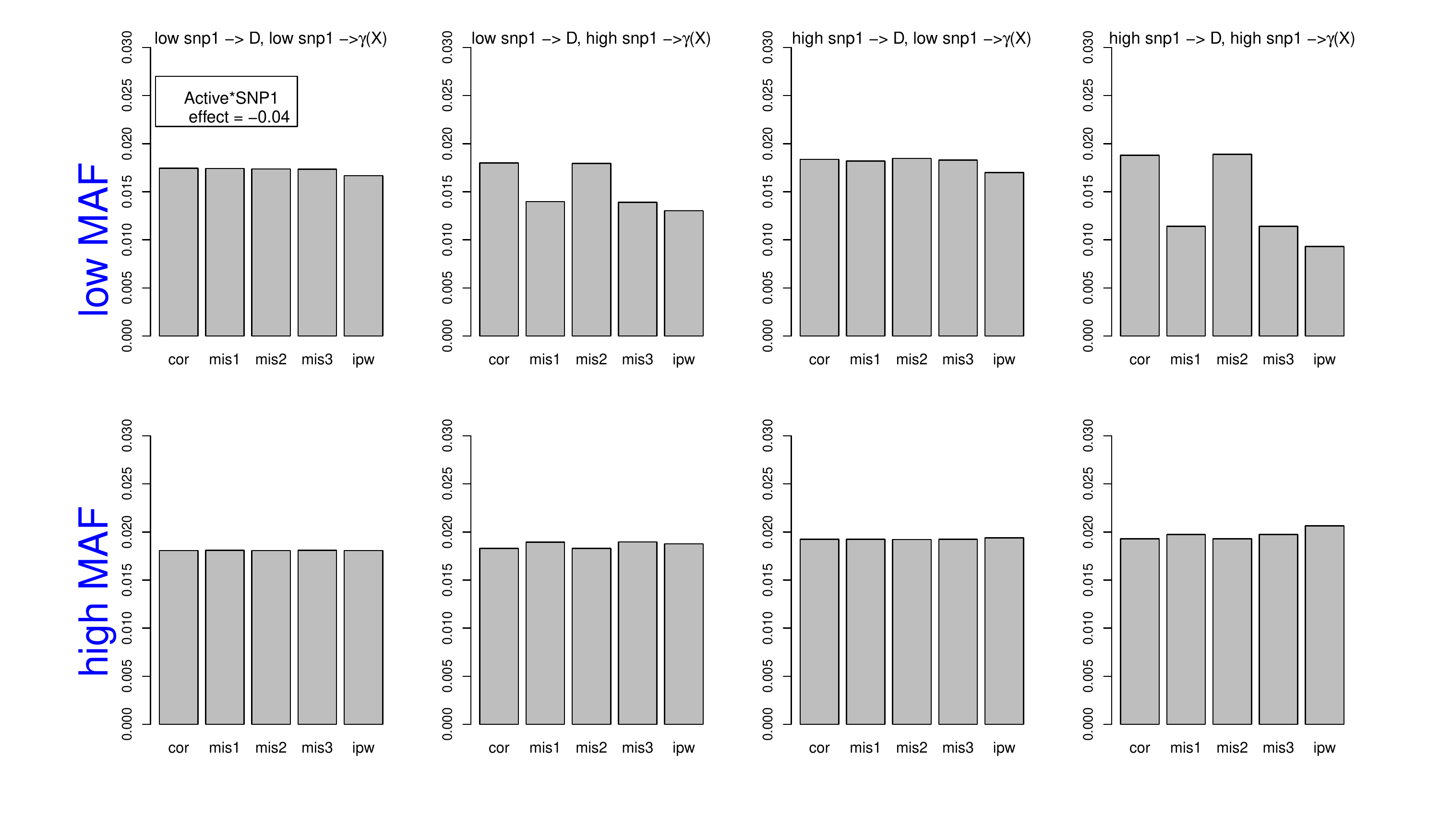}
\end{figure}

\begin{figure}\caption{Comparison between the  Mean Square Error (MSE) of the interaction Active$\times$SNP1 effect, over 1000 simulations, of the control-function estimator under various forms of mispecification (mis1, mis2, mis3) and under correct specification (cor) of the selection bias function, and of the IPW.  We compare between all combinations in which SNP1 and SNP2 have either low or high MAF, the effect of SNP1 on the disease model is either low or high, and the effect of SNP1 on the selection bias model ($\gamma(\bX)$) is either low or high. }
\includegraphics[scale = 0.6]{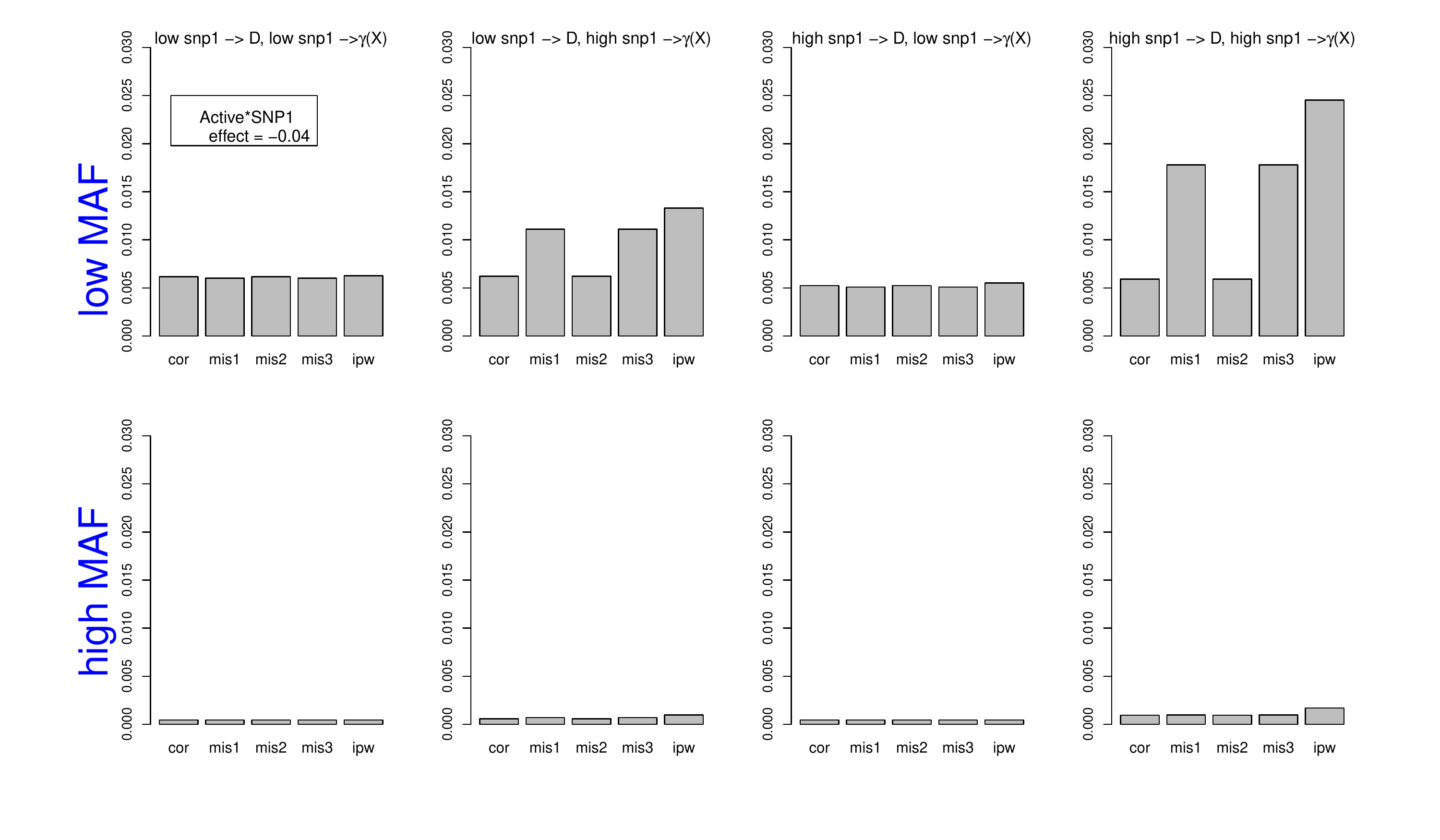}
\end{figure}

\begin{figure}\caption{Comparison between the estimated bias of SNP2 effect, over 1000 simulations, of the control-function estimator under various forms of mispecification (mis1, mis2, mis3) and under correct specification (cor) of the selection bias function, and of the IPW.  We compare between all combinations in which SNP1 and SNP2 have either low or high MAF, the effect of SNP1 on the disease model is either low or high, and the effect of SNP1 on the selection bias model ($\gamma(\bX)$) is either low or high. }
\includegraphics[scale = 0.6]{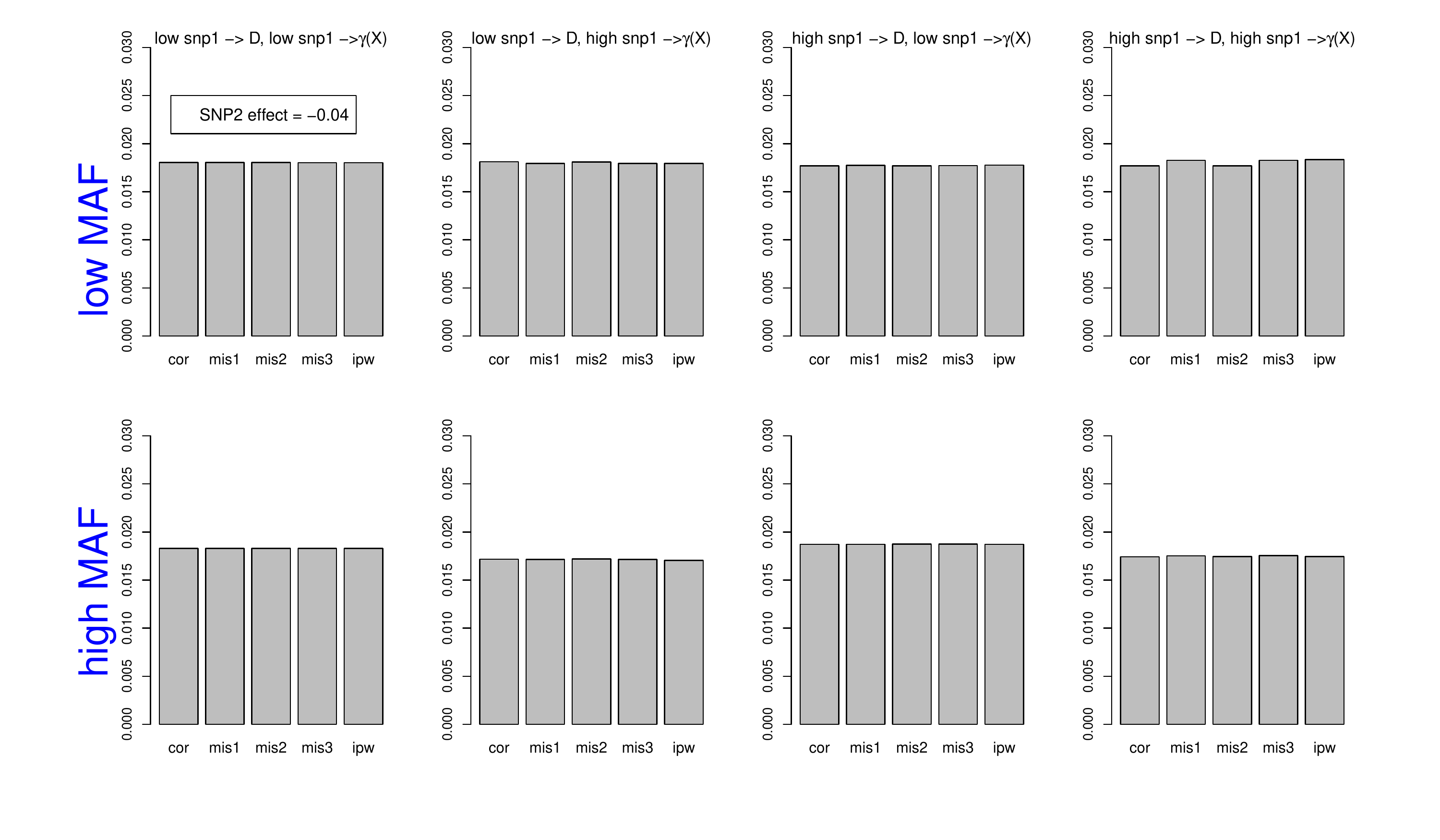}
\end{figure}

\begin{figure}\caption{Comparison between the Mean Square Error (MSE) of SNP2 effect, over 1000 simulations, of the control-function estimator under various forms of mispecification (mis1, mis2, mis3) and under correct specification (cor) of the selection bias function, and of the IPW.  We compare between all combinations in which SNP1 and SNP2 have either low or high MAF, the effect of SNP1 on the disease model is either low or high, and the effect of SNP1 on the selection bias model ($\gamma(\bX)$) is either low or high. }
\includegraphics[scale = 0.6]{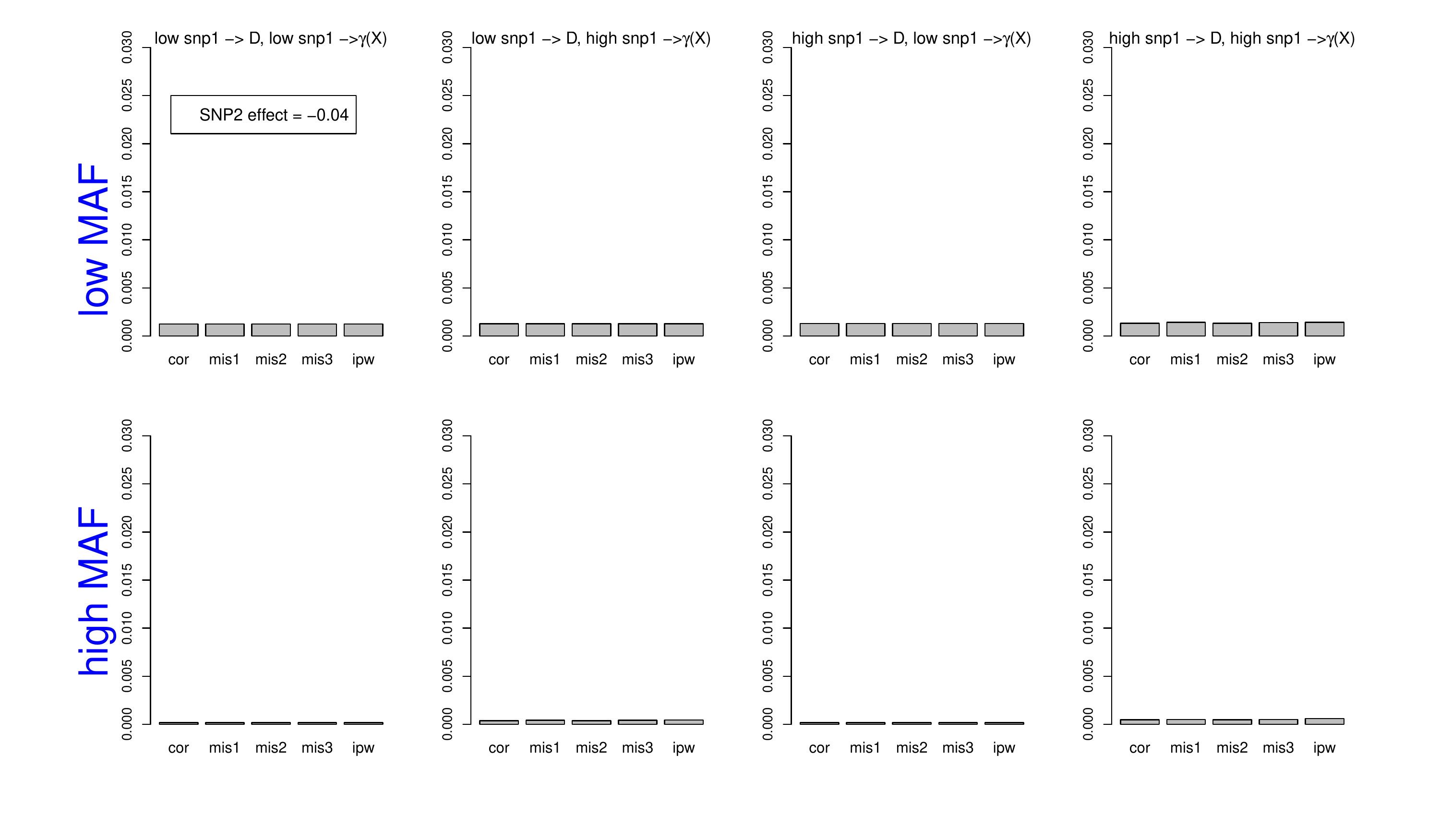}
\end{figure}

\clearpage

\begin{table}[htbp] 
\caption{Simulation results, averaged over 1000 simulations, for estimating the effect of covariates on a the simulated log(BMI) outcomes. The SNPs used had \textbf{low} MAF,
the effect of SNP1 on the disease distribution was \textbf{low}, and its effect on  the selection bias function was \textbf{low}.} 
\label{tab:log}
\begin{center}
 \scalebox{0.7}{
\begin{tabular}{lrrrrr}
\hline
\multicolumn{1}{l}{Estimator}&\multicolumn{1}{c}{Bias}&\multicolumn{1}{c}{MSE}&\multicolumn{1}{c}{emp sd}&\multicolumn{1}{c}{est sd}&\multicolumn{1}{c}{coverage}\tabularnewline
\hline
\multicolumn{6}{c}{Intercept, $\beta_0 = 3.077$}\\ \hline 
ours-cor&$ 0.001$&$0.002$&$0.047$&$0.048$&$0.952$\tabularnewline
ours-mis1&$ 0.001$&$0.002$&$0.046$&$0.048$&$0.951$\tabularnewline
ours-mis2&$ 0.001$&$0.002$&$0.046$&$0.048$&$0.951$\tabularnewline
ours-mis3&$ 0.001$&$0.002$&$0.046$&$0.048$&$0.951$\tabularnewline
ipw&$ 0.001$&$0.002$&$0.047$&$0.048$&$0.949$\tabularnewline
pooled&$-0.061$&$0.005$&$0.040$&$0.040$&$0.667$\tabularnewline
dind&$ 0.012$&$0.002$&$0.037$&$0.037$&$0.945$\tabularnewline\hline
\multicolumn{6}{c}{Age, $\beta_1 = 0.002$}\\ \hline
ours-cor&$ 0.000$&$0.000$&$0.001$&$0.001$&$0.943$\tabularnewline
ours-mis1&$ 0.000$&$0.000$&$0.001$&$0.001$&$0.944$\tabularnewline
ours-mis2&$ 0.000$&$0.000$&$0.001$&$0.001$&$0.943$\tabularnewline
ours-mis3&$ 0.000$&$0.000$&$0.001$&$0.001$&$0.943$\tabularnewline
ipw&$ 0.000$&$0.000$&$0.001$&$0.001$&$0.939$\tabularnewline
pooled&$ 0.000$&$0.000$&$0.001$&$0.001$&$0.949$\tabularnewline
dind&$ 0.000$&$0.000$&$0.001$&$0.001$&$0.953$\tabularnewline
\hline
\multicolumn{6}{c}{Smoker, $\beta_2 = -0.012$}\\ \hline
ours-cor&$ 0.001$&$0.000$&$0.017$&$0.017$&$0.933$\tabularnewline
ours-mis1&$ 0.001$&$0.000$&$0.017$&$0.017$&$0.932$\tabularnewline
ours-mis2&$ 0.001$&$0.000$&$0.017$&$0.017$&$0.933$\tabularnewline
ours-mis3&$ 0.001$&$0.000$&$0.017$&$0.017$&$0.934$\tabularnewline
ipw&$ 0.001$&$0.000$&$0.017$&$0.017$&$0.932$\tabularnewline
pooled&$-0.003$&$0.000$&$0.013$&$0.013$&$0.944$\tabularnewline
dind&$ 0.018$&$0.000$&$0.012$&$0.012$&$0.684$\tabularnewline
\hline
\multicolumn{6}{c}{Physically active, $\beta_3 = -0.032$}\\ \hline
ours-cor&$ 0.000$&$0.000$&$0.015$&$0.015$&$0.942$\tabularnewline
ours-mis1&$ 0.000$&$0.000$&$0.015$&$0.015$&$0.942$\tabularnewline
ours-mis2&$ 0.000$&$0.000$&$0.015$&$0.015$&$0.942$\tabularnewline
ours-mis3&$ 0.000$&$0.000$&$0.015$&$0.015$&$0.943$\tabularnewline
ipw&$ 0.000$&$0.000$&$0.015$&$0.015$&$0.943$\tabularnewline
pooled&$ 0.006$&$0.000$&$0.013$&$0.013$&$0.929$\tabularnewline
dind&$-0.003$&$0.000$&$0.012$&$0.012$&$0.954$\tabularnewline
\hline
\multicolumn{6}{c}{SNP1, $\beta_4 = -0.032$}\\ \hline
ours-cor&$-0.001$&$0.002$&$0.049$&$0.045$&$0.916$\tabularnewline
ours-mis1&$-0.001$&$0.002$&$0.049$&$0.045$&$0.910$\tabularnewline
ours-mis2&$-0.001$&$0.002$&$0.049$&$0.045$&$0.915$\tabularnewline
ours-mis3&$-0.001$&$0.002$&$0.049$&$0.045$&$0.913$\tabularnewline
ipw&$-0.002$&$0.002$&$0.049$&$0.045$&$0.914$\tabularnewline
pooled&$-0.027$&$0.002$&$0.040$&$0.039$&$0.883$\tabularnewline
dind&$-0.023$&$0.002$&$0.035$&$0.036$&$0.902$\tabularnewline
\hline
\multicolumn{6}{c}{SNP2, $\beta_5 = -0.040$}\\ \hline
ours-cor&$ 0.001$&$0.001$&$0.035$&$0.034$&$0.931$\tabularnewline
ours-mis1&$ 0.001$&$0.001$&$0.035$&$0.034$&$0.930$\tabularnewline
ours-mis2&$ 0.001$&$0.001$&$0.035$&$0.034$&$0.930$\tabularnewline
ours-mis3&$ 0.001$&$0.001$&$0.035$&$0.034$&$0.929$\tabularnewline
ipw&$ 0.001$&$0.001$&$0.035$&$0.034$&$0.930$\tabularnewline
pooled&$ 0.006$&$0.001$&$0.031$&$0.031$&$0.942$\tabularnewline
dind&$-0.001$&$0.001$&$0.028$&$0.028$&$0.946$\tabularnewline
\hline
\multicolumn{6}{c}{Active$\times$SNP1, $\beta_6 = -0.021$}\\ \hline
ours-cor&$-0.001$&$0.006$&$0.078$&$0.072$&$0.926$\tabularnewline
ours-mis1&$-0.001$&$0.006$&$0.078$&$0.071$&$0.926$\tabularnewline
ours-mis2&$-0.001$&$0.006$&$0.079$&$0.072$&$0.926$\tabularnewline
ours-mis3&$-0.001$&$0.006$&$0.078$&$0.071$&$0.926$\tabularnewline
ipw&$-0.002$&$0.006$&$0.079$&$0.071$&$0.926$\tabularnewline
pooled&$ 0.001$&$0.004$&$0.067$&$0.063$&$0.940$\tabularnewline
dind&$ 0.001$&$0.003$&$0.059$&$0.058$&$0.948$\tabularnewline

\hline
\end{tabular} }
\end{center}
\end{table}
 \thispagestyle{empty}
 \clearpage

\begin{table}[htbp] 
\caption{Simulation results, averaged over 1000 simulations, for estimating the effect of covariates on a the simulated log(BMI) outcomes. The SNPs used had \textbf{high} MAF, 
the effect of SNP1 on the disease distribution was \textbf{low}, and its effect on  the selection bias function was \textbf{low}. }
\label{tab:log}
\begin{center}
 \scalebox{0.7}{
\begin{tabular}{lrrrrr}
\hline
\multicolumn{1}{l}{Estimator}&\multicolumn{1}{c}{Bias}&\multicolumn{1}{c}{MSE}&\multicolumn{1}{c}{emp sd}&\multicolumn{1}{c}{est sd}&\multicolumn{1}{c}{coverage}\tabularnewline
\hline
\multicolumn{6}{c}{Intercept, $\beta_0 = 3.077$}\\ \hline 
ours-cor&$-0.001$&$0.003$&$0.056$&$0.055$&$0.943$\tabularnewline
ours-mis1&$-0.001$&$0.003$&$0.056$&$0.055$&$0.942$\tabularnewline
ours-mis2&$-0.001$&$0.003$&$0.056$&$0.055$&$0.943$\tabularnewline
ours-mis3&$-0.001$&$0.003$&$0.056$&$0.055$&$0.942$\tabularnewline
ipw&$-0.001$&$0.003$&$0.057$&$0.055$&$0.946$\tabularnewline
pooled&$-0.069$&$0.007$&$0.050$&$0.049$&$0.699$\tabularnewline
dind&$ 0.043$&$0.004$&$0.044$&$0.043$&$0.834$\tabularnewline

\hline
\multicolumn{6}{c}{Age, $\beta_1 = 0.002$}\\ \hline
ours-cor&$ 0.000$&$0.000$&$0.001$&$0.001$&$0.944$\tabularnewline
ours-mis1&$ 0.000$&$0.000$&$0.001$&$0.001$&$0.944$\tabularnewline
ours-mis2&$ 0.000$&$0.000$&$0.001$&$0.001$&$0.944$\tabularnewline
ours-mis3&$ 0.000$&$0.000$&$0.001$&$0.001$&$0.944$\tabularnewline
ipw&$ 0.000$&$0.000$&$0.001$&$0.001$&$0.941$\tabularnewline
pooled&$ 0.000$&$0.000$&$0.001$&$0.001$&$0.939$\tabularnewline
dind&$ 0.000$&$0.000$&$0.001$&$0.001$&$0.941$\tabularnewline

\hline
\multicolumn{6}{c}{Smoker, $\beta_2 = -0.012$}\\ \hline
ours-cor&$ 0.000$&$0.000$&$0.017$&$0.017$&$0.950$\tabularnewline
ours-mis1&$ 0.000$&$0.000$&$0.017$&$0.017$&$0.953$\tabularnewline
ours-mis2&$ 0.000$&$0.000$&$0.017$&$0.017$&$0.953$\tabularnewline
ours-mis3&$ 0.000$&$0.000$&$0.017$&$0.017$&$0.953$\tabularnewline
ipw&$ 0.000$&$0.000$&$0.017$&$0.017$&$0.949$\tabularnewline
pooled&$-0.009$&$0.000$&$0.014$&$0.014$&$0.918$\tabularnewline
dind&$ 0.021$&$0.001$&$0.013$&$0.012$&$0.613$\tabularnewline

\hline
\multicolumn{6}{c}{Physically active, $\beta_3 = -0.032$}\\ \hline
ours-cor&$ 0.001$&$0.001$&$0.027$&$0.027$&$0.953$\tabularnewline
ours-mis1&$ 0.001$&$0.001$&$0.027$&$0.027$&$0.952$\tabularnewline
ours-mis2&$ 0.001$&$0.001$&$0.027$&$0.027$&$0.952$\tabularnewline
ours-mis3&$ 0.001$&$0.001$&$0.027$&$0.027$&$0.951$\tabularnewline
ipw&$ 0.001$&$0.001$&$0.027$&$0.027$&$0.959$\tabularnewline
pooled&$ 0.006$&$0.001$&$0.024$&$0.025$&$0.953$\tabularnewline
dind&$-0.006$&$0.001$&$0.022$&$0.021$&$0.947$\tabularnewline

\hline
\multicolumn{6}{c}{SNP1, $\beta_4 = -0.032$}\\ \hline
ours-cor&$ 0.000$&$0.000$&$0.014$&$0.013$&$0.944$\tabularnewline
ours-mis1&$ 0.000$&$0.000$&$0.014$&$0.013$&$0.944$\tabularnewline
ours-mis2&$ 0.000$&$0.000$&$0.014$&$0.013$&$0.944$\tabularnewline
ours-mis3&$ 0.000$&$0.000$&$0.014$&$0.013$&$0.944$\tabularnewline
ipw&$ 0.000$&$0.000$&$0.014$&$0.013$&$0.942$\tabularnewline
pooled&$-0.026$&$0.001$&$0.012$&$0.012$&$0.426$\tabularnewline
dind&$-0.021$&$0.001$&$0.010$&$0.010$&$0.490$\tabularnewline

\hline
\multicolumn{6}{c}{SNP2, $\beta_5 = -0.040$}\\ \hline
ours-cor&$ 0.000$&$0.000$&$0.014$&$0.014$&$0.943$\tabularnewline
ours-mis1&$ 0.000$&$0.000$&$0.014$&$0.014$&$0.943$\tabularnewline
ours-mis2&$ 0.000$&$0.000$&$0.014$&$0.014$&$0.943$\tabularnewline
ours-mis3&$ 0.000$&$0.000$&$0.014$&$0.014$&$0.944$\tabularnewline
ipw&$ 0.001$&$0.000$&$0.014$&$0.014$&$0.945$\tabularnewline
pooled&$ 0.006$&$0.000$&$0.012$&$0.012$&$0.919$\tabularnewline
dind&$-0.003$&$0.000$&$0.010$&$0.010$&$0.937$\tabularnewline

\hline
\multicolumn{6}{c}{Active$\times$SNP1, $\beta_6 = -0.021$}\\ \hline
ours-cor&$-0.001$&$0.000$&$0.021$&$0.021$&$0.949$\tabularnewline
ours-mis1&$-0.001$&$0.000$&$0.021$&$0.021$&$0.949$\tabularnewline
ours-mis2&$-0.001$&$0.000$&$0.021$&$0.021$&$0.949$\tabularnewline
ours-mis3&$-0.001$&$0.000$&$0.021$&$0.021$&$0.949$\tabularnewline
ipw&$-0.001$&$0.000$&$0.021$&$0.021$&$0.951$\tabularnewline
pooled&$ 0.002$&$0.000$&$0.019$&$0.019$&$0.949$\tabularnewline
dind&$ 0.002$&$0.000$&$0.016$&$0.016$&$0.936$\tabularnewline

\hline
\end{tabular} }
\end{center}
\end{table}
 \thispagestyle{empty}
 \clearpage

\begin{table}[htbp] 
\caption{Simulation results, averaged over 1000 simulations, for estimating the effect of covariates on a the simulated log(BMI) outcomes. The SNPs used had \textbf{low} MAF, the effect of SNP1 on the disease distribution was \textbf{high}, and its effect on  the selection bias function was \textbf{low}.  }
\label{tab:log}
\begin{center}
 \scalebox{0.7}{
\begin{tabular}{lrrrrr}
\hline
\multicolumn{1}{l}{Estimator}&\multicolumn{1}{c}{Bias}&\multicolumn{1}{c}{MSE}&\multicolumn{1}{c}{emp sd}&\multicolumn{1}{c}{est sd}&\multicolumn{1}{c}{coverage}\tabularnewline
\hline
\multicolumn{6}{c}{Intercept, $\beta_0 = 3.077$}\\ \hline 
ours-cor&$-0.003$&$0.002$&$0.049$&$0.048$&$0.947$\tabularnewline
ours-mis1&$-0.003$&$0.002$&$0.049$&$0.048$&$0.948$\tabularnewline
ours-mis2&$-0.003$&$0.002$&$0.049$&$0.048$&$0.947$\tabularnewline
ours-mis3&$-0.003$&$0.002$&$0.049$&$0.048$&$0.948$\tabularnewline
ipw&$-0.003$&$0.002$&$0.050$&$0.048$&$0.947$\tabularnewline
pooled&$-0.064$&$0.006$&$0.040$&$0.040$&$0.650$\tabularnewline
dind&$ 0.008$&$0.001$&$0.037$&$0.037$&$0.945$\tabularnewline

\hline
\multicolumn{6}{c}{Age, $\beta_1 = 0.002$}\\ \hline
ours-cor&$ 0.000$&$0.000$&$0.001$&$0.001$&$0.944$\tabularnewline
ours-mis1&$ 0.000$&$0.000$&$0.001$&$0.001$&$0.945$\tabularnewline
ours-mis2&$ 0.000$&$0.000$&$0.001$&$0.001$&$0.945$\tabularnewline
ours-mis3&$ 0.000$&$0.000$&$0.001$&$0.001$&$0.946$\tabularnewline
ipw&$ 0.000$&$0.000$&$0.001$&$0.001$&$0.939$\tabularnewline
pooled&$ 0.000$&$0.000$&$0.001$&$0.001$&$0.945$\tabularnewline
dind&$ 0.000$&$0.000$&$0.001$&$0.001$&$0.951$\tabularnewline

\hline
\multicolumn{6}{c}{Smoker, $\beta_2 = -0.012$}\\ \hline
ours-cor&$ 0.000$&$0.000$&$0.017$&$0.017$&$0.947$\tabularnewline
ours-mis1&$ 0.000$&$0.000$&$0.017$&$0.017$&$0.947$\tabularnewline
ours-mis2&$ 0.000$&$0.000$&$0.017$&$0.017$&$0.948$\tabularnewline
ours-mis3&$ 0.000$&$0.000$&$0.017$&$0.017$&$0.947$\tabularnewline
ipw&$ 0.000$&$0.000$&$0.017$&$0.017$&$0.944$\tabularnewline
pooled&$-0.003$&$0.000$&$0.013$&$0.013$&$0.944$\tabularnewline
dind&$ 0.018$&$0.000$&$0.012$&$0.012$&$0.690$\tabularnewline

\hline
\multicolumn{6}{c}{Physically active, $\beta_3 = -0.032$}\\ \hline
ours-cor&$ 0.000$&$0.000$&$0.015$&$0.015$&$0.945$\tabularnewline
ours-mis1&$ 0.000$&$0.000$&$0.015$&$0.015$&$0.945$\tabularnewline
ours-mis2&$ 0.000$&$0.000$&$0.015$&$0.015$&$0.945$\tabularnewline
ours-mis3&$ 0.000$&$0.000$&$0.015$&$0.015$&$0.945$\tabularnewline
ipw&$ 0.000$&$0.000$&$0.015$&$0.015$&$0.947$\tabularnewline
pooled&$ 0.006$&$0.000$&$0.013$&$0.013$&$0.925$\tabularnewline
dind&$-0.003$&$0.000$&$0.012$&$0.012$&$0.929$\tabularnewline

\hline
\multicolumn{6}{c}{SNP1, $\beta_4 = -0.032$}\\ \hline
ours-cor&$-0.003$&$0.002$&$0.045$&$0.043$&$0.924$\tabularnewline
ours-mis1&$-0.003$&$0.002$&$0.044$&$0.041$&$0.921$\tabularnewline
ours-mis2&$-0.003$&$0.002$&$0.045$&$0.043$&$0.926$\tabularnewline
ours-mis3&$-0.003$&$0.002$&$0.044$&$0.041$&$0.918$\tabularnewline
ipw&$-0.004$&$0.002$&$0.045$&$0.043$&$0.924$\tabularnewline
pooled&$-0.044$&$0.003$&$0.028$&$0.029$&$0.676$\tabularnewline
dind&$-0.002$&$0.001$&$0.026$&$0.027$&$0.951$\tabularnewline

\hline
\multicolumn{6}{c}{SNP2, $\beta_5 = -0.040$}\\ \hline
ours-cor&$ 0.001$&$0.001$&$0.036$&$0.035$&$0.927$\tabularnewline
ours-mis1&$ 0.001$&$0.001$&$0.036$&$0.035$&$0.927$\tabularnewline
ours-mis2&$ 0.001$&$0.001$&$0.036$&$0.035$&$0.927$\tabularnewline
ours-mis3&$ 0.001$&$0.001$&$0.036$&$0.035$&$0.927$\tabularnewline
ipw&$ 0.001$&$0.001$&$0.036$&$0.035$&$0.930$\tabularnewline
pooled&$ 0.005$&$0.001$&$0.031$&$0.031$&$0.943$\tabularnewline
dind&$-0.002$&$0.001$&$0.029$&$0.028$&$0.954$\tabularnewline

\hline
\multicolumn{6}{c}{Active$\times$SNP1, $\beta_6 = -0.021$}\\ \hline
ours-cor&$ 0.000$&$0.005$&$0.073$&$0.067$&$0.908$\tabularnewline
ours-mis1&$-0.001$&$0.005$&$0.071$&$0.064$&$0.907$\tabularnewline
ours-mis2&$ 0.000$&$0.005$&$0.072$&$0.067$&$0.906$\tabularnewline
ours-mis3&$ 0.000$&$0.005$&$0.071$&$0.064$&$0.906$\tabularnewline
ipw&$-0.002$&$0.006$&$0.074$&$0.067$&$0.896$\tabularnewline
pooled&$-0.003$&$0.002$&$0.047$&$0.047$&$0.946$\tabularnewline
dind&$-0.001$&$0.002$&$0.043$&$0.044$&$0.954$\tabularnewline

\hline
\end{tabular} }
\end{center}
\end{table}
 \thispagestyle{empty}
 \clearpage

\begin{table}[htbp] 
\caption{Simulation results, averaged over 1000 simulations, for estimating the effect of covariates on a the simulated log(BMI) outcomes. The SNPs used had \textbf{high} MAF, the effect of SNP1 on the disease distribution was \textbf{high}, and its effect on  the selection bias function was \textbf{low}. }
\label{tab:log}
\begin{center}
 \scalebox{0.7}{
\begin{tabular}{lrrrrr}
\hline
\multicolumn{1}{l}{Estimator}&\multicolumn{1}{c}{Bias}&\multicolumn{1}{c}{MSE}&\multicolumn{1}{c}{emp sd}&\multicolumn{1}{c}{est sd}&\multicolumn{1}{c}{coverage}\tabularnewline
\hline
\multicolumn{6}{c}{Intercept, $\beta_0 = 3.077$}\\ \hline 
ours-cor&$ 0.001$&$0.003$&$0.056$&$0.056$&$0.956$\tabularnewline
ours-mis1&$ 0.001$&$0.003$&$0.056$&$0.056$&$0.956$\tabularnewline
ours-mis2&$ 0.001$&$0.003$&$0.056$&$0.056$&$0.956$\tabularnewline
ours-mis3&$ 0.001$&$0.003$&$0.056$&$0.056$&$0.956$\tabularnewline
ipw&$ 0.001$&$0.003$&$0.056$&$0.056$&$0.954$\tabularnewline
pooled&$-0.023$&$0.003$&$0.048$&$0.050$&$0.923$\tabularnewline
dind&$ 0.017$&$0.002$&$0.042$&$0.043$&$0.939$\tabularnewline

\hline
\multicolumn{6}{c}{Age, $\beta_1 = 0.002$}\\ \hline
ours-cor&$ 0.000$&$0.000$&$0.001$&$0.001$&$0.953$\tabularnewline
ours-mis1&$ 0.000$&$0.000$&$0.001$&$0.001$&$0.954$\tabularnewline
ours-mis2&$ 0.000$&$0.000$&$0.001$&$0.001$&$0.953$\tabularnewline
ours-mis3&$ 0.000$&$0.000$&$0.001$&$0.001$&$0.954$\tabularnewline
ipw&$ 0.000$&$0.000$&$0.001$&$0.001$&$0.953$\tabularnewline
pooled&$ 0.000$&$0.000$&$0.001$&$0.001$&$0.958$\tabularnewline
dind&$ 0.000$&$0.000$&$0.001$&$0.001$&$0.967$\tabularnewline

\hline
\multicolumn{6}{c}{Smoker, $\beta_2 = -0.012$}\\ \hline
ours-cor&$ 0.000$&$0.000$&$0.017$&$0.017$&$0.945$\tabularnewline
ours-mis1&$ 0.000$&$0.000$&$0.017$&$0.017$&$0.946$\tabularnewline
ours-mis2&$ 0.000$&$0.000$&$0.017$&$0.017$&$0.945$\tabularnewline
ours-mis3&$ 0.000$&$0.000$&$0.017$&$0.017$&$0.948$\tabularnewline
ipw&$ 0.000$&$0.000$&$0.017$&$0.017$&$0.943$\tabularnewline
pooled&$-0.003$&$0.000$&$0.014$&$0.014$&$0.950$\tabularnewline
dind&$ 0.023$&$0.001$&$0.012$&$0.012$&$0.527$\tabularnewline

\hline
\multicolumn{6}{c}{Physically active, $\beta_3 = -0.032$}\\ \hline
ours-cor&$-0.001$&$0.001$&$0.027$&$0.028$&$0.948$\tabularnewline
ours-mis1&$-0.001$&$0.001$&$0.027$&$0.028$&$0.949$\tabularnewline
ours-mis2&$-0.001$&$0.001$&$0.027$&$0.028$&$0.947$\tabularnewline
ours-mis3&$-0.001$&$0.001$&$0.027$&$0.028$&$0.949$\tabularnewline
ipw&$-0.001$&$0.001$&$0.027$&$0.028$&$0.943$\tabularnewline
pooled&$ 0.006$&$0.001$&$0.026$&$0.028$&$0.968$\tabularnewline
dind&$-0.002$&$0.001$&$0.024$&$0.024$&$0.948$\tabularnewline

\hline
\multicolumn{6}{c}{SNP1, $\beta_4 = -0.032$}\\ \hline
ours-cor&$ 0.000$&$0.000$&$0.014$&$0.014$&$0.938$\tabularnewline
ours-mis1&$ 0.000$&$0.000$&$0.014$&$0.013$&$0.939$\tabularnewline
ours-mis2&$ 0.000$&$0.000$&$0.014$&$0.014$&$0.939$\tabularnewline
ours-mis3&$ 0.000$&$0.000$&$0.014$&$0.013$&$0.938$\tabularnewline
ipw&$ 0.000$&$0.000$&$0.014$&$0.014$&$0.940$\tabularnewline
pooled&$-0.058$&$0.003$&$0.012$&$0.012$&$0.002$\tabularnewline
dind&$ 0.006$&$0.000$&$0.011$&$0.011$&$0.919$\tabularnewline

\hline
\multicolumn{6}{c}{SNP2, $\beta_5 = -0.040$}\\ \hline
ours-cor&$ 0.000$&$0.000$&$0.014$&$0.014$&$0.951$\tabularnewline
ours-mis1&$ 0.000$&$0.000$&$0.014$&$0.014$&$0.953$\tabularnewline
ours-mis2&$ 0.000$&$0.000$&$0.014$&$0.014$&$0.954$\tabularnewline
ours-mis3&$ 0.000$&$0.000$&$0.014$&$0.014$&$0.954$\tabularnewline
ipw&$ 0.000$&$0.000$&$0.014$&$0.014$&$0.955$\tabularnewline
pooled&$ 0.004$&$0.000$&$0.012$&$0.012$&$0.925$\tabularnewline
dind&$-0.004$&$0.000$&$0.010$&$0.010$&$0.926$\tabularnewline

\hline
\multicolumn{6}{c}{Active$\times$SNP1, $\beta_6 = -0.021$}\\ \hline
ours-cor&$ 0.000$&$0.000$&$0.021$&$0.021$&$0.948$\tabularnewline
ours-mis1&$ 0.000$&$0.000$&$0.021$&$0.021$&$0.947$\tabularnewline
ours-mis2&$ 0.000$&$0.000$&$0.021$&$0.021$&$0.946$\tabularnewline
ours-mis3&$ 0.000$&$0.000$&$0.021$&$0.021$&$0.946$\tabularnewline
ipw&$ 0.001$&$0.000$&$0.022$&$0.021$&$0.943$\tabularnewline
pooled&$ 0.000$&$0.000$&$0.019$&$0.019$&$0.961$\tabularnewline
dind&$-0.002$&$0.000$&$0.017$&$0.017$&$0.943$\tabularnewline

\hline
\end{tabular} }
\end{center}
\end{table}
 \thispagestyle{empty}
 \clearpage

\begin{table}[htbp] 
\caption{Simulation results, averaged over 1000 simulations, for estimating the effect of covariates on a the simulated log(BMI) outcomes. The SNPs used had \textbf{low} MAF, the effect of SNP1 on the disease distribution was \textbf{low}, and its effect on  the selection bias function was \textbf{high}.  }
\label{tab:log}
\begin{center}
 \scalebox{0.7}{
\begin{tabular}{lrrrrr}
\hline
\multicolumn{1}{l}{Estimator}&\multicolumn{1}{c}{Bias}&\multicolumn{1}{c}{MSE}&\multicolumn{1}{c}{emp sd}&\multicolumn{1}{c}{est sd}&\multicolumn{1}{c}{coverage}\tabularnewline
\hline
\multicolumn{6}{c}{Intercept, $\beta_0 = 3.077$}\\ \hline 
ours-cor&$ 0.001$&$0.002$&$0.047$&$0.048$&$0.952$\tabularnewline
ours-mis1&$ 0.001$&$0.002$&$0.047$&$0.049$&$0.948$\tabularnewline
ours-mis2&$ 0.001$&$0.002$&$0.047$&$0.048$&$0.952$\tabularnewline
ours-mis3&$ 0.001$&$0.002$&$0.047$&$0.049$&$0.947$\tabularnewline
ipw&$ 0.001$&$0.002$&$0.048$&$0.049$&$0.946$\tabularnewline
pooled&$-0.060$&$0.006$&$0.047$&$0.047$&$0.753$\tabularnewline
dind&$ 0.031$&$0.003$&$0.043$&$0.043$&$0.885$\tabularnewline

\hline
\multicolumn{6}{c}{Age, $\beta_1 = 0.002$}\\ \hline
ours-cor&$ 0.000$&$0.000$&$0.001$&$0.001$&$0.944$\tabularnewline
ours-mis1&$ 0.000$&$0.000$&$0.001$&$0.001$&$0.949$\tabularnewline
ours-mis2&$ 0.000$&$0.000$&$0.001$&$0.001$&$0.944$\tabularnewline
ours-mis3&$ 0.000$&$0.000$&$0.001$&$0.001$&$0.949$\tabularnewline
ipw&$ 0.000$&$0.000$&$0.001$&$0.001$&$0.944$\tabularnewline
pooled&$ 0.000$&$0.000$&$0.001$&$0.001$&$0.944$\tabularnewline
dind&$ 0.000$&$0.000$&$0.001$&$0.001$&$0.946$\tabularnewline

\hline
\multicolumn{6}{c}{Smoker, $\beta_2 = -0.012$}\\ \hline
ours-cor&$ 0.001$&$0.000$&$0.017$&$0.017$&$0.934$\tabularnewline
ours-mis1&$ 0.001$&$0.000$&$0.017$&$0.017$&$0.942$\tabularnewline
ours-mis2&$ 0.001$&$0.000$&$0.017$&$0.017$&$0.934$\tabularnewline
ours-mis3&$ 0.001$&$0.000$&$0.017$&$0.017$&$0.943$\tabularnewline
ipw&$ 0.001$&$0.000$&$0.017$&$0.017$&$0.940$\tabularnewline
pooled&$-0.006$&$0.000$&$0.015$&$0.015$&$0.938$\tabularnewline
dind&$ 0.020$&$0.001$&$0.014$&$0.014$&$0.694$\tabularnewline

\hline
\multicolumn{6}{c}{Physically active, $\beta_3 = -0.032$}\\ \hline
ours-cor&$ 0.000$&$0.000$&$0.015$&$0.015$&$0.942$\tabularnewline
ours-mis1&$ 0.000$&$0.000$&$0.015$&$0.015$&$0.947$\tabularnewline
ours-mis2&$ 0.000$&$0.000$&$0.015$&$0.015$&$0.942$\tabularnewline
ours-mis3&$ 0.000$&$0.000$&$0.015$&$0.015$&$0.947$\tabularnewline
ipw&$ 0.000$&$0.000$&$0.015$&$0.015$&$0.945$\tabularnewline
pooled&$ 0.006$&$0.000$&$0.013$&$0.015$&$0.970$\tabularnewline
dind&$-0.005$&$0.000$&$0.012$&$0.013$&$0.960$\tabularnewline

\hline
\multicolumn{6}{c}{SNP1, $\beta_4 = -0.032$}\\ \hline
ours-cor&$-0.004$&$0.004$&$0.060$&$0.064$&$0.966$\tabularnewline
ours-mis1&$-0.008$&$0.004$&$0.066$&$0.055$&$0.900$\tabularnewline
ours-mis2&$-0.004$&$0.004$&$0.060$&$0.064$&$0.965$\tabularnewline
ours-mis3&$-0.008$&$0.004$&$0.066$&$0.055$&$0.902$\tabularnewline
ipw&$-0.009$&$0.005$&$0.069$&$0.065$&$0.945$\tabularnewline
pooled&$-0.445$&$0.212$&$0.121$&$0.045$&$0.004$\tabularnewline
dind&$-0.440$&$0.204$&$0.104$&$0.041$&$0.001$\tabularnewline

\hline
\multicolumn{6}{c}{SNP2, $\beta_5 = -0.040$}\\ \hline
ours-cor&$ 0.001$&$0.001$&$0.036$&$0.035$&$0.932$\tabularnewline
ours-mis1&$ 0.001$&$0.001$&$0.036$&$0.035$&$0.935$\tabularnewline
ours-mis2&$ 0.001$&$0.001$&$0.036$&$0.035$&$0.933$\tabularnewline
ours-mis3&$ 0.001$&$0.001$&$0.036$&$0.035$&$0.936$\tabularnewline
ipw&$ 0.001$&$0.001$&$0.036$&$0.035$&$0.934$\tabularnewline
pooled&$ 0.008$&$0.001$&$0.035$&$0.035$&$0.943$\tabularnewline
dind&$-0.001$&$0.001$&$0.032$&$0.032$&$0.947$\tabularnewline

\hline
\multicolumn{6}{c}{Active$\times$SNP1, $\beta_6 = -0.021$}\\ \hline
ours-cor&$-0.001$&$0.006$&$0.079$&$0.098$&$0.975$\tabularnewline
ours-mis1&$-0.005$&$0.011$&$0.105$&$0.087$&$0.918$\tabularnewline
ours-mis2&$-0.001$&$0.006$&$0.079$&$0.098$&$0.978$\tabularnewline
ours-mis3&$-0.005$&$0.011$&$0.105$&$0.087$&$0.918$\tabularnewline
ipw&$-0.006$&$0.013$&$0.115$&$0.101$&$0.941$\tabularnewline
pooled&$ 0.034$&$0.041$&$0.200$&$0.073$&$0.497$\tabularnewline
dind&$ 0.035$&$0.031$&$0.172$&$0.066$&$0.509$\tabularnewline

\hline
\end{tabular} }
\end{center}
\end{table}
 \thispagestyle{empty}
 \clearpage

\begin{table}[htbp] 
\caption{Simulation results, averaged over 1000 simulations, for estimating the effect of covariates on a the simulated log(BMI) outcomes. The SNPs used had \textbf{high} MAF, the effect of SNP1 on the disease distribution was \textbf{low}, and its effect on  the selection bias function was \textbf{high}.  }
\label{tab:log}
\begin{center}
 \scalebox{0.7}{
\begin{tabular}{lrrrrr}
\hline
\multicolumn{1}{l}{Estimator}&\multicolumn{1}{c}{Bias}&\multicolumn{1}{c}{MSE}&\multicolumn{1}{c}{emp sd}&\multicolumn{1}{c}{est sd}&\multicolumn{1}{c}{coverage}\tabularnewline
\hline
\multicolumn{6}{c}{Intercept, $\beta_0 = 3.077$}\\ \hline 
ours-cor&$-0.003$&$0.004$&$0.061$&$0.073$&$0.984$\tabularnewline
ours-mis1&$-0.003$&$0.005$&$0.068$&$0.081$&$0.981$\tabularnewline
ours-mis2&$-0.003$&$0.004$&$0.061$&$0.073$&$0.984$\tabularnewline
ours-mis3&$-0.003$&$0.005$&$0.068$&$0.081$&$0.982$\tabularnewline
ipw&$-0.005$&$0.007$&$0.082$&$0.079$&$0.927$\tabularnewline
pooled&$-0.092$&$0.039$&$0.175$&$0.175$&$0.904$\tabularnewline
dind&$ 0.574$&$0.340$&$0.101$&$0.094$&$0.000$\tabularnewline

\hline
\multicolumn{6}{c}{Age, $\beta_1 = 0.002$}\\ \hline
ours-cor&$ 0.000$&$0.000$&$0.001$&$0.001$&$0.985$\tabularnewline
ours-mis1&$ 0.000$&$0.000$&$0.001$&$0.002$&$0.980$\tabularnewline
ours-mis2&$ 0.000$&$0.000$&$0.001$&$0.001$&$0.985$\tabularnewline
ours-mis3&$ 0.000$&$0.000$&$0.001$&$0.002$&$0.980$\tabularnewline
ipw&$ 0.000$&$0.000$&$0.002$&$0.002$&$0.937$\tabularnewline
pooled&$ 0.000$&$0.000$&$0.004$&$0.003$&$0.938$\tabularnewline
dind&$ 0.000$&$0.000$&$0.002$&$0.002$&$0.931$\tabularnewline

\hline
\multicolumn{6}{c}{Smoker, $\beta_2 = -0.012$}\\ \hline
ours-cor&$-0.001$&$0.001$&$0.023$&$0.024$&$0.964$\tabularnewline
ours-mis1&$-0.001$&$0.001$&$0.025$&$0.026$&$0.959$\tabularnewline
ours-mis2&$-0.001$&$0.001$&$0.023$&$0.025$&$0.968$\tabularnewline
ours-mis3&$-0.001$&$0.001$&$0.025$&$0.026$&$0.964$\tabularnewline
ipw&$-0.001$&$0.001$&$0.025$&$0.027$&$0.958$\tabularnewline
pooled&$-0.100$&$0.012$&$0.049$&$0.051$&$0.502$\tabularnewline
dind&$ 0.075$&$0.006$&$0.028$&$0.027$&$0.224$\tabularnewline

\hline
\multicolumn{6}{c}{Physically active, $\beta_3 = -0.032$}\\ \hline
ours-cor&$ 0.001$&$0.001$&$0.027$&$0.029$&$0.967$\tabularnewline
ours-mis1&$ 0.000$&$0.001$&$0.031$&$0.046$&$0.996$\tabularnewline
ours-mis2&$ 0.001$&$0.001$&$0.027$&$0.029$&$0.967$\tabularnewline
ours-mis3&$ 0.000$&$0.001$&$0.031$&$0.046$&$0.996$\tabularnewline
ipw&$ 0.000$&$0.001$&$0.031$&$0.031$&$0.958$\tabularnewline
pooled&$ 0.004$&$0.003$&$0.055$&$0.088$&$0.995$\tabularnewline
dind&$-0.067$&$0.008$&$0.062$&$0.047$&$0.653$\tabularnewline

\hline
\multicolumn{6}{c}{SNP1, $\beta_4 = -0.032$}\\ \hline
ours-cor&$ 0.000$&$0.000$&$0.017$&$0.019$&$0.968$\tabularnewline
ours-mis1&$-0.001$&$0.000$&$0.019$&$0.021$&$0.967$\tabularnewline
ours-mis2&$ 0.000$&$0.000$&$0.017$&$0.019$&$0.968$\tabularnewline
ours-mis3&$-0.001$&$0.000$&$0.019$&$0.021$&$0.967$\tabularnewline
ipw&$-0.001$&$0.000$&$0.020$&$0.021$&$0.961$\tabularnewline
pooled&$-0.442$&$0.197$&$0.041$&$0.042$&$0.000$\tabularnewline
dind&$-0.411$&$0.170$&$0.026$&$0.022$&$0.000$\tabularnewline

\hline
\multicolumn{6}{c}{SNP2, $\beta_5 = -0.040$}\\ \hline
ours-cor&$ 0.002$&$0.000$&$0.019$&$0.019$&$0.940$\tabularnewline
ours-mis1&$ 0.002$&$0.000$&$0.021$&$0.020$&$0.945$\tabularnewline
ours-mis2&$ 0.002$&$0.000$&$0.019$&$0.019$&$0.940$\tabularnewline
ours-mis3&$ 0.002$&$0.000$&$0.021$&$0.020$&$0.945$\tabularnewline
ipw&$ 0.002$&$0.000$&$0.021$&$0.021$&$0.944$\tabularnewline
pooled&$ 0.037$&$0.003$&$0.042$&$0.042$&$0.867$\tabularnewline
dind&$-0.020$&$0.001$&$0.023$&$0.023$&$0.861$\tabularnewline

\hline
\multicolumn{6}{c}{Active$\times$SNP1, $\beta_6 = -0.021$}\\ \hline
ours-cor&$ 0.000$&$0.001$&$0.024$&$0.029$&$0.984$\tabularnewline
ours-mis1&$ 0.000$&$0.001$&$0.027$&$0.031$&$0.983$\tabularnewline
ours-mis2&$ 0.000$&$0.001$&$0.024$&$0.029$&$0.984$\tabularnewline
ours-mis3&$ 0.000$&$0.001$&$0.027$&$0.031$&$0.983$\tabularnewline
ipw&$ 0.000$&$0.001$&$0.031$&$0.031$&$0.953$\tabularnewline
pooled&$ 0.039$&$0.006$&$0.070$&$0.068$&$0.893$\tabularnewline
dind&$ 0.039$&$0.004$&$0.049$&$0.036$&$0.735$\tabularnewline

\hline
\end{tabular} }
\end{center}
\end{table}
 \thispagestyle{empty}
 \clearpage

\begin{table}[htbp] 
\caption{Simulation results, averaged over 1000 simulations, for estimating the effect of covariates on a the simulated log(BMI) outcomes. The SNPs used had \textbf{low} MAF, the effect of SNP1 on the disease distribution was \textbf{high}, and its effect on  the selection bias function was \textbf{high}.  }
\label{tab:log}
\begin{center}
 \scalebox{0.7}{
\begin{tabular}{lrrrrr}
\hline
\multicolumn{1}{l}{Estimator}&\multicolumn{1}{c}{Bias}&\multicolumn{1}{c}{MSE}&\multicolumn{1}{c}{emp sd}&\multicolumn{1}{c}{est sd}&\multicolumn{1}{c}{coverage}\tabularnewline
\hline
\multicolumn{6}{c}{Intercept, $\beta_0 = 3.077$}\\ \hline 
ours-cor&$-0.003$&$0.002$&$0.050$&$0.048$&$0.947$\tabularnewline
ours-mis1&$-0.003$&$0.003$&$0.051$&$0.050$&$0.941$\tabularnewline
ours-mis2&$-0.003$&$0.002$&$0.050$&$0.048$&$0.947$\tabularnewline
ours-mis3&$-0.003$&$0.003$&$0.051$&$0.050$&$0.943$\tabularnewline
ipw&$-0.003$&$0.003$&$0.052$&$0.050$&$0.935$\tabularnewline
pooled&$-0.064$&$0.006$&$0.049$&$0.048$&$0.729$\tabularnewline
dind&$ 0.031$&$0.003$&$0.044$&$0.044$&$0.885$\tabularnewline

\hline
\multicolumn{6}{c}{Age, $\beta_1 = 0.002$}\\ \hline

ours-cor&$ 0.000$&$0.000$&$0.001$&$0.001$&$0.945$\tabularnewline
ours-mis1&$ 0.000$&$0.000$&$0.001$&$0.001$&$0.944$\tabularnewline
ours-mis2&$ 0.000$&$0.000$&$0.001$&$0.001$&$0.945$\tabularnewline
ours-mis3&$ 0.000$&$0.000$&$0.001$&$0.001$&$0.942$\tabularnewline
ipw&$ 0.000$&$0.000$&$0.001$&$0.001$&$0.934$\tabularnewline
pooled&$ 0.000$&$0.000$&$0.001$&$0.001$&$0.947$\tabularnewline
dind&$ 0.000$&$0.000$&$0.001$&$0.001$&$0.952$\tabularnewline

\hline
\multicolumn{6}{c}{Smoker, $\beta_2 = -0.012$}\\ \hline

ours-cor&$ 0.000$&$0.000$&$0.017$&$0.017$&$0.946$\tabularnewline
ours-mis1&$ 0.000$&$0.000$&$0.018$&$0.018$&$0.948$\tabularnewline
ours-mis2&$ 0.000$&$0.000$&$0.017$&$0.017$&$0.948$\tabularnewline
ours-mis3&$ 0.000$&$0.000$&$0.018$&$0.018$&$0.949$\tabularnewline
ipw&$ 0.000$&$0.000$&$0.018$&$0.018$&$0.949$\tabularnewline
pooled&$-0.002$&$0.000$&$0.016$&$0.016$&$0.950$\tabularnewline
dind&$ 0.026$&$0.001$&$0.014$&$0.015$&$0.568$\tabularnewline

\hline
\multicolumn{6}{c}{Physically active, $\beta_3 = -0.032$}\\ \hline
ours-cor&$ 0.000$&$0.000$&$0.015$&$0.015$&$0.945$\tabularnewline
ours-mis1&$ 0.000$&$0.000$&$0.015$&$0.015$&$0.946$\tabularnewline
ours-mis2&$ 0.000$&$0.000$&$0.015$&$0.015$&$0.945$\tabularnewline
ours-mis3&$ 0.000$&$0.000$&$0.015$&$0.015$&$0.947$\tabularnewline
ipw&$ 0.000$&$0.000$&$0.015$&$0.015$&$0.945$\tabularnewline
pooled&$ 0.006$&$0.000$&$0.013$&$0.016$&$0.967$\tabularnewline
dind&$-0.006$&$0.000$&$0.013$&$0.014$&$0.952$\tabularnewline

\hline
\multicolumn{6}{c}{SNP1, $\beta_4 = -0.032$}\\ \hline
ours-cor&$-0.012$&$0.007$&$0.081$&$0.093$&$0.975$\tabularnewline
ours-mis1&$-0.019$&$0.009$&$0.095$&$0.067$&$0.839$\tabularnewline
ours-mis2&$-0.012$&$0.007$&$0.081$&$0.093$&$0.975$\tabularnewline
ours-mis3&$-0.019$&$0.009$&$0.094$&$0.068$&$0.842$\tabularnewline
ipw&$-0.020$&$0.011$&$0.101$&$0.096$&$0.954$\tabularnewline
pooled&$-0.529$&$0.286$&$0.073$&$0.035$&$0.000$\tabularnewline
dind&$-0.474$&$0.228$&$0.063$&$0.032$&$0.000$\tabularnewline

\hline
\multicolumn{6}{c}{SNP2, $\beta_5 = -0.040$}\\ \hline
ours-cor&$ 0.001$&$0.001$&$0.036$&$0.035$&$0.931$\tabularnewline
ours-mis1&$ 0.001$&$0.001$&$0.038$&$0.036$&$0.935$\tabularnewline
ours-mis2&$ 0.001$&$0.001$&$0.036$&$0.035$&$0.928$\tabularnewline
ours-mis3&$ 0.001$&$0.001$&$0.038$&$0.036$&$0.935$\tabularnewline
ipw&$ 0.000$&$0.001$&$0.038$&$0.036$&$0.935$\tabularnewline
pooled&$ 0.005$&$0.001$&$0.038$&$0.037$&$0.948$\tabularnewline
dind&$-0.004$&$0.001$&$0.035$&$0.033$&$0.944$\tabularnewline

\hline
\multicolumn{6}{c}{Active$\times$SNP1, $\beta_6 = -0.021$}\\ \hline
ours-cor&$ 0.000$&$0.006$&$0.077$&$0.141$&$0.999$\tabularnewline
ours-mis1&$-0.007$&$0.018$&$0.133$&$0.106$&$0.885$\tabularnewline
ours-mis2&$ 0.000$&$0.006$&$0.077$&$0.141$&$0.999$\tabularnewline
ours-mis3&$-0.007$&$0.018$&$0.133$&$0.106$&$0.889$\tabularnewline
ipw&$-0.009$&$0.025$&$0.156$&$0.147$&$0.944$\tabularnewline
pooled&$-0.008$&$0.016$&$0.127$&$0.057$&$0.625$\tabularnewline
dind&$-0.004$&$0.012$&$0.108$&$0.051$&$0.659$\tabularnewline

\hline
\end{tabular} }
\end{center}
\end{table}
 \thispagestyle{empty}
 \clearpage

\begin{table}[htbp] 
\caption{Simulation results, averaged over 1000 simulations, for estimating the effect of covariates on a the simulated log(BMI) outcomes. The SNPs used had \textbf{high} MAF, the effect of SNP1 on the disease distribution was \textbf{high}, and its effect on  the selection bias function was \textbf{high}.  }
\label{tab:log}
\begin{center}
 \scalebox{0.7}{
\begin{tabular}{lrrrrr}
\hline
\multicolumn{1}{l}{Estimator}&\multicolumn{1}{c}{Bias}&\multicolumn{1}{c}{MSE}&\multicolumn{1}{c}{emp sd}&\multicolumn{1}{c}{est sd}&\multicolumn{1}{c}{coverage}\tabularnewline
\hline
\multicolumn{6}{c}{Intercept, $\beta_0 = 3.077$}\\ \hline 
ours-cor&$ 0.000$&$0.004$&$0.065$&$0.084$&$0.989$\tabularnewline
ours-mis1&$ 0.001$&$0.005$&$0.068$&$0.088$&$0.988$\tabularnewline
ours-mis2&$ 0.000$&$0.004$&$0.065$&$0.084$&$0.989$\tabularnewline
ours-mis3&$ 0.001$&$0.005$&$0.068$&$0.088$&$0.988$\tabularnewline
ipw&$-0.001$&$0.008$&$0.092$&$0.094$&$0.951$\tabularnewline
pooled&$ 0.077$&$0.037$&$0.176$&$0.186$&$0.941$\tabularnewline
dind&$ 0.351$&$0.130$&$0.083$&$0.083$&$0.019$\tabularnewline

\hline
\multicolumn{6}{c}{Age, $\beta_1 = 0.002$}\\ \hline
ours-cor&$ 0.000$&$0.000$&$0.001$&$0.002$&$0.996$\tabularnewline
ours-mis1&$ 0.000$&$0.000$&$0.001$&$0.002$&$0.994$\tabularnewline
ours-mis2&$ 0.000$&$0.000$&$0.001$&$0.002$&$0.996$\tabularnewline
ours-mis3&$ 0.000$&$0.000$&$0.001$&$0.002$&$0.995$\tabularnewline
ipw&$ 0.000$&$0.000$&$0.002$&$0.002$&$0.954$\tabularnewline
pooled&$ 0.000$&$0.000$&$0.004$&$0.004$&$0.952$\tabularnewline
dind&$ 0.000$&$0.000$&$0.002$&$0.002$&$0.960$\tabularnewline

\hline
\multicolumn{6}{c}{Smoker, $\beta_2 = -0.012$}\\ \hline
ours-cor&$ 0.000$&$0.001$&$0.029$&$0.028$&$0.937$\tabularnewline
ours-mis1&$ 0.000$&$0.001$&$0.031$&$0.029$&$0.930$\tabularnewline
ours-mis2&$ 0.000$&$0.001$&$0.029$&$0.029$&$0.938$\tabularnewline
ours-mis3&$ 0.000$&$0.001$&$0.030$&$0.030$&$0.940$\tabularnewline
ipw&$-0.001$&$0.001$&$0.034$&$0.032$&$0.930$\tabularnewline
pooled&$-0.057$&$0.006$&$0.054$&$0.053$&$0.802$\tabularnewline
dind&$ 0.117$&$0.014$&$0.025$&$0.024$&$0.003$\tabularnewline

\hline
\multicolumn{6}{c}{Physically active, $\beta_3 = -0.032$}\\ \hline
ours-cor&$-0.001$&$0.001$&$0.028$&$0.033$&$0.974$\tabularnewline
ours-mis1&$-0.001$&$0.001$&$0.029$&$0.038$&$0.987$\tabularnewline
ours-mis2&$-0.001$&$0.001$&$0.028$&$0.033$&$0.974$\tabularnewline
ours-mis3&$-0.001$&$0.001$&$0.029$&$0.038$&$0.987$\tabularnewline
ipw&$-0.002$&$0.001$&$0.034$&$0.035$&$0.948$\tabularnewline
pooled&$ 0.017$&$0.004$&$0.065$&$0.106$&$0.997$\tabularnewline
dind&$-0.040$&$0.005$&$0.059$&$0.047$&$0.818$\tabularnewline

\hline
\multicolumn{6}{c}{SNP1, $\beta_4 = -0.032$}\\ \hline
ours-cor&$-0.002$&$0.001$&$0.023$&$0.027$&$0.978$\tabularnewline
ours-mis1&$-0.002$&$0.001$&$0.023$&$0.021$&$0.937$\tabularnewline
ours-mis2&$-0.002$&$0.001$&$0.023$&$0.027$&$0.978$\tabularnewline
ours-mis3&$-0.002$&$0.001$&$0.023$&$0.021$&$0.937$\tabularnewline
ipw&$-0.003$&$0.001$&$0.027$&$0.028$&$0.955$\tabularnewline
pooled&$-0.596$&$0.356$&$0.038$&$0.045$&$0.000$\tabularnewline
dind&$-0.168$&$0.029$&$0.027$&$0.021$&$0.000$\tabularnewline

\hline
\multicolumn{6}{c}{SNP2, $\beta_5 = -0.040$}\\ \hline
ours-cor&$ 0.001$&$0.000$&$0.022$&$0.022$&$0.970$\tabularnewline
ours-mis1&$ 0.001$&$0.000$&$0.022$&$0.023$&$0.968$\tabularnewline
ours-mis2&$ 0.001$&$0.000$&$0.022$&$0.022$&$0.970$\tabularnewline
ours-mis3&$ 0.001$&$0.000$&$0.022$&$0.023$&$0.969$\tabularnewline
ipw&$ 0.001$&$0.001$&$0.025$&$0.025$&$0.959$\tabularnewline
pooled&$ 0.025$&$0.002$&$0.043$&$0.045$&$0.927$\tabularnewline
dind&$-0.033$&$0.002$&$0.021$&$0.020$&$0.614$\tabularnewline

\hline
\multicolumn{6}{c}{Active$\times$SNP1, $\beta_6 = -0.021$}\\ \hline
ours-cor&$ 0.001$&$0.001$&$0.031$&$0.040$&$0.991$\tabularnewline
ours-mis1&$ 0.001$&$0.001$&$0.031$&$0.032$&$0.953$\tabularnewline
ours-mis2&$ 0.001$&$0.001$&$0.031$&$0.040$&$0.991$\tabularnewline
ours-mis3&$ 0.001$&$0.001$&$0.031$&$0.032$&$0.953$\tabularnewline
ipw&$ 0.002$&$0.002$&$0.041$&$0.042$&$0.954$\tabularnewline
pooled&$ 0.012$&$0.004$&$0.064$&$0.072$&$0.967$\tabularnewline
dind&$ 0.001$&$0.002$&$0.039$&$0.032$&$0.892$\tabularnewline

\hline
\end{tabular} }
\end{center}
\end{table}
 \thispagestyle{empty}
\clearpage

\end{document}